\documentclass[10pt,journal,compsoc]{IEEEtran}

\newif\iftr
\newif\ifcnf

\trfalse
\cnftrue

\trtrue
\cnffalse

\usepackage{etex}

\newif\ifall    
\newif\ifsq     
\newif\ifnonb   
\newif\iftodos

\newif\ifsqCAP
\newif\ifsqVS
\newif\ifsqEN
\newif\ifsqTIT

\ifcnf
\sqtrue
\sqCAPtrue
\sqENtrue
\sqVStrue
\sqTITtrue

\else

\sqfalse
\sqCAPfalse
\sqENfalse
\sqVSfalse
\sqTITfalse
\fi

\newcommand{\vspaceSQ}[1]{\ifsqVS\vspace{#1}\fi}
\newcommand{\enlargeSQ}[1]{\ifsqEN\enlargethispage{\baselineskip}\fi}

\newcommand{\tr}[1]{\iftr #1 \fi}
\newcommand{\all}[1]{\ifall #1 \fi}

\ifcnf
\usepackage[font={scriptsize}]{caption}
\usepackage[font={scriptsize}]{subcaption}
\else
\usepackage{caption}
\usepackage{subcaption}
\fi

\usepackage{graphicx}
\usepackage{booktabs}
\usepackage{epstopdf}
\usepackage{float}
\usepackage{url}
\usepackage{placeins}
\usepackage{amsmath,amssymb,amsfonts}
\usepackage{mathtools,mathrsfs}
\usepackage{multirow}
\usepackage{rotating}
\usepackage{pbox}
\usepackage[normalem]{ulem}

\usepackage{inconsolata}
\usepackage{listings}

\newcommand{\subparagraph}{}
\usepackage[compact]{titlesec}

\usepackage{etoolbox}

\makeatletter
\patchcmd{\ttlh@hang}{\parindent\z@}{\parindent\z@\leavevmode}{}{}
\patchcmd{\ttlh@hang}{\noindent}{}{}{}
\makeatother

\usepackage{cleveref}
\usepackage[utf8]{inputenc}
\crefname{section}{§}{§§}
\Crefname{section}{§}{§§}

\usepackage[10pt]{moresize}

\usepackage{xcolor}
\definecolor{darkgrey}{RGB}{70,70,70}
\definecolor{lightgrey}{RGB}{200,200,200}
\definecolor{lyellow}{RGB}{255,255,200}

\usepackage{enumitem}

\usepackage{makecell}

\usepackage[linesnumbered,ruled,vlined]{algorithm2e}
\SetAlFnt{\scriptsize\tt}
\SetAlCapFnt{\scriptsize\sf}
\SetAlCapNameFnt{\scriptsize\sf}

\usepackage{algpseudocode}
\algblock{ParFor}{EndParFor}
\algnewcommand\algorithmicparfor{\textbf{parfor}}
\algnewcommand\algorithmicpardo{\textbf{do}}
\algnewcommand\algorithmicendparfor{\textbf{end\ parfor}}
\algrenewtext{ParFor}[1]{\algorithmicparfor\ #1\ \algorithmicpardo}
\algrenewtext{EndParFor}{\algorithmicendparfor}

\newcommand{\maciej}[1]{\textcolor{blue}{[Maciej: #1]}}

\newcommand{\marc}[1]{\textcolor{blue}{[Marc: #1]}}

\newcommand{\macb}[1]{\textbf{\textsf{#1}}}

\usepackage[hang,flushmargin]{footmisc}

\usepackage{bm}

\usepackage{scalerel,stackengine}
\stackMath
\newcommand\rwh[1]{%
\savestack{\tmpbox}{\stretchto{%
  \scaleto{%
      \scalerel*[\widthof{\ensuremath{#1}}]{\kern-.6pt\bigwedge\kern-.6pt}%
          {\rule[-\textheight/2]{1ex}{\textheight}}
            }{\textheight}%
}{0.5ex}}%
\stackon[1pt]{#1}{\tmpbox}%
}

\def\HiLiGA{\leavevmode\rlap{\hbox to \hsize{\color{black!10}\leaders\hrule height 1\baselineskip depth 1ex\hfill}}}
\def\HiLiGB{\leavevmode\rlap{\hbox to \hsize{\color{black!25}\leaders\hrule height 1\baselineskip depth 1ex\hfill}}}
\def\HiLiGC{\leavevmode\rlap{\hbox to \hsize{\color{black!40}\leaders\hrule height 1\baselineskip depth 1ex\hfill}}}
\def\HiLiGD{\leavevmode\rlap{\hbox to \hsize{\color{black!55}\leaders\hrule height 1\baselineskip depth 1ex\hfill}}}
\def\HiLiGE{\leavevmode\rlap{\hbox to \hsize{\color{black!70}\leaders\hrule height 1\baselineskip depth 1ex\hfill}}}
\def\HiLiGF{\leavevmode\rlap{\hbox to \hsize{\color{black!85}\leaders\hrule height 1\baselineskip depth 1ex\hfill}}}

\usepackage{algpseudocode}
\usepackage{tikz}
\usetikzlibrary{calc}
\usepackage{xcolor}
\makeatletter

\usepackage{fontawesome}
\usepackage{pifont}
\usepackage{textcomp}

\usepackage{xcolor}
\usepackage{soul}
\usepackage{dblfloatfix}

\usepackage{colortbl}

\ifcnf
\usepackage[firstinits=true]{biblatex}
\fi
\ifcnf

\fi

\newcommand{\noAnswer}{\textcolor{lightgray}{\faQuestionCircle}}
\newcommand{\comment}[1]{\ignorespaces}

\iftr

\renewcommand{\hl}[1]{#1}
\renewcommand{\rowcolor}[1]{}
\renewcommand{\marginpar}[1]{}
\renewcommand{\colorbox}[2]{#2}

\fi

\newcolumntype{y}{>{}l}

\IEEEaftertitletext{\vspace{-2\baselineskip}}

\newif\ifHL
\HLfalse
\newcommand{\HL}[1]{\ifHL #1 \fi}

\begin{document}

\title{\vspaceSQ{-0.5em}Practice of Streaming Processing of Dynamic Graphs: Concepts, Models, and Systems\vspaceSQ{-0.5em}}

\author{Maciej Besta$^1$, Marc Fischer$^2$, Vasiliki Kalavri$^3$, Michael Kapralov$^4$, Torsten Hoefler$^1$\\
{\footnotesize\vspaceSQ{0.25em}{$^1$ETH Zurich\quad\quad
$^2$PRODYNA (Schweiz) AG\quad\quad $^3$Boston University\quad\quad
$^4$School of Computer and Communication Sciences, EPFL\vspaceSQ{-0.25em}}}}

\IEEEtitleabstractindextext{%
\begin{abstract}
Graph processing has become an important part of various areas of computing,
including machine learning, medical applications, social network analysis,
computational sciences, and others. A growing amount of the associated graph
processing workloads are \emph{dynamic}, with millions of edges added or
removed per second. Graph streaming frameworks are specifically crafted to
enable the processing of such highly dynamic workloads. Recent years have seen
the development of many such frameworks. However, they differ in their general
architectures (with key details such as the support for the concurrent execution
of graph updates and queries, or the incorporated graph data organization), the types of
updates and workloads allowed, and many others. To facilitate the understanding
of this growing field, we provide the first analysis and taxonomy of dynamic
and streaming graph processing. We focus on identifying the fundamental system
designs and on understanding their support for concurrency, and
for different graph updates as well as analytics workloads. We also crystallize
  the meaning of different concepts associated with streaming graph processing,
  such as dynamic, temporal, online, and time-evolving graphs, edge-centric
  processing, models for the maintenance of updates, and graph databases.
  Moreover, we provide a bridge with the very rich landscape of graph streaming
  theory by giving a broad overview of recent theoretical related advances, and
  by discussing which graph streaming models and settings could be helpful in
  developing more powerful streaming frameworks and designs. We also outline
  graph streaming workloads and research challenges. 
%
%
\vspace{-0.5em}
\end{abstract}

}

\maketitle

\IEEEdisplaynontitleabstractindextext
\IEEEpeerreviewmaketitle

\iftr
\else
{\vspace{-1.0em}\noindent \textbf{An extended version of this paper is available at\\ \url{https://arxiv.org/abs/1912.12740}}\vspace{1em}}
\fi

\vspaceSQ{-1em}
\section{Introduction}
\label{sec:intro}
\vspaceSQ{-0.5em}

Analyzing massive graphs has become an important task. Example applications
are investigating the Internet structure~\cite{boldi2004webgraph}, analyzing
social or neural relationships~\cite{ben2019modular}, or capturing the behavior
of proteins~\cite{di2012protein}.
Efficient processing of such graphs is challenging. First, these graphs are
large, {reaching even tens of trillions of edges~\mbox{\cite{ching2015one,
lin2018shentu, maass2017mosaic, liu2017graphene, kumar2016g, roy2015chaos,
wu2015gram, shao2013trinity}}}. Second, the graphs in question are
\emph{dynamic}: new friendships appear, novel links are created, or protein
interactions change.  For example, 500 million new tweets in the Twitter social
network appear per day, or billions of transactions in retail transaction
graphs are generated every year~\cite{ammar2016techniques}.

\marginpar{\vspace{1em}\colorbox{yellow}{\textbf{R-4}}\\\colorbox{yellow}{\textbf{(1)}}}

\emph{Graph streaming frameworks} such as \hl{GraphOne~\mbox{\cite{kumar2019graphone}}} or
Aspen~\cite{dhulipala2019low} emerged to enable processing and analyzing
dynamically evolving graphs. Contrarily to static frameworks such as
Ligra~\cite{shun2013ligra, han2015giraph}, such systems execute graph analytics
algorithms (e.g., PageRank) \emph{concurrently} with graph updates (e.g., edge
insertions). Thus, these frameworks must tackle unique challenges, for example
effective modeling and storage of dynamic datasets, efficient ingestion of a
stream of graph updates concurrently with graph queries, or support for
effective programming model. In this work, we present the first taxonomy and
analysis of such system aspects of the streaming processing of dynamic graphs.

\sethlcolor{lyellow}

{We also crystallize the meaning of different concepts in streaming and
dynamic graph processing. We investigate the notions of \emph{temporal},
\emph{time-evolving}, \emph{online}, and \emph{dynamic} graphs, as well as
the differences between graph streaming frameworks and a related class of \emph{graph database
systems}.}

\sethlcolor{yellow}

We also analyze relations between the practice and the theory of streaming
graph processing to facilitate incorporating recent theoretical advancements
into the practical setting, to enable more powerful streaming frameworks. There
exist different related theoretical settings, such as \emph{streaming
graphs}~\cite{mcgregor2014graph} or \emph{dynamic
graphs}~\cite{bhattacharya2019new} that come with different goals and
techniques. Moreover, each of these settings comes with different
\emph{models}, for example the \emph{dynamic graph stream}
model~\cite{kane2012counting} or the \emph{semi-streaming}
model~\cite{feigenbaum2005graph}. These models assume different features of
the processed streams, and they are used to develop provably efficient
streaming algorithms. We analyze which theoretical settings and models are best
suited for different practical scenarios, providing guidelines for architects
and developers on what concepts could be useful for different classes of
systems.

Next, we outline \emph{models for the maintenance of updates}, such as the edge
decay model~\cite{xie2015dynamic}.  These models are independent of the
above-mentioned models for developing streaming algorithms.  Specifically, they
aim to define the way in which edge insertions and deletions are considered for
updating different maintained structural graph properties such as distances between vertices. For example, the edge decay
model captures the fact that edge updates from the past should \emph{gradually}
be made less relevant for the current status of a given structural graph property. 

%
%

Finally, there are \emph{general-purpose} dataflow systems such as Apache
Flink~\cite{carbone2015apache} or Differential
Dataflow~\cite{mcsherry2013differential}. We discuss the support for graph
processing in such designs. 

In general, we provide the following contributions:

\begin{itemize}
\item We crystallize the meaning of different concepts in dynamic and streaming
graph processing, and we analyze the connections to the areas of graph
databases and to the theory of streaming and dynamic graph algorithms.
\item We provide the first taxonomy of graph streaming frameworks, identifying
and analyzing key dimensions in their design, including data models and
organization, concurrent execution, data distribution, targeted architecture, and
others.
\item We use our taxonomy to survey, categorize, and compare over graph streaming
frameworks.
\item We discuss in detail the design of selected frameworks.
%
%
%
\end{itemize}

\macb{Complementary Surveys and Analyses}
\emph{We provide the first taxonomy and survey on general streaming and
dynamic graph processing}. We complement related surveys on the \emph{theory}
of graph streaming models and algorithms~\cite{mcgregor2014graph,
zhang2010survey, aggarwal2014evolutionary, o2009survey}, analyses on \emph{static} graph
processing~\cite{han2014experimental, doekemeijer2014survey, shi2018graph,
batarfi2015large, mccune2015thinking, besta2017push}, and on \emph{general}
streaming~\cite{kamburugamuve2016survey}. 
Finally, only one prior work summarized types of graph updates, 
partitioning of dynamic graphs, and some
challenges~\cite{vaquero2014systems}.

\section{Background and Notation}
\label{section:definitions}
\vspaceSQ{-0.25em}

We first present concepts used in all the sections.
\iftr
We summarize the key
symbols in Table~\ref{tab:symbols}.
\fi

\iftr 

\begin{table}[h]
\vspaceSQ{-0.25em}
\centering
\footnotesize
\scriptsize
\sf
\begin{tabular}{ll@{}}
\toprule
                    $G = (V,E)$ & \makecell[l]{An unweighted graph; $V$ and $E$ are sets of vertices and edges.}\\
                    $w(e)$ & \makecell[l]{The weight of an edge $e = (u,v)$.}\\
                    $n,m$&Numbers of vertices and edges in $G$; $|V| = n, |E| = m$.\\
                    $N_v$ & The set of vertices adjacent to vertex $v$ ($v$'s neighbors).\\
                    $d_v, d$ & The degree of a vertex $v$, the maximum degree in a graph.\\
\bottomrule
\end{tabular}
\vspaceSQ{-1em}
\caption{The most important symbols used in the paper.}
\vspaceSQ{-1em}
\label{tab:symbols}
\end{table}

\fi


\macb{Graph Model}
We model an undirected graph $G$ as a tuple $(V,E)$; $V = \{v_1, ..., v_n\}$ is
a set of vertices and $E = \{e_1, ..., e_m\} \subseteq V \times V$ is a set of
edges; $|V|=n$ and $|E|=m$. If $G$ is directed, we use the name \emph{arc} to
refer to an edge with a direction.  $N_v$ denotes the set of vertices adjacent
to vertex~$v$, $d_v$ is $v$'s degree, and $d$ is the maximum degree in $G$. If
$G$ is weighted, it is modeled by a tuple $(V,E,w)$. Then, $w(e)$ is the weight
of an edge $e \in E$. 
{A weight is a single arbitrary number (e.g., an integer or a float).}
%


\ifcnf
\else

\begin{figure*}[t]
\vspaceSQ{-0.5em}
\centering
\includegraphics[width=0.8\textwidth]{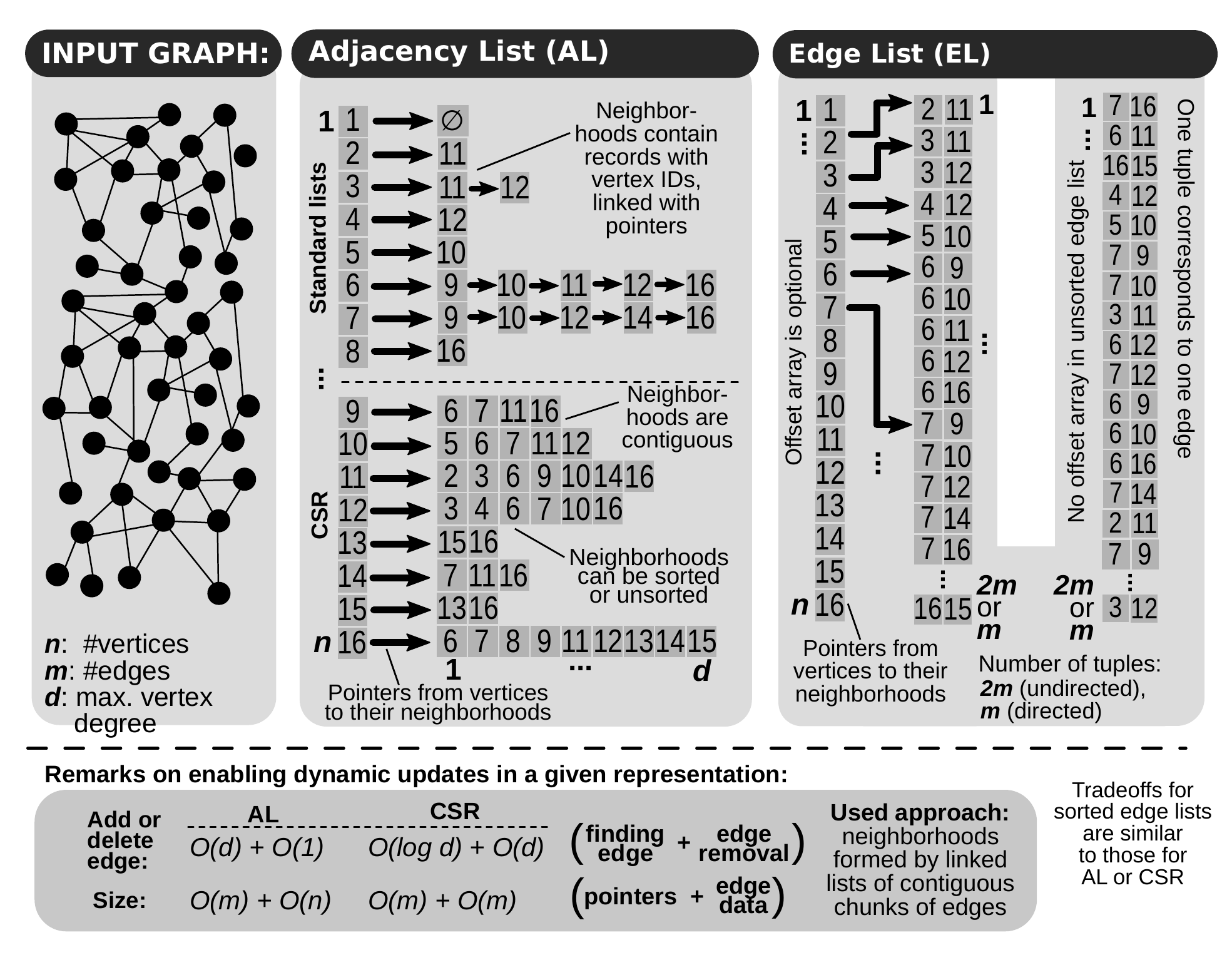}
\vspaceSQ{-2em}
\caption{{\textbf{Illustration of fundamental graph representations}.}
}
%
%
\label{fig:reps}
\end{figure*}

\fi

\macb{Graph Representations}
We also summarize fundamental \emph{static} graph \emph{representations}; they
are used as a basis to develop dynamic graph representations in different
frameworks. These are \tr{the \emph{adjacency matrix} (AM),} the \emph{adjacency
list} (AL), the \emph{edge list} (EL), {and the Compressed Sparse Row (CSR,
sometimes referred to as Adjacency
Array~\mbox{\cite{busato2018hornet}})}\footnote{{Some works use CSR to
describe a graph representation where all neighborhoods form a single
contiguous array~\mbox{\cite{kumar2019graphone}}. In this work, we use CSR to
indicate a representation where each neighborhood is contiguous, but not
necessarily all of them together.}}. 
We illustrate these representations and we
provide remarks on their dynamic variants in Figure~\ref{fig:reps}.
\tr{In \textbf{AM}, a matrix $\mathbf{M} \in \{0,1\}^{n,n}$ determines the
connectivity of vertices: 
$\mathbf{M}_{u,v}=1 \Leftrightarrow (u,v)\in E$.}
In \textbf{AL}, each vertex $u$ has an associated adjacency list
$\mathnormal{A}_{u}$. This adjacency list maintains the IDs of all vertices
adjacent to $u$. We have 
$
v\in \mathnormal{A}_{u} \Leftrightarrow (u,v)\in E.
$
\tr{AM uses $\mathcal{O}\left(n^2\right)$ space and can check connectivity of two
vertices in $\mathcal{O}\left(1\right)$ time. }
AL requires $\mathcal{O}\left(n
+ m\right)$ space and it can check connectivity in
$\mathcal{O}\left(|\mathnormal{A}_u|\right) \subseteq
\mathcal{O}\left(d\right)$ time.
\textbf{EL} is similar to AL in the asymptotic time and space complexity as
well as the general design. The main difference is that each edge is stored
explicitly, with both its source and destination vertex. In AL and EL, a
potential cause for inefficiency is scanning all edges to find neighbors of a
given vertex. To alleviate this, index structures are
employed~\cite{besta2018log}.
Finally, \textbf{CSR} resembles AL but it consists of $n$ \emph{contiguous
arrays} with neighborhoods of vertices. Each array is usually sorted by vertex
IDs. CSR also contains a structure with offsets (or pointers) to each
neighborhood array.

\macb{Graph Accesses}
We often distinguish between \emph{graph queries} and \emph{graph updates}. A
graph query (also called a \emph{read}) may perform some computation on a graph
and it {returns information} about the graph {without modifying its 
structure}. 
\all{This information can be simple (e.g., the degree of a given vertex) or
complex (e.g., some subgraph)}
{Such query can be \emph{local}, also referred to as \emph{fine} (e.g.,
accessing a single vertex or edge) or \emph{global} (e.g., a PageRank analytics
computation returning ranks of vertices)}. 
A graph update, {also called a \emph{mutation},} \emph{modifies} the
graph structure and/or attached labels or values (e.g., edge
weights).

\section{Clarification of Concepts and Areas}

The term ``graph streaming'' has been used in different ways and has different
meanings, depending on the context. We first extensively discuss and clarify
these meanings, and we use this discussion to precisely illustrate the scope of
our taxonomy and analyses. 
We illustrate all the considered concepts in Figure~\ref{fig:concepts}. 
To foster developing more powerful and versatile systems for dynamic and
streaming graph processing, we also {summarize} theoretical concepts.

\iftr
\begin{figure*}[hbtp]
\else
\begin{figure}[h]
\fi
\vspaceSQ{-1em}
\centering
\iftr
\includegraphics[width=1.0\textwidth]{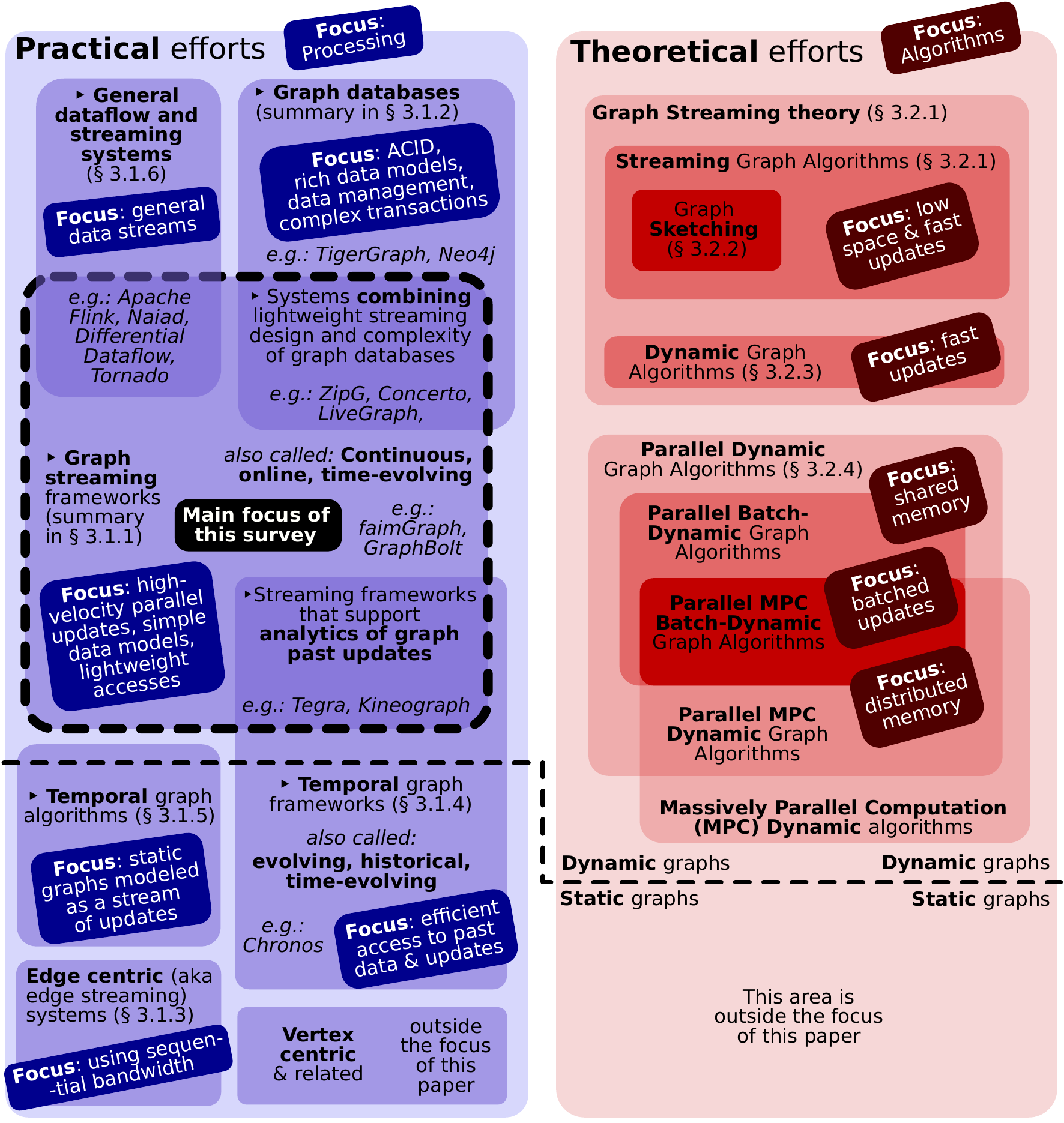}
\else
\includegraphics[width=1\columnwidth]{concepts_5.pdf}
\fi
\vspaceSQ{-2em}
\caption{
Overview of the \textbf{domains and concepts in the practice and theory of
streaming and dynamic graph processing and algorithms.} This work focuses on
\textbf{streaming graph processing} and its relations to other domains.}
\vspaceSQ{-1em}
\label{fig:concepts}
\iftr
\end{figure*}
\else
\end{figure}
\fi

\subsection{Applied Dynamic and Streaming Graph Processing}
\vspaceSQ{-0.15em}

We first outline the applied aspects and areas of dynamic and streaming graph
processing.

\subsubsection{Streaming, Dynamic, and Time-Evolving Graphs}
\vspaceSQ{-0.15em}

Many works~\cite{ediger2012stinger, dhulipala2019low} use a term
``streaming'' or ``streaming graphs'' to refer to a setting in which a graph is
\emph{dynamic}~\cite{sha2017accelerating} (also referred to as
\emph{time-evolving}~\cite{iyer2016time}, \emph{continuous}~\cite{dexter2019formal}, or
\emph{online}~\cite{filippidou2015online}) and it can be modified with updates
such as edge insertions/deletions. \textbf{This setting is the primary focus of
this survey.}
{In the work, we use ``dynamic'' to refer to the graph dataset being
modified, and we reserve ``streaming'' to refer to the form of incoming graph
accesses or updates.}
{The time window of the associated queries in the online setting is of the form \mbox{$\left[\text{Now}
- \delta, \text{Now}\right]$}~\mbox{\cite{iyer2019tegra}}.}

{Closely related terms are \emph{batch analytics} or \emph{stream
analytics}, used in relation to the computations and/or the \emph{computation
model}~\mbox{\cite{akidau2015dataflow}}. They refer to, respectively, running
graph analytics \emph{from scratch} (on static or dynamic data), and to running
such analytics \emph{incrementally}, on dynamic data.  In this work, to comply
with naming used in numerous works on dynamic graph processing, unless stated
otherwise, we use the term ``batch'' to refer to \emph{the ingestion of a
certain number of graph updates together.}}

\vspaceSQ{-0.15em}
\subsubsection{Graph Databases and NoSQL Stores}
\vspaceSQ{-0.25em}

Graph databases~\cite{besta2019demystifying} are related to streaming and
dynamic graph processing in that they support graph updates.  Graph databases
(both ``native'' graph database systems and NoSQL stores used as graph
databases (e.g., RDF stores or document stores)) were described in detail in a
recent work~\cite{besta2019demystifying} and are beyond the main focus of this
paper. However, there are numerous fundamental differences and similarities
between graph databases and graph streaming frameworks, and we discuss these
aspects in Section~\ref{sec:gdb}.

\vspaceSQ{-0.15em}
\subsubsection{Streaming Processing of Static Graphs}
\vspaceSQ{-0.25em}

Some works~\cite{zhou2018fpga, besta2019graph, neuendorffer2008streaming,
roy2013x} use ``streaming'' (also referred to as \emph{edge-centric}) to
indicate a setting in which the input graph is \emph{static} but its edges are
processed in a streaming fashion (as opposed to an approach based on random
accesses into the graph data). Example associated frameworks are
X-Stream~\cite{roy2013x}, ShenTu~\cite{lin2018shentu},
{RStream~\mbox{\cite{wang2018rstream}}}, and several FPGA
designs~\cite{besta2019graph}. Such designs are outside the main focus of this
survey; some of them were described by other works dedicated to static graph
processing~\cite{besta2019graph, doekemeijer2014survey}.
%

\vspaceSQ{-0.15em}
\subsubsection{{Historical} Graph Processing}
\vspaceSQ{-0.25em}

\iftr
There exist efforts into analyzing {\emph{historical} } (also referred to as --
somewhat confusingly -- \emph{temporal} or \emph{[time]-evolving})
graphs~\cite{vora2016synergistic, then2017automatic, ren2011querying,
michail2016introduction, miao2015immortalgraph, hartmann2017analyzing,
han2014chronos, fouquet2018enabling, khurana2015storing, khurana2013efficient,
semertzidis2016time, xiangyu2020efficient, byun2019chronograph, han2018auxo,
lightenberg2018tink, steinbauer2016dynamograph, moffitt2017towards,
zaki2016comprehensive, pitoura2017historical, moffitt2017temporal,
fard2012towards, sun2007graphscope}.
\else
There exist efforts into analyzing \emph{temporal} (also referred to as
{historical} or -- somewhat confusingly -- as \emph{[time]-evolving})
graphs~\cite{han2014chronos}.
\fi
As noted by Dhulipala et al.~\cite{dhulipala2019low}, these efforts differ from
streaming/dynamic/time-evolving graph analysis in that \emph{one stores all
past (historical) graph data to be able to query the graph as it appeared at
any point in the past}. Contrarily, in streaming/dynamic/time-evolving graph
processing, one focuses on keeping a graph in one (present) state.  Additional
snapshots are mainly dedicated to more efficient ingestion of graph updates,
and \emph{not} to preserving historical data for time-related analytics. 
Moreover, almost all works that focus solely on temporal graph analysis,
{for example the Chronos system~\mbox{\cite{han2014chronos}}}, are
\emph{not} dynamic {(i.e., they are \emph{offline})}: {there is no notion
of new incoming updates, but solely a series of past graph snapshots
(instances)}. 
{The time window of queries in historical graph processing is of the form \mbox{$\left[T
- \delta, T+\delta\right]$}~\mbox{\cite{iyer2019tegra}}, where \mbox{$T$} is some
selected arbitrary point in the past.}
\textbf{These efforts are outside the focus of this survey} {(we
exclude these efforts, because they come with numerous challenges and design
decisions (e.g., temporal graph models~\mbox{\cite{zaki2016comprehensive}},
temporal algebra~\mbox{\cite{moffitt2017temporal}}, strategies for snapshot
retrieval~\mbox{\cite{xiangyu2020efficient}}) that require separate extensive
treatment, while being unrelated to the streaming and dynamic graph
processing)}.
Still, \emph{we describe concepts and systems that -- while focusing on
streaming processing of dynamic graphs, also enable keeping and processing
historical data}. {One such example is Tegra~\mbox{\cite{iyer2019tegra}}}.
 
\all{
\begin{figure*}[t]
\vspace{-1.5em}
\centering
\includegraphics[width=0.94\textwidth]{taxonomy_three-domains.pdf}
\vspace{-0.75em}
\caption{
    Overview of the \textbf{identified taxonomy of dynamic graph
    streaming frameworks}
    \marc{Would be great if there are references to the according chapter
    (e.g. "System Type \$4.2", "Accepted Streams \$4.1", etc.)}
    \marc{Others: wrong ")" in "... reducing used storage\textbf{)},
    Missing "the" in "What is \textbf{the} support for rich edge data}}
\vspace{-1.5em}
\label{fig:taxonomy}
\end{figure*}
}

\subsubsection{{Temporal Graph Algorithms}}

{Certain works analyze graphs where edges carry timing information, e.g.,
the order of communication between entities~\mbox{\cite{wu2016reachability,
wu2014path}}. One method to process such graphs is to \emph{model them as a
stream of incoming edges}, with the arrival time based on temporal information
attached to edges. Thus, while being static graphs, their representation is
dynamic. Thus, we picture these schemes as being partially in the dynamic
setting in Figure~\mbox{\ref{fig:concepts}}. These works come with no
\emph{frameworks}, and are outside the focus of our work.}

\subsubsection{{General Dataflow and Streaming Systems}}

{General streaming and dataflow systems, such as Apache
Flink~\mbox{\cite{carbone2015apache}},
Naiad~\mbox{\cite{murray2016incremental}},
Tornado~\mbox{\cite{shi2016tornado}}, or Differential
Dataflow~\mbox{\cite{mcsherry2013differential}}, can also be used to process
dynamic graphs. However, most of the dimensions of our taxonomy are not
well-defined for these general purpose systems.
Overall, these systems provide a very general programming model and impose no
restrictions on the format of streaming updates or graph state that the users
construct.  Thus, in principle, they could process queries and updates
concurrently, support rich attached data, or even use transactional semantics.
However, they do not come with pre-built features specifically targeting
graphs.}

\subsection{Theory of Streaming and Dynamic Graphs}

We next proceed to outline concepts in the theory of dynamic and streaming
graph models and algorithms. Despite the fact that detailed descriptions are
outside the scope of this paper, we firmly believe that explaining the
associated general theoretical concepts and crystallizing their relations to
the applied domain may facilitate developing more powerful streaming systems by
-- for example -- incorporating efficient algorithms with provable bounds on
their performance. In this section, we outline different theoretical areas and
their focus. 
%
%
In general, in all the following theoretical settings, one is
interested in maintaining  (sometimes approximations to) a structural graph property of
interest, such as connectivity structure, spectral structure, or shortest path
distance metric, for graphs that are being modified by incoming updates (edge
insertions and deletions).

\subsubsection{Streaming Graph Algorithms}
\vspaceSQ{-0.25em}

\sethlcolor{lyellow}

{In \emph{streaming graph algorithms}~\mbox{\cite{feigenbaum2005graph,
datar2002maintaining}}, one usually starts with an empty graph with no edges
(but with a fixed set of vertices). Then, at each algorithm
step, a new edge is inserted into the graph, or an existing edge is deleted.
Each such algorithm is parametrized by (1) \emph{space complexity} (space
used by a data structure that maintains a graph being updated), (2)
\emph{update time} (time to execute an update), (3) \emph{query time} (time to
compute an estimate of a given structural graph property), (4) \emph{accuracy of the
computed structural property}, and (5) \emph{preprocessing time} (time to
construct the initial graph data structure)~\mbox{\cite{bhattacharya2015space}}.
Different streaming models can introduce additional assumptions, for example
the Sliding Window Model provides restrictions on the \emph{number of previous
edges in the stream, considered for estimating the
property}~\mbox{\cite{datar2002maintaining}}.}

{The goal is to develop algorithms that minimize different parameter values,
with a special focus on \emph{minimizing the storage for the graph data
structure}. While space complexity is the main focus, significant effort is
devoted to optimizing the runtime of streaming algorithms, specifically the
time to process an edge update, as well as the time to recover the final
solution (see, e.g., \mbox{\cite{LarsenNNT19}} and~\mbox{\cite{KMMMNST19}} for some recent
developments). Typically the space requirement of graph streaming algorithms is
\mbox{$O(n\ \text{polylog}\ n)$} (this is known as the semi-streaming
model~\mbox{\cite{feigenbaum2005graph}}), i.e., about the space needed to store a few
spanning trees of the graph. Some recent works achieve "truly sublinear" space
\mbox{$o(n)$}, which is sublinear in the number of vertices of the graph and is
particularly good for sparse
graphs~\mbox{\cite{KapralovKS14,EsfandiariHLMO18,BuryGMMSVZ19,AssadiKLY16,AssadiKL17,PengS18,KMNT19}}.
The reader is referred to surveys on graph streaming
algorithms~\mbox{\cite{muthukrishnan2005data, GuhaM12, mcgregor2014graph}} for more
references.}

\sethlcolor{yellow}

\tr{\macb{Applicability in Practical Settings}
Streaming algorithms can be used when there are hard limits on the maximum
space allowed for keeping the processed graph, as well as a need for very fast
updates per edge.
Moreover, one should bear in mind that many of these algorithms provide
approximate outcomes.  Finally, the majority of these algorithms assumes the
knowledge of certain structural graph properties in advance, most often the number of
vertices~$n$.}


\subsubsection{Graph Sketching and Dynamic Graph Streams}
\vspaceSQ{-0.25em}

Graph sketching~\cite{ahn2012graph} is an influential technique
for processing graph streams with both insertions and deletions. The
idea is to apply classical sketching techniques such as
\textsc{CountSketch}~\cite{minton2014improved} or distinct elements sketch
(e.g., \textsc{HyperLogLog}~\cite{flajolet2007hyperloglog})
to the edge incidence matrix of the input graph. 
%
%
Existing results show how to approximate the connectivity and cut
structure~\cite{ahn2012graph,AndoniCKQWZ16}, spectral
structure~\cite{KNST19,KMMMNST19}, shortest path
metric~\cite{ahn2012graph,KapralovW14}, or subgraph
counts~\cite{kane2012counting,KallaugherKP18} using small sketches. Extensions
to some of these techniques to hypergraphs were also proposed~\cite{GuhaMT15}.


\iftr
\all{\vspaceSQ{-0.15em}
\subsubsection{Multi-Pass Streaming Graph Algorithms}
\vspaceSQ{-0.35em}}

Some streaming graph algorithms use the notion of a \emph{bounded stream},
i.e., the number of graph updates is bounded. Streaming and applying all such
updates once is referred to as a \emph{single pass}. Now, some streaming graph
algorithms allow for \emph{multiple passes}, i.e., streaming all edge updates
\emph{more than once}. This is often used to improve the approximation quality
of the computed solution~\cite{feigenbaum2005graph}.

%

There exist numerous other works in the theory of streaming graphs.
Variations of the semi-streaming model allow stream
manipulations across passes, (also known as the \emph{W-Stream}
model~\cite{demetrescu2009trading}) or stream sorting passes (known as the
\emph{Stream-Sort} model~\cite{aggarwal2004streaming}).
We omit these efforts are they are outside the scope of this paper.
\fi

\vspaceSQ{-0.15em}
\subsubsection{Dynamic Graph Algorithms}
\vspaceSQ{-0.35em}


In the related area of \emph{dynamic graph algorithms} one is interested in
developing algorithms that approximate a combinatorial property of the input graph of interest (e.g., connectivity, shortest path distance, cuts, spectral properties) under
edge insertions and deletions. Contrarily to graph streaming, in dynamic graph algorithms one puts less
focus on minimizing space needed to store graph data. Instead, the primary goal
is to \emph{minimize time to conduct graph updates}. This has led to several
very fast algorithms that provide updates with amortized poly-logarithmic
update time complexity. See~\cite{bhattacharya2019new,ChechikZhang19,behnezhad19,danupon19,DurfeeGGP19,ForsterG19,DuanHZ19} and references within for some of the most recent developments.

\tr{\macb{Applicability in Practical Settings}
Dynamic graph algorithms can match settings where primary focus is
on fast updates, without severe limitations on the available space.}

\subsubsection{Parallel Dynamic Graph Algorithms}

Many algorithms were developed under the \emph{\textbf{parallel dynamic
model}}, in which a graph undergoes a series of
incoming \emph{parallel} updates.
Next, the \emph{\textbf{parallel batch-dynamic model}} is a recent development
in the area of parallel dynamic graph
algorithms~\cite{Acar,SimsiriTTW16,acar2019parallel, tseng2019batch}. In this
model, a graph is modified by updates coming \emph{in batches}. A batch size is
usually a function of $n$, for example $\log n$ or $\sqrt{n}$. Updates from
each batch can be applied to a graph \emph{in parallel}. The motivation for
using batches is twofold: (1) incorporating parallelism into ingesting updates,
and (2) reducing the cost per update. The associated algorithms focus on
minimizing time to ingest updates into the graph while accurately maintaining a
given structural graph property.  

A \emph{\textbf{variant}}~\cite{durfee2019parallel} that \emph{combines the
parallel batch-dynamic model with the Massively Parallel Computation (MPC)
model}~\cite{karloff2010model} was also recently described. The MPC model is
motivated by distributed frameworks such as MapReduce~\cite{dean2008mapreduce}.
In this model, the maintained graph is stored on a certain number of machines
(additionally assuming that the data in one batch fits into one machine).  Each
machine has a certain amount of space sublinear with respect to $n$. The main
goal of MPC algorithms is to solve a given problem using $O(1)$ communication
rounds while minimizing the volume of data communicated between the
machines~\cite{karloff2010model}. 
%

Finally, \emph{\textbf{another variant}} of the MPC model that addresses
dynamic graph algorithms \emph{but without considering batches}, was also
recently developed~\cite{italiano2019dynamic}.

\iftr
\macb{Applicability in Practical Settings}
Algorithms developed in the above models may be well-suited for enhancing
streaming graph frameworks as these algorithms explicitly (1)
maximize the amount of parallelism by using the concept of batches, and (2)
minimize time to ingest updates.
\fi

\begin{figure*}[t]
\vspaceSQ{-0.5em}
\centering
\includegraphics[width=1.0\textwidth]{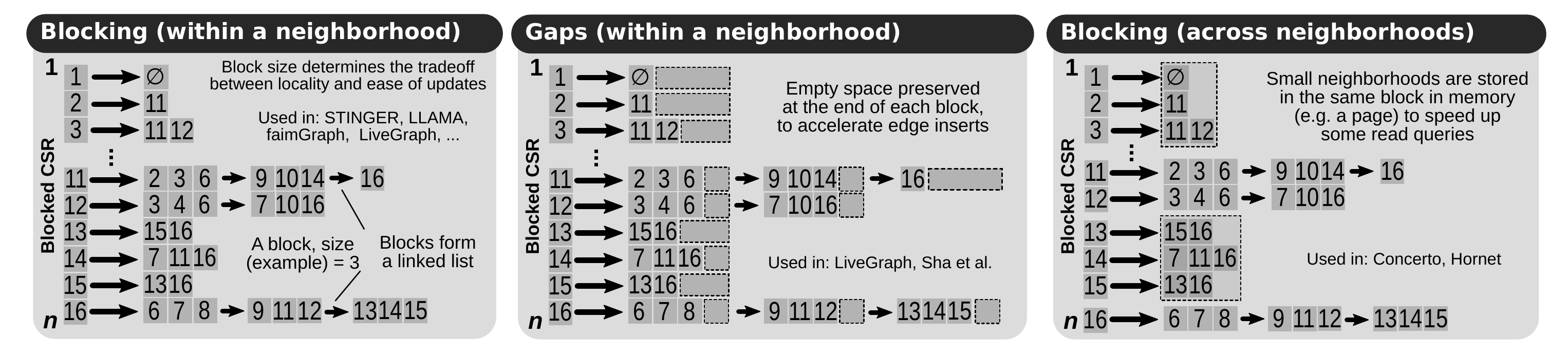}
\vspaceSQ{-1.75em}
\caption{{\textbf{Illustration of blocking-related optimizatios in dynamic graph representations}.}}
\vspaceSQ{-1.5em}
\label{fig:reps-opt}
\end{figure*}

\section{Taxonomy of Frameworks}
\label{sec:taxo}
\vspaceSQ{-0.25em}

We identify a taxonomy of graph streaming frameworks.
{We offer a detailed analysis of concrete frameworks using the taxonomy in
Section~\mbox{\ref{sec:frameworks}} and in Tables~\mbox{\ref{tab:1}}--\mbox{\ref{tab:2}}.
Overall, the identified taxonomy divides all the associated aspects into
six classes: 
\emph{ingesting updates (\mbox{\cref{sec:arch_updates}})}, \emph{historical
data maintenance (\mbox{\cref{sec:historical}})}, \emph{dynamic graph
representation (\mbox{\cref{sec:rep}})}, \emph{incremental changes
(\mbox{\cref{sec:incremental_comp}})}, \emph{programming API and models
(\mbox{\cref{sec:apis}})}, and \emph{general architectural features
(\mbox{\cref{sec:general}})}.}
%
%
{Due to space constraints, we focus on the \emph{details of the system
architecture} and we only sketch the \emph{straightforward} taxonomy aspects
(e.g., whether a system targets CPUs or GPUs) and list\footnote{{More details are in the extended
paper version (see the link on page~1)}} them
in~\mbox{\cref{sec:general}}.}

\subsection{{Architecture of Dynamic Graph Representation}}
\label{sec:rep}
\vspaceSQ{-0.25em}

\sethlcolor{lyellow}

{A core aspect of a streaming framework is
the used representation of the maintained graph.}

\subsubsection{{Used Fundamental Graph Representations}}
\vspaceSQ{-0.25em}

{While the details of how each system maintains the graph dataset usually
vary, the used representations can be grouped into a small set of fundamental
types.}
{Some frameworks use one of the \textbf{basic graph representations} (AL, EL,
CSR, or AM) which are described in Section~\mbox{\ref{section:definitions}}.}
{Other graph representations are \textbf{based on trees}, where there is
some additional hierarchical data structure imposed on the otherwise flat
connectivity data; this hierarchical information is used to accelerate dynamic
queries.}
{Finally, frameworks constructed on top of more general infrastructure use a
\textbf{representation provided by the underlying system}.}

\subsubsection{Blocking Within and Across Neighborhoods}

{In the taxonomy, we distinguish a common design choice in systems based on CSR
or its variants. Specifically, one can combine the key design principles of AL
and CSR by dividing each neighborhood into contiguous \emph{blocks} (also
referred to as \emph{chunks}) that are larger than a single vertex ID (as in a
basic AL) but smaller than a whole neighborhood (as in a basic CSR). This
offers a tradeoff between flexible modifications in AL and more locality (and
thus more efficient neighborhood traversals) in CSR~\mbox{\cite{neo4j_book}}. 
Now, this blocking scheme is applied \textbf{within} each single
neighborhood. We also distinguish a variant where multiple neighborhoods are
grouped inside one block. We will refer to this scheme as blocking
\textbf{across} neighborhoods.
An additional optimization in the blocking scheme is to pre-allocate some
reserved space at the end of each such contiguous block, to offer some number
of fast edge insertions that do not require block reallocation.
All these schemes are pictured in Figure~\mbox{\ref{fig:reps-opt}}.}

\subsubsection{{Supported Types of Vertex and Edge Data}}

{Contrarily to graph databases that heavily use rich graph models such as the
Labeled Property Graph~\mbox{\cite{gdb_query_language_Angles}}, graph streaming
frameworks usually offer simple data models, focusing on the \emph{graph
structure} and not on \emph{rich data attached to vertices or edges}. Still,
different frameworks support basic additional vertex or edge data, most often
\textbf{weights}. Next, in certain systems, both an edge and a vertex can
have a \textbf{type} or an \textbf{attached property}. Finally, an edge
can also have a \textbf{timestamp} that indicates the time of inserting this
edge into the graph. A timestamp can also indicate a modification (e.g., an
update of a weight of an existing edge). 
Details of such rich data are specific to each framework.}

\subsubsection{{Other Indexing Structures}}

{One uses indexing structures to accelerate different queries. In our
taxonomy, we distinguish indices that speed up queries related to the
\textbf{graph structure}, \textbf{rich data} (i.e., vertex or edge properties
or labels), and \textbf{historic (temporal) aspects} (e.g., indices for edge
timestamps).}

\sethlcolor{yellow}

\subsection{Graph Storage Architecture and Mutations}
\label{sec:arch_updates}
\vspaceSQ{-0.25em}

{The first core architectural aspect of any graph streaming framework are the
details of {its graph storage engines, and how incoming updates are ingested
into it}.

\subsubsection{{Concurrent Queries and Updates}}

We start with achieving concurrency between
queries and updates (mutations).
\sethlcolor{yellow}
{One approach is based on \textbf{coarse-grained} synchronization
(also referred to as \textbf{discretization}).}
\sethlcolor{lyellow}
{Here, one popular method is based on \textbf{snapshots}. Updates and
queries are isolated from each other by making them execute on two different
copies (snapshots) of the graph data. At some point, such snapshots are merged
together. Depending on a system, the scope of data duplication (i.e., only a
part of the graph may be copied into a new snapshot) and the details of merging
may differ.}
\sethlcolor{yellow}
{Snapshots can be created in different ways, for example with the well-known
\textbf{copy-on-write} scheme, or \textbf{periodically} as determined by the
underlying system details, or using \textbf{tombstones}.}

\sethlcolor{lyellow}
{In coarse-grained synchronization, one ingests updates, or resolves
queries, \emph{in batches}, i.e., multiple at a time, to amortize overheads
from ensuring consistency of the maintained graph. We distinguish this design
choice in the taxonomy because of its widespread use.
Moreover, we identify a popular optimization in which a batch of edges to be
removed or inserted is first \textbf{sorted} based on the ID of adjacent
vertices. This introduces a certain overhead, but it also facilitates parallel
ingestion of updates: updates associated with different vertices can be easier
identified.}

{In \textbf{fine-grained} synchronization} \sethlcolor{yellow}{(also
referred to as \textbf{continuous} updates)}, \sethlcolor{lyellow}{in
contrast to coarse-grained synchronization (where updates are merged with the
main graph representation during dedicated phases), updates are incorporated
into the main dataset as soon as they arrive, often interleaved with queries,
using synchronization protocols based on fine-grained locks and/or atomic
operations.}
{A variant of fine-grained synchronization is \textbf{Differential
Dataflow}~\mbox{\cite{mcsherry2013differential}}, where the ingestion strategy allows
for concurrent updates and queries by relying on a combination of logical time,
maintaining the knowledge of updates (referred to as deltas), and progress
tracking. Specifically, the differential dataflow design operates on
collections of key-value pairs enriched with timestamps and delta values. It
views dynamic data as additions to or removals from input collections and
tracks their evolution using logical time.}

{Finally, as also noted in past work~\mbox{\cite{dhulipala2019low}}, a system may
simply not enable concurrency of queries and updates, and instead
alternate between incorporating batches of graph updates and graph queries
(i.e., updates are being applied to the graph structure while queries wait, and
vice versa). This type of architecture may enable a high ratio of digesting
updates as it does not have to resolve the problem of the consistency of graph
queries running interleaved, concurrently, with updates being digested.}

\sethlcolor{yellow}

\subsubsection{{Transactional Support}}

{We distinguish systems that support \textbf{transactions}, understood as
units of work that enable isolation between concurrent accesses and correct
recovery from potential failures. Moreover, some (but not all) systems ensure
the \textbf{ACID semantics} of transactions.}

\subsection{{Architecture of Historical Data Maintenance}}
\label{sec:historical}
\vspaceSQ{-0.25em}

{While we do not focus on systems \emph{solely} dedicated to the off-line
analysis of historical graph data, some streaming systems \emph{enable
different forms of accessing/analyzing such data}.}

\subsubsection{{Storing Past Snapshots}}

{In general, a streaming system may enable \textbf{storing past snapshots},
i.e., consistent past views (instances) of the whole dataset. 
%
%
{Two general approaches for maintaining such past instances are (1)
keeping snapshots themselves and (2) maintaining changes to the graph.
The former approach makes deriving a given snapshot very efficient.
However, it may come with storage overheads
if many snapshots are maintained. The latter scheme
reduces storage overheads, but it may be time-consuming because
one has to reapply graph changes to construct a snapshot on demand.}

\subsubsection{Visibility of Past Graph Updates}
\label{sec:maintaining_updates}

{There are several ways in which the information about past updates can be
stored.}
Most systems only maintain a ``live'' version of the graph, where information
about the past updates is not maintained\footnote{\scriptsize {This approach is
sometimes referred to as the \textbf{``snapshot'' model}.} Here, the word
``snapshot'' means ``a complete view of the graph, with all its updates''. This
naming is somewhat confusing, as ``snapshot'' can also mean ``a specific copy
of the graph generated for concurrent processing of updates and queries'',
cf.~\cref{sec:arch_updates}.}, in which all incoming graph updates are being
incorporated into the structure of the maintained graph and they are all used
to update or derive maintained structural graph properties. For example, if a
user is interested in distances between vertices, then -- in the snapshot model
-- the derived distances use \emph{all} past graph updates.  Formally, if we
define the maintained graph at a given time~$t$ as $G_t = (V, E_t)$, then we
have $E_t = \{e\ |\ e \in E \land t(e) \le t \}$, where $E$ are all graph edges
and $t(e)$ is the timestamp of $e \in E$~\cite{xie2015dynamic}.

Some streaming systems use the \textbf{sliding window model}, in which edges
beyond certain moment in the past are being omitted when computing graph
properties.  Using the same notation as above, the maintained graph can be
modeled as $G_{t, t'} = (V, E_{t, t'})$, where $E_{t, t'} = \{e\ |\ e \in E
\land t \le t(e) \le t' \}$.  Here, $t$ and $t'$ are moments in time that
define the width of the \emph{sliding window}, i.e., a span of time with graph
updates that are being used for deriving certain query
answers~\cite{xie2015dynamic}. 

Both the snapshot model and the sliding window model do not reflect certain
important aspects of the changing reality. The former takes into account all
relationships \emph{equally}, without distinguishing between the older and more
recent ones. The latter enables omitting old relationships but does it
\emph{abruptly}, without considering the fact that certain connections may
become \emph{less relevant} in time but still \emph{be present}. To alleviate
these issues, the \textbf{edge decay model} was proposed~\cite{xie2015dynamic}.
In this model, each edge~$e$ (with a timestamp $t(e) \le t$) has an independent
probability~$P^f(e)$ of being included in an analysis. $P^f(e) = f(t - t(e))$
is a non-decreasing \emph{decay function} that determines \emph{how fast edges
age}. The authors of the edge decay model set $f$ to be decreasing
exponentially, with the resulting model being called the \textbf{probabilistic
edge decay model}.

\subsection{Architecture of Incremental {Computation}}
\label{sec:incremental_comp}

A streaming framework may support an approach called ``incremental changes''
for faster convergence of graph algorithms.
{Assume that a certain graph algorithm is executed and produces some results,
for example page ranks of each vertex. Now, the key observation behind the
  incremental changes is that the subsequent graph updates may not necessarily
  result in large changes to the derived page rank values.  Thus, instead of
  recomputing the ranks from scratch, one can attempt to minimize the scope of
  recomputation, resulting in ``incremental'' changes to the ranking results.
In our taxonomy, we will distinguish between supporting incremental changes in
{the \textbf{offline} (``\textbf{recomputation}'') or the \textbf{online}
(``\textbf{refinement}'') mode. 
In the former, one updates analytics outcomes with \emph{recomputation}.  In
the latter, one tracks dependencies (on-the-fly) between modified values and
uses these dependencies to simply adjust the values that must be updated, from
the point where the values become affected, without restarting computation.
Recomputation based schemes may further differ in the amount of data that must
be recomputed.  For example, one may restart the computation from scratch for
the whole graph upon mutations, or identify which vertices changed, and
recompute precisely the values associated with these vertices.}

\subsection{{Supported Programming APIs and Models}}
\label{sec:apis}

{The final part of our taxonomy is the offered programming model and API. 
We identify two key classes of designs.

{\textbf{Graph Mutations}}
First, a framework may offer a
selection of functions for \textbf{modifying the maintained graph}; such API
may consist of \emph{simple} basic functions (e.g., insert an edge) or \emph{complex} ones
(e.g., merge two graphs). Here, we additionally identify APIs for \emph{triggered}
events taking place upon specific updates, and for accessing and manipulating
the \emph{pending} graph updates (that await being ingested into the graph
representation).}

{\textbf{API for Graph Analytics}}
The second key API that a framework may support consists of functions for
\textbf{running graph computations} on top of the maintained graph. Here, we
identify specific APIs for controlling graph algorithms (e.g., PageRank)
processing the \emph{main (i.e., ``live'')} graph snapshot, or for controlling such
computations running on top of \emph{past} snapshots. Moreover, our taxonomy
includes an API or models for \emph{incremental} processing of the outcomes of graph algorithms
(cf.~\mbox{\cref{sec:incremental_comp}}).

\all{Systems such as Kineograph~\cite{cheng2012kineograph} \marc{This is true
that Kineograph relies on snapshots. At the same time, "Kineograph adopts a
vertex-based computation model" and "Kineograph supports the push and the pull
models in the computations" - not sure how this relates precisely to the BSP
model mentioned below}, CellIQ~\cite{iyer2015celliq}, and
Chronos~\cite{han2014chronos} operate on a \textbf{snapshot-based model}. They
use time discretization to express an evolving graph $G_S$ as a series of
\emph{static} graphs $G_1, G_2, \ldots$, each of which represents a snapshot of
$G_S$ at a particular point in time.  Computations are defined via a BSP model
which operates independently per snapshot. In some cases, cross-snapshot
analytics, such as sliding window computations
(cf.~\cref{sec:maintaining_updates}) are also supported. Snapshots are created
by either making a copy of the entire graph periodically or by maintaining a
graph data structure that can be updated incrementally to reflect a state of
the graph at a given point in time.}

\subsection{General Architectural Features of Frameworks}
\label{sec:general}
\vspaceSQ{-0.25em}

The general features are the \textbf{location} of the maintained
graph data (e.g., main memory or GPU memory), whether it is
\textbf{distributed}, what is the \textbf{targeted hardware architecture}
(general CPUs or GPUs), and whether a system is \textbf{general-purpose} or is
it developed specifically for {graph analytics}.

\section{{Analysis of} Frameworks}
\label{sec:frameworks}
\vspaceSQ{-0.25em}

{We now analyze existing frameworks using our taxonomy
(cf.~Section~\mbox{\ref{sec:taxo}}) in Tables~\mbox{\ref{tab:1}} --
\mbox{\ref{tab:2}}, and in the following text.} 
\tr{We also describe selected frameworks in more detail. }
%
%
%
We use symbols ``\faBatteryFull'', ``\faBatteryHalf'', and ``\faTimes'' to
indicate that a given system offers a given feature, offers a given feature in
a limited way, and does not offer a given feature,
respectively\footnote{\scriptsize We encourage participation in this survey.
In case the reader possesses additional information relevant for the tables,
the authors would welcome the input. We also encourage the reader to send us
any other information that they deem important, e.g., details of systems not
mentioned in the current survey version.}.
\tr{``\noAnswer'' indicates we were unable to infer this information based on
the available documentation.}

\begin{table*}[hbtp]
\centering
\setlength{\tabcolsep}{0.5pt}
\ifcnf
\else
\renewcommand{\arraystretch}{1.2}
\fi
\scriptsize
\ssmall
\ifHL
\begin{tabular}{@{}llllylylyyyyyylll@{}}
\else
\begin{tabular}{@{}llllyylllllllllll@{}}
\fi
\toprule
\makecell[c]{\textbf{Reference}} &
\makecell[c]{{\textbf{Ds?}}} &
\makecell[c]{\textbf{Data} \textbf{location}} &
\makecell[c]{\textbf{Arch.}} &
\makecell[c]{\textbf{F?}} &
\makecell[l]{\textbf{Con?}} &
{\textbf{B?}} &
{\textbf{sB?}} &
{\textbf{T?}} &
{\textbf{acid?}} &
{\textbf{P?}} &
{\textbf{L?}} &
{\textbf{S?}} &
{\textbf{D?}} &
\makecell[c]{\textbf{Edge}\\\textbf{updates}} &
\makecell[c]{\textbf{Vertex}\\\textbf{updates}} &
\makecell[c]{\textbf{Remarks}} \\
\midrule
STINGER~\cite{ediger2012stinger} & \faTimes & \makecell[l]{M-mem.} & {CPU} & S & \faTimes & \faBatteryFull & \faBatteryFull & \faTimes & \faTimes & \faBatteryHalf & \faBatteryFull & \faTimes & \faTimes & \faBatteryFull\ (A/R) & \faBatteryHalf$^*$\ (A/R) & \makecell[l]{$^*$Removal is unclear} \\
UNICORN~\cite{suzumura2014towards}  & \faBatteryFull & M-mem. & CPU & C & \faTimes & \faBatteryFull & \faTimes & \faTimes & \faTimes & \faTimes & \faBatteryFull & \faTimes & \faTimes   & \faBatteryFull\ (A/R) & \faBatteryFull\ (A/R) & \makecell[l]{{\textbf{Extends IBM InfoSphere Streams~\mbox{\cite{biem2010ibm}}}}} \comment{The paper is of low quality, many grammatical errors} \\
DISTINGER~\cite{feng2015distinger} & \faBatteryFull  & \makecell[l]{M-mem.} & \makecell[l]{CPU} & S & \faTimes & \faBatteryFull & \faTimes & \faTimes & \faTimes & \faTimes & \faBatteryFull & \faTimes & \faTimes  & \faBatteryFull\ (A/R) & \faBatteryFull\ (A/R) & \textbf{Extends STINGER~\cite{ediger2012stinger}} \\
cuSTINGER~\cite{green2016custinger} & \faTimes & \makecell[l]{GPU mem.} & \makecell[l]{{GPU$^*$}} & S & \faTimes & \faBatteryFull & \faTimes & \faTimes & \faTimes & \faTimes & \faBatteryFull & \faTimes & \faTimes  & \faBatteryFull\ (A/R) & \faBatteryFull\ (A/R)  & \textbf{Extends STINGER~\cite{ediger2012stinger}}. \makecell[l]{$^*$Single GPU.} \\
EvoGraph~\cite{sengupta2017evograph} & \faTimes & M-mem. \comment{High performance due to asynchronous mode and deep copy operations between host and GPU} & GPU$^*$ & C & \faTimes & \faBatteryFull & \faTimes & \faTimes & \faTimes  & \faBatteryHalf & \faBatteryFull & \faTimes & \faTimes  & \faBatteryFull\ (A/R) & \faBatteryFull\ (A/R) & \makecell[l]{Supports multi-tenancy to share GPU resources. $^*$Single GPU.} \\
Hornet~\cite{busato2018hornet} & \faTimes & \makecell[l]{GPU, M-mem.} & GPU$^\dagger$ & S & \faTimes$^*$ & \faBatteryFull & \faBatteryFull & \faTimes & \faTimes \comment{A part of the algorithm is in parallel during bulk update} & \faTimes & \faBatteryFull & \faTimes & \faTimes  & \faBatteryFull\ (A/R/U) & \faBatteryFull\ (A/R/U) & \makecell[l]{$^*$Not mentioned. $^\dagger$Single GPU} \\
\HL{\rowcolor{yellow}} GraPU~\cite{sheng2018grapu, sheng2020exploiting} & \faBatteryFull & M-mem., disk & CPU & C & \faTimes & \faBatteryFull & \faTimes$^*$ & \faTimes & \faTimes & \faTimes & \faBatteryFull & \faTimes & \faTimes & \faBatteryFull\ (A/R) & \faTimes & $^*$Batches are processed with non-straightforward schemes \\
\HL{\rowcolor{yellow}} Grace~\cite{prabhakaran2012managing} & \faTimes & M-mem. & CPU & S+C & \faBatteryFull\ {(s:C)} & \faBatteryFull & \noAnswer & \faBatteryFull & \faBatteryFull & \faBatteryHalf$^\dagger$ & \faBatteryFull & \faTimes & \faTimes  & \faBatteryFull\ (A/R/U) & \faBatteryFull\ (A/R) & {$^\dagger$To implement transactions} \\
Kineograph~\cite{cheng2012kineograph}  & \faBatteryFull & M-mem. & CPU & C+S & \faBatteryFull\ {(s:P)} & \faBatteryFull & \faTimes & \faBatteryFull & \faTimes & \faBatteryFull & \faBatteryFull & \faBatteryFull & \faBatteryFull & \faBatteryFull\ (A/U$^*$) & \faBatteryFull\ (A/U$^*$) & \makecell[l]{$^*$Custom update functions are possible} \\
LLAMA~\cite{macko2015llama} & \faTimes & \makecell[l]{M-mem., disk} & {CPU} & S & \faBatteryFull\ {(s:C)} & \faBatteryFull & \faBatteryFull & \faTimes & \faTimes & \faBatteryFull & \faBatteryFull & \faTimes & \faTimes  & \faBatteryFull\ (A/R) & \faBatteryFull\ (A/R) & ---  \\
CellIQ~\cite{iyer2015celliq} & \faBatteryFull & \makecell[l]{Disk (HDFS)} & CPU & C & \faBatteryFull\ {(s)} & \noAnswer & \faTimes & \faTimes & \faTimes & \faBatteryFull & \faBatteryFull & \faBatteryFull & \faTimes & \faBatteryFull\ (A/R) & \faBatteryFull\ (A/R) & \makecell[l]{{\textbf{Extends GraphX~\mbox{\cite{gonzalez2014graphx}} and Spark~\mbox{\cite{zaharia2012resilient}}}}. $^*$No details.} \\ \comment{Yes, has support for window computation! Input is a stream of "GeoGraph", but it is not mentioned how one can construct such an object}
GraphTau~\cite{iyer2016time} & \faBatteryFull & \makecell[l]{M-mem., disk} & CPU & C & \faBatteryFull\ {(s)$^*$} & \faBatteryFull & \faTimes & \faTimes & \faTimes & \faBatteryFull & \faBatteryFull & \faBatteryFull & \faTimes & \faBatteryFull\ (A/R) & \faBatteryFull\ (A/R) & {\textbf{Extends Spark}. $^*$Offers \textbf{more} than simple snapshots.} \comment{section 4.1 says that the RDD are elements of vertex and edge updates} \\
\HL{\rowcolor{yellow}} DeltaGraph~\cite{dexter2016lazy} & \faTimes & M-mem. & CPU & C & \faBatteryFull\ (s:C)$^*$ & \faTimes & \faTimes & \faTimes & \faTimes & \faTimes & \faBatteryFull & \faTimes & \faTimes  & \faBatteryFull\ (A/R) & \faBatteryFull\ (A/R) & $^*$Relies on Haskell's features to create snapshots \\
GraphIn~\cite{sengupta2016graphin} & \faTimes$^*$ & M-mem. & CPU & C+S & \faBatteryHalf\ {(s)} & \faBatteryFull & \faTimes & \faTimes & \faTimes  & \faTimes$^\dagger$ & \faBatteryFull & \faTimes & \faTimes  & \faBatteryHalf$^*$\ (A/R) & \faBatteryHalf$^*$\ (A/R) & $^*$Details are unclear. {$^\dagger$Only mentioned} \\
Aspen~\cite{dhulipala2019low} & \faTimes & M-mem., disk & CPU & S+C & \faBatteryFull\ {(s:C)$^*$} & \noAnswer & \noAnswer & \faTimes & \faTimes & \faTimes & \faBatteryFull & \faTimes & \faTimes  & \faBatteryFull\ (A/R) & \faBatteryFull\ (A/R) & {$^*$Focus on lightweight snapshots; enables \textbf{serializability}} \\ 
Tegra~\cite{iyer2019tegra} & \faBatteryFull & \makecell[l]{M-mem., disk} & CPU & C+S & \faBatteryFull\ {(s)} & \noAnswer & \noAnswer & \faTimes & \faTimes & \faBatteryFull & \faBatteryHalf$^*$ & \faBatteryFull & \noAnswer & \faBatteryFull\ (A/R) & \faBatteryFull\ (A/R) & {\textbf{Extends Spark}. $^*$Live updates are considered but outside core focus.} \\
GraphInc~\cite{cai2012facilitating} & \faBatteryFull & M-mem., disk & CPU & C & \faBatteryFull\ {(s)$^*$} & \noAnswer & \noAnswer & \faTimes & \faTimes \comment{Section "Handling update streams" mentions that it is possible} & \faTimes & \faBatteryFull & \faTimes & \faTimes & \faBatteryFull\ (A/R/U) & \faBatteryFull\ (A/R/U) & \makecell[l]{{\textbf{Extends Apache Giraph~\mbox{\cite{apache_giraph}}}. $^*$Keeps separate storage for the graph}\\ {structure and for Pregel computations, but no details are provided.}} \\
ZipG~\cite{khandelwal2017zipg} & \faBatteryFull & M-mem. & CPU & S+C & \faBatteryFull\ {(s)} & \noAnswer & \noAnswer & \faTimes & \faTimes & \faBatteryHalf\ & \faBatteryFull & \faTimes & \faTimes  & \faBatteryFull\ (A/R/U) & \faBatteryFull\ (A/R/U) & {\textbf{Extends Spark \& Succinct~\mbox{\cite{agarwal2015succinct}}}} \comment{The framework allows to get only the changes from a given time} \\ 
GraphOne~\cite{kumar2019graphone} & \faTimes & M-mem. & CPU & S+C & \faBatteryFull\ {(s:T)} \comment{The paper says that snapshots "can be created in real-time using create-static-view()"} & \faBatteryFull & \faBatteryFull & \faTimes & \faTimes & \faBatteryHalf & \faBatteryFull & \faTimes & \faTimes  & \faBatteryFull\ (A/R) & \faBatteryFull\ (A/R) & Updates of weights are possible \\
LiveGraph~\cite{zhu2019livegraph} & \faTimes & M-mem., disk & CPU & S+C & \faBatteryFull\ {(s:C)} & \faTimes & na & \faBatteryFull & \faBatteryFull & \faTimes & \faBatteryFull & \faTimes & \faTimes  & \faBatteryFull\ (A/R/U) & \faBatteryFull\ (A/R/U) & --- \\
Concerto~\cite{lee2013views} & \faBatteryFull & M-mem. & CPU & S+C & \faBatteryFull\ {(f)$^*$} & \faBatteryFull & \faTimes & \faBatteryFull & \faBatteryFull & \faTimes & \faBatteryFull & \faTimes & \faTimes  & \faBatteryHalf\ (A/U) & \faBatteryHalf\ (A/U) & \makecell[l]{{$^*$A two-phase commit protocol based on fine-grained atomics}} \\
aimGraph~\cite{winter2017autonomous} & \faTimes & \makecell[l]{GPU mem.} & \makecell[l]{{GPU$^*$}} & S+C & \faBatteryHalf\ {(f)$^\dagger$} & \faBatteryFull & \noAnswer & \faTimes & \faTimes & \faTimes & \faBatteryFull & \faTimes & \faTimes  & \faBatteryFull\ (A/R) & \faTimes & $^*$Single GPU. {$^\dagger$Only \textbf{fine} reads/updates are considered.} \\
faimGraph~\cite{winter2018faimgraph} & \faTimes & \makecell[l]{GPU, M-mem.} & GPU$^*$ & S+C & \faBatteryHalf\ {(f)$^\dagger$} & \faBatteryFull & \faBatteryFull & \faTimes & \faTimes & \faTimes & \faBatteryFull & \faTimes & \faTimes  & \faBatteryFull\ (A/R) & \faBatteryFull\ (A/R) & $^*$Single GPU. {$^\dagger$Only \textbf{fine} reads/updates, using locks/atomics.} \comment{the stream type is not mentioned explicitly, but one can infer that by the given algorithm outline} \\
\HL{\rowcolor{yellow}} GraphBolt~\cite{mariappan2019graphbolt} & \faTimes & M-mem. & CPU & C+S & \faBatteryHalf\ {(f)$^*$} & \faBatteryFull & \faBatteryFull & \faTimes & \faTimes & \faTimes & \faBatteryFull & \faTimes & \faTimes & \faBatteryFull\ (A/R) & \faBatteryFull\ (A/R) & \textbf{Uses Ligra~\cite{shun2013ligra}}. $^*$Fine edge updates are supported. \\
\rowcolor{yellow} DZiG~\cite{mariappan2021dzig} & \faTimes & M-mem. & CPU & C+S & \faBatteryFull\ (f) & \faBatteryFull & \noAnswer & \faTimes & \faTimes & \faTimes & \faBatteryFull & \faTimes & \faTimes & \faBatteryFull\ (A/R) & \faBatteryFull\ (A/R) & \\
%
%
\HL{\rowcolor{yellow}} RisGraph~\cite{feng2020risgraph} & \faTimes & M-mem. & CPU & C/S & \faBatteryFull\ (sc)$^*$ & \faBatteryFull$^\dagger$ & \noAnswer & \faTimes & \faTimes & \faBatteryFull & \faBatteryFull & \faTimes & \faTimes & \faBatteryFull\ (A/R) & \faBatteryHalf\ (A/R) & $^*$Details in~\cref{sec:analysis_conc}.  \\
GPMA (Sha~\cite{sha2017accelerating}) & \faBatteryHalf$^*$ & \makecell[l]{GPU mem.} & \makecell[l]{GPU$^*$} & S & \faBatteryHalf\ {(o)$^\dagger$} & \faBatteryFull & \noAnswer & \faTimes & \faTimes & \faTimes & \faBatteryFull & \faBatteryFull & \faTimes & \faBatteryFull\ (A/R) & \faTimes& \makecell[l]{$^*$Multiple GPUs within one server. {$^\dagger$Details in~\mbox{\cref{sec:analysis_conc}}.}} \\
KickStarter~\cite{vora2017kickstarter}$^*$ & \faBatteryFull \comment{KickStarter runs on ASPIRE, which is a distributed graph processing system}  & M-mem. & CPU & C & na$^*$ & \faBatteryFull & na$^*$ & na$^*$ & \comment{Trimming can be done in parallel} na$^*$ & na$^*$ & \faBatteryFull & na$^*$ & na$^*$  & \faBatteryFull\ (A/R) & \noAnswer & {\textbf{Uses ASPIRE~\mbox{\cite{vora2014aspire}}}. $^*$It is a \emph{runtime technique}.} \\
%
%
Mondal et al.~\cite{mondal2012managing} & \faBatteryFull & M-mem.$^*$ & CPU & C+S & \faBatteryHalf$^\dagger$ & \noAnswer$^\dagger$ & \noAnswer$^\dagger$ & \faBatteryFull & \faBatteryFull & \faTimes & \faBatteryFull & \noAnswer$^\dagger$ & \noAnswer$^\dagger$  & \faBatteryHalf$^\dagger$\ (A) & \faBatteryHalf$^\dagger$\ (A) & \makecell[l]{$^*$\textbf{{Uses CouchDB as backend~\mbox{\cite{anderson2010couchdb}}}, $^\dagger$Unclear (relies on CouchDB)}} \\
\HL{\rowcolor{yellow}} iGraph~\cite{ju2016igraph} & \faBatteryFull & M-mem. & CPU & C & \noAnswer & \faBatteryFull & \faTimes & \faTimes & \faTimes & \faTimes & \faBatteryFull & \faTimes & \faTimes  & \faBatteryHalf\ (A/U) & \faBatteryHalf\ (A/U) & {\textbf{Extends Spark}} \\
\HL{\rowcolor{yellow}} Sprouter~\cite{abughofa2018sprouter} & \faBatteryFull & M-mem., disk & CPU & C & \noAnswer & \noAnswer & \faTimes & \faTimes & \faTimes & \faTimes & \faBatteryFull & \faTimes & \faTimes  & \faBatteryHalf\ (A) & \noAnswer  & \textbf{Extends Spark} \\
\bottomrule
\end{tabular}
\vspaceSQ{-1em}
\caption{Comparison of selected representative works. {They are grouped by the method of achieving concurrency 
between queries and updates (mutations). Within each group, the systems are sorted by publication date.}
{``\textbf{Ds?}''} (distributed): does a design target distributed environments such as clusters, supercomputers, or data centers?
``\textbf{Data location}'': the location of storing the processed dataset (``M-mem.'': main memory; a system is primarily in-memory).
``\textbf{Arch.}'': targeted architecture.
{``\textbf{F}'': focus on: computation (C), storage (S), computation and storage (C/S), mainly computation
with some focus on storage (C+S), or mainly storage with some focus on computation (S+C).}
``\textbf{Con?}'' (a method of achieving concurrent updates and queries): does a
design support updates (e.g., edge insertions and removals) proceeding
concurrently with queries that access the graph structure (e.g., edge lookups or PageRank computation).
{Whenever supported, we detail the method used for maintaining this
concurrency: {(s): snapshots (method unknown), (s:C): snapshots created with
copy-on-write, (s:P): snapshots created periodically, (s:T): snapshots created
with tombstones,} (f): fine-grained synchronization, (sc): scheduling, (o):
overlap.}
{``\textbf{B?}'' (batches): are updates batched? {Batching entails support for data mutations at coarse granularity.}}
{``\textbf{sB?}'' (sorted batches): can batches of updates be sorted for more performance?}
{``\textbf{T?}'' (transactions): are transactions supported?}
{``\textbf{acid?}'': are ACID transaction properties offered?}
``\textbf{P}'': Does the system enable storing past graph snapshots?
%
%
{``\textbf{L?}'' (live): are live updates supported (i.e., does a system maintain a graph snapshot
that is ``up-to-date'': it continually ingests incoming updates)?}
{``\textbf{S?}'' (sliding): does a system support the Sliding Window Model for accessing past updates?}
{``\textbf{D?}'' (decay): does a system support the Decay Model for accessing past updates?}
``\textbf{Vertex / edge updates}'': support for inserting and/or removing edges and/or vertices;
``\textbf{A}'': add, ``\textbf{R}'': remove, ``\textbf{U}'': update.
``\faBatteryFull'': Support.
``\faBatteryHalf'': Partial / limited support.
``\faTimes'': No support.
``\noAnswer'': Unknown.
}
\label{tab:1}
\vspaceSQ{-1.75em}
\end{table*}

\begin{table*}[hbtp]
\centering
\setlength{\tabcolsep}{0.2pt}
\ifcnf
\else
\renewcommand{\arraystretch}{1.2}
\fi
\scriptsize
\ssmall
\ifHL
\begin{tabular}{@{}lllllyyyyyyy@{}}
\else
\begin{tabular}{@{}llllllllylll@{}}
\fi
\toprule
\makecell[c]{\textbf{Reference}} & \makecell[c]{\textbf{Rich}\\\textbf{edge
data}} & \makecell[c]{\textbf{Rich}\\\textbf{vertex data}} &
\makecell[c]{\textbf{Tested analytics}\\ \textbf{workloads}} &
\makecell[c]{\textbf{{Fundamental}}\\ \textbf{representation}} & \textbf{iB?} &
\textbf{aB?} & \textbf{Id?} & 
\makecell[c]{\textbf{Ic?}} & 
\textbf{PrM?} & \textbf{PrC?} & \textbf{Remarks} \\
\midrule
STINGER~\cite{ediger2012stinger} & \makecell[l]{\faBatteryFull\ (T, W, TS)} & \faBatteryFull\ (T) & \makecell[l]{CL, BC, BFS, CC, $k$-core} & CSR & \faBatteryFull & \faTimes & \faBatteryFull\ (a, d) & \faTimes & \faBatteryFull\ (sm) & \faTimes \\
\HL{\rowcolor{yellow}} Grace~\cite{prabhakaran2012managing} & \faBatteryHalf\ (W) & \faTimes & PR, CC, SSSP, BFS, DFS & CSR & \faBatteryHalf$^*$ & \faTimes & \faBatteryFull\ (a) & \faTimes   & \faBatteryFull\ (sm) & \faTimes & \makecell[l]{$^*$Due to partitioning of neighborhoods.}  \\
Concerto~\cite{lee2013views} & \faBatteryFull\ (P) & \faBatteryFull\ (P) & $k$-hop, $k$-core & CSR & \noAnswer & \faBatteryFull & \faBatteryFull & \faTimes & \faBatteryFull\ (sm, tr$^*$) & \faBatteryFull\ (sa, i)$^*$ & $^*$Graph views \& event-driven processing \\
LLAMA~\cite{macko2015llama}  & \faBatteryFull\ (P) & \faBatteryFull\ (P) & \makecell[l]{PR, BFS, TC} & CSR$^*$ & \faBatteryFull & \faTimes & \faBatteryFull\ (a, t) & \faTimes  & \faBatteryFull\ (sm) & \faTimes & $^*$multiversioned \\
DISTINGER~\cite{feng2015distinger}  & \makecell[l]{\faBatteryFull\ (T, W, TS)} & \makecell[l]{\faBatteryFull\ (T, W)} & PR & CSR & \faBatteryFull & \faTimes & \faBatteryFull\ (a, d) & \faTimes & \faBatteryFull\ (sm) & \faTimes & --- \\
cuSTINGER~\cite{green2016custinger}    & \faBatteryHalf$^*$ (W, P, TS) & \faBatteryHalf$^*$ (W, P) & TC & CSR & \faTimes & \faTimes & \faBatteryFull\ (a, d) & \faTimes &  \faBatteryFull\ (sm) & \faTimes & \makecell[l]{$^*$No details} \\
aimGraph~\cite{winter2017autonomous}  & \faBatteryHalf$^*$\ (W) & \faBatteryHalf$^*$\ (W) & --- & CSR$^*$ & \faBatteryFull & \faTimes & \faBatteryFull\ (a) & \faTimes & \faBatteryFull\ (sm) & \faTimes & $^*$Resembles CSR. \\
Hornet~\cite{busato2018hornet}  & \faBatteryHalf\ (W) & \faTimes & BFS, SpMV, $k$-Truss & CSR & \faTimes & \faBatteryFull & \faBatteryFull\ (a) & \faTimes & \faBatteryFull\ (sm) & \faTimes & --- \\ 
faimGraph~\cite{winter2018faimgraph} & \faBatteryFull\ (W, P) & \faBatteryFull\ (W, P) & \makecell[l]{PR, TC} & CSR$^*$ & \faBatteryFull & \faTimes & \faBatteryFull\ (a) &  \faTimes & \faBatteryFull\ (sm) & \faTimes & $^*$Resembles CSR \\ 
LiveGraph~\cite{zhu2019livegraph} & \faBatteryFull\ (T, P) & \faBatteryFull\ (P) & PR, CC & CSR & \faBatteryFull\ (g) & \faTimes & \faBatteryFull\ (a) & \faTimes &  \faBatteryFull\ (sm)$^*$ & \faTimes & $^*$Primarily a data store \\
\HL{\rowcolor{yellow}} GraphBolt~\cite{mariappan2019graphbolt} & \faBatteryFull\ (W) & \faTimes & PR, BP, LP, CoEM, CF, TC & CSR & \faBatteryFull & \faTimes & \faBatteryFull\ (a) & \faBatteryFull\ {(Rf/m,n)} &  \faBatteryFull\ (sm) & \faBatteryFull\ (sa$^*$, i) & $^*$Relies on BSP and Ligra's mappings \\
GraphIn~\cite{sengupta2016graphin}  & \faTimes & \faBatteryHalf\ (P) & \makecell[l]{BFS, CC, CL} & CSR + EL & \faTimes & \faTimes & \noAnswer & \faBatteryFull\ {(Rc/\mbox{\noAnswer})} & \faBatteryFull\ (sm) & \faBatteryFull\ (sa, i)$^\dagger$ & $^\dagger$Relies on GAS. \\
EvoGraph~\cite{sengupta2017evograph} & \faTimes & \faTimes & TC, CC, BFS & CSR + EL & \faTimes & \faTimes & \noAnswer & \faBatteryFull\ {(Rc/\mbox{\noAnswer})} &  \faBatteryFull\ (sm) & \faBatteryFull\ (sa, i) & --- \\ 
GraphOne~\cite{kumar2019graphone} & \faBatteryHalf\ (W, {T, P}) & \faTimes & \makecell[l]{BFS, PR, 1-Hop-query} & CSR + EL & \faBatteryFull & \faTimes & \faBatteryFull\ (a, t) & {\mbox{\faBatteryFull\ }(\mbox{\noAnswer}/\mbox{\noAnswer}) } & \faBatteryFull\ (sm, ss) & \faBatteryFull\ (sa, p) & ---  \\
\HL{\rowcolor{yellow}} GraPU~\cite{sheng2018grapu, sheng2020exploiting} & \faBatteryHalf\ (W) & \faTimes & BFS, SSSP, SSWP & AL$^*$ & \faTimes & \faTimes & \faBatteryFull\ (a) & \faBatteryFull\ {(Rf\mbox{$^\dagger$}/m)} & \faBatteryFull\ (sm) & \faBatteryFull\ (sa, i, sai) & $^*$Relies on GoFS. {$^\dagger$Relies on KickStarter} \\
\HL{\rowcolor{yellow}} RisGraph~\cite{feng2020risgraph} & \faBatteryFull\ (W) & \faTimes & CC, BFS, SSSP, SSWP & AL & \noAnswer & \noAnswer & \faBatteryFull\ (a) & \faBatteryFull\ {(Rf\mbox{$^*$}/m)} & \faBatteryFull\ (sm) & \faBatteryFull\ (sa, p) & {\mbox{$^*$}Inspired by KickStarter}  \\
\rowcolor{yellow} DZiG~\cite{mariappan2021dzig} & \faBatteryFull\ (W) & \faTimes & PR, BP, CoEM, CF, LP & AL & \noAnswer & \faTimes & \faBatteryFull\ (a) & \faBatteryFull\ (Rf/m,n) & \faBatteryFull\ (sm) & \faBatteryFull\ (sa, i) & \\
Kineograph~\cite{cheng2012kineograph}  & \faTimes & \faTimes & \makecell[l]{TR, SSSP, $k$-exposure} & KV store + AL$^*$ & \noAnswer & \faTimes & \faBatteryFull\ (a) & \faBatteryFull\ {(Rc/\mbox{\noAnswer})} &  \faBatteryFull\ (sm) & \faBatteryHalf\ (sa)$^\dagger$ & $^*$Details are unclear. $^\dagger$Uses vertex-centric  \\
Mondal et al.~\cite{mondal2012managing} & \faTimes & \faTimes & --- & KV store + documents$^*$ & \faTimes & \faTimes & \faBatteryFull\ (a) & \faTimes & \faBatteryFull\ (sm)$^*$ & \noAnswer$^*$ & $^*$Relies on CouchDB \\
CellIQ~\cite{iyer2015celliq}  & \faBatteryFull\ (P) & \faBatteryFull\ (P) & \makecell[l]{Cellular specific} & Collections (series)$^*$ & \faTimes & \faTimes & \faBatteryFull\ (a, d) & \faBatteryFull\ (Rc/\noAnswer) & \faBatteryFull\ (sm) & \faBatteryFull\ (sa, i)$^\dagger$ & $^*$Uses RDDs. $^\dagger$Focus on geopartitioning \\ 
\HL{\rowcolor{yellow}} iGraph~\cite{ju2016igraph} & \noAnswer & \faTimes & PR & {RDDs} & \faTimes & \faTimes & \faBatteryFull & \faBatteryFull\ (Rc/\noAnswer) & \faBatteryFull\ (sm) & \faBatteryFull\ (sa, i)$^*$ & $^*$Relies on vertex-centric \& BSP \\
GraphTau~\cite{iyer2016time} & \faTimes & \faTimes & PR, CC & {RDDs (series)} & \faTimes & \faTimes & \noAnswer & \faBatteryFull\ (Rc/\noAnswer) $^*$ & \faBatteryFull\ (sm) & \faBatteryFull\ (sa, i, p)$^\dagger$ & $^\dagger$Relies on BSP \& vertex-centric. \\
ZipG~\cite{khandelwal2017zipg} & \faBatteryFull\ (T, P, TS) & \faBatteryFull\ (P) & TAO \& LinkBench & \makecell[l]{Compressed flat files} & \faTimes & \faTimes & \faBatteryFull\ (a) & \faTimes & \faBatteryFull\ (sm) & \faTimes & --- \\
\HL{\rowcolor{yellow}} Sprouter~\cite{abughofa2018sprouter} & \noAnswer & \noAnswer & PR & Tables$^*$ & \faTimes & \faTimes & \noAnswer & \faTimes & \faBatteryFull\ (sm) & \faTimes & $^*$Relies on HGraphDB \\
\HL{\rowcolor{yellow}} DeltaGraph~\cite{dexter2016lazy} & \faTimes & \faTimes & --- & {Inductive graphs}$^*$ & \faTimes & \faTimes & \faTimes & \faTimes & \faBatteryFull\ (sm, am) & \faBatteryFull\ (sa)$^\dagger$ & \makecell[l]{$^*$Specific to functional languages~\cite{dexter2016lazy}.\\ $^\dagger$Mappings of vertices/edges} \\
GPMA (Sha~\cite{sha2017accelerating})  & \faBatteryHalf\ (TS) & \faTimes & PR, BFS, CC & \makecell[l]{Tree-based (PMA)} & \faBatteryHalf\ (g)$^*$ & \faTimes & \faBatteryFull\ (a) & \faTimes & \faBatteryFull\ (sm) & \faTimes & $^*$A contiguous array with gaps in it \\
%
%
Aspen~\cite{dhulipala2019low} & \faTimes &  \faTimes$^*$ & \makecell[l]{BFS, BC, MIS, 2-hop, CL} & \makecell[l]{Tree-based (C-Trees)} & \faBatteryFull & \faTimes & \faBatteryFull\ (a) & \faTimes  & \faBatteryFull\ (sm) & \faBatteryHalf\ (sa)$^*$ & $^*$Relies on Ligra \\ 
Tegra~\cite{iyer2019tegra} & \faBatteryFull\ (P) & \faBatteryFull\ (P) & PR, CC & Tree-based (PART~\cite{davepersistent}) & \faBatteryFull$^*$ & \faTimes & \faBatteryFull\ (a) & \faBatteryFull\ (Rc/m,n) & \faBatteryFull\ (sm) & \faBatteryFull\ (sa$^\dagger$, i, p) & $^*$For properties. $^\dagger$Relies on GAS \\
%
%
GraphInc~\cite{cai2012facilitating} & \faBatteryFull\ (P) & \faBatteryFull\ (P) & SSSP, CC, PR & \noAnswer$^*$ & \faTimes & \faTimes & \faBatteryHalf\ (a) & \faBatteryFull\ (Rc/\noAnswer)  & \faBatteryFull\ (sm) & \faBatteryHalf\ (sa)$^*$ & $^*$Uses Giraph's structures and model \\ 
UNICORN~\cite{suzumura2014towards} & \faTimes & \faTimes & \makecell[l]{PR, RW} & \noAnswer$^*$ & \faTimes & \faTimes & \noAnswer & \faBatteryFull & \faBatteryFull\ (sm) & \faBatteryFull\ (sa, i) & $^*$Uses InfoSphere \\ 
KickStarter~\cite{vora2017kickstarter}  & \faBatteryHalf\ (W) & \faTimes & SSWP, CC, SSSP, BFS & na$^*$ & na$^*$ & na$^*$ & na$^*$ & \faBatteryFull\ (Rf/m) &  \faBatteryFull\ (sm) & na$^*$ & $^*$Kickstarter is a \emph{runtime technique} \\ 
\bottomrule
\end{tabular}
\vspaceSQ{-1em}
\caption{
Comparison of selected representative works. {They are grouped by the used
fundamental graph representation (within each group, by
publication date).}
``\textbf{Rich edge/vertex data}'': enabling additional data to be attached to
an edge or a vertex (\ul{``T''}: type, \ul{``P''}: property, \ul{``W''}:
weight, \ul{``TS''}: timestamp).
``\textbf{Tested analytics workloads}'': evaluated workloads beyond simple
queries {(\textbf{PR}: PageRank, \textbf{TR}: TunkRank, \textbf{CL}:
clustering, \textbf{BC}: Betweenness Centrality, \textbf{CC}: Connected
Components, \textbf{BFS}: Breadth-First Search, \textbf{SSSP}: Single Source
Shortest Paths, \textbf{DFS}: Depth-First Search, \textbf{TC}: Triangle
Counting, \textbf{SpMV}: Sparse matrix-vector multiplication, \textbf{BP}:
Belief Propagation, \textbf{LP}: Label Propagation, \textbf{CoEM}: Co-Training
Expectation Maximization, \textbf{CF}: Collaborative Filtering, \textbf{SSWP}:
Single Source Widest Path, \textbf{TAO \& LinkBench}: workloads used in
Facebook's TAO and in LinkBench~\mbox{\cite{armstrong2013linkbench}},
\textbf{MIS}: Maximum Independent Set), \textbf{RW}: Random Walk}.
{``\textbf{Fundamental Representation}'': A key representation used to store
the graph structure; all representation are explained in
Section~\mbox{\ref{sec:taxo}}.}
{
``\textbf{iB}'': Is blocking used to increase the locality of edges
\emph{within} the representation of a single neighborhood?
``\textbf{(g)}'': one uses empty \emph{gaps} at the ends of blocks, to provide
pre-allocated empty storage for faster edge insertions.
``\textbf{aB}'': Is blocking used to increase the locality of edges
\emph{across} different neighborhoods (i.e., can one store different
neighborhoods within \emph{one block})?
``\textbf{Id}'': Is indexing used?
``\textbf{(a)}'': Indexing of the graph adjacency data, ``\textbf{(d)}'': Indexing of
rich edge/vertex data, ``\textbf{(t)}'': Indexing of different graph snapshots, in the time dimension?
``\textbf{Ic}'': Are incremental changes supported?
{``\textbf{Rc}'': incremental changes based on recomputation (the ``offline approach''). ``\textbf{Rf}'': incremental changes based on refinement (the ``online approach'').}
``\textbf{(m)}'': Explicit support for \emph{monotonic} algorithms in the context of incremental changes.
{``\textbf{(m,n)}'': Explicit support for both \emph{monotonic} and \emph{non-monotonic} algorithms in the context of incremental changes.}
%
%
``\textbf{PrM}'': Does the system offer a dedicated programming model (or API) related to graph \emph{modifications}?
``\textbf{(sm)}'': API for simple graph modifications.
``\textbf{(am)}'': API for advanced graph modifications.
``\textbf{(tr)}'': API for triggered reactions to graph modifications.
``\textbf{(ss)}'': API for manipulating with the updates awaiting being ingested (e.g., stored in the log). 
``\textbf{PrC}'': Does the system offer a dedicated programming model (or API) related to graph \emph{computations}
(i.e., analytics running on top of the graph being modified)?
``\textbf{(sa)}'': API for graph algorithms / analytics (e.g., PageRank) processing the main (i.e., up-to-date) graph snapshot.
``\textbf{(p)}'': API for graph algorithms / analytics (e.g., PageRank) processing the \emph{past} graph snapshots.
``\textbf{(i)}'': API for incremental processing of graph algorithms / analytics.
``\textbf{(sai)}'' (i.e., (sa) + (i)): API for graph algorithms / analytics processing the \emph{incremental changes themselves}.
}
``\faBatteryFull'', ``\faBatteryHalf'', ``\faTimes'': A design offers a given
feature, offers it in a limited way, and does not offer it,
respectively.
``\noAnswer'': Unknown.
%
} 
\label{tab:2}
\vspaceSQ{-1.75em}
\end{table*}

\subsection{{{Graph Storage Architecture and Mutations}}}
\label{sec:analysis_conc}

{We start with analyzing the method for achieving concurrency between
updates and queries. 
Note that, with queries, we mean both local (fine) reads (e.g., fetching a
weight of a given edge), but also global analytics (e.g., running
PageRank) that also do not modify the graph structure.}

{First, most frameworks use snapshots. We observe that such frameworks have
also some other snapshot-related design feature, for example Grace (uses
snapshots also to implement transactions), GraphTau and Tegra (both support
storing past snapshots), or DeltaGraph (harnesses Haskell's feature to create
snapshots).}
{The most popular mechanism for creating snapshots is copy-on-write, used
in Grace, LLAMA, and others. The details (e.g., layouts or structures
being copied) heavily depend on a specific system. Kineograph uses snapshots created periodically.
GraphOne uses tombstones that mark which edges are to be removed, and thus could
be swapped with new edges to be inserted.}
\sethlcolor{lyellow}{Second, a certain group of frameworks use fine-grained
synchronization. The interleaving of updates and read queries is supported only
with respect to fine reads (i.e., parallel \emph{ingestion} of updates
\emph{while} running global analytics such as PageRank are not supported in the
considered systems).}\sethlcolor{yellow}
{Furthermore, two interesting methods for efficient concurrent ingestion of
updates and queries have recently been proposed in the RisGraph
system~\mbox{\cite{feng2020risgraph}} and by Sha et
al.~\mbox{\cite{sha2017accelerating}}. The former uses \textbf{scheduling of
updates}, i.e., the system uses fine-grained synchronization enhanced with a
specialized scheduler that manipulates the ordering and timing of applying incoming
updates to maximize throughput and minimize latency (different
timings of applying updates may result in different performance
penalties).
In the latter, one \textbf{overlaps} the ingestion of updates 
with transferring the information about queries (e.g., over PCIe).}

{We observe that, while almost all systems use batching, only a few sort
batches; the sorting overhead often exceeds benefits from faster ingestion.
Next, only five frameworks support transactions, and four in total offer the
ACID semantics of transactions. This illustrates that performance and high
ingestion ratios are prioritized in the design of streaming frameworks over
overall system robustness. Some frameworks that support ACID transactions rely
with this respect on some underlying data store infrastructure: Sinfonia (for
Concerto) and CouchDB (for the system by Mondal et al.).  Others (Grace and
LiveGraph) provide their own implementations of ACID.}

\subsection{{Analysis of Support for Keeping Historical Data}}

Our analysis shows that reasonably many systems ($>$10) support keeping
past data in some way. 
The details heavily depend on a given system. 
{For example, Kineograph focuses
on keeping past snapshots created periodically by the underlying runtime.
Tegra enhances this approach by enabling the user to additionally create
snapshots at arbitrary times.}

{To reduce both storage and performance overheads, 
the authors of Tegra observe that
one could employ some combination of keeping snapshots {and} maintaining graph changes.
Thus, performance would be improved as one would not have to start from scratch
to arrive at a certain snapshot. Simultaneously,
the memory pressure is reduced because not all snapshots are stored explicitly.
However, this approach is not heavily explored in the literature.
Systems such as STINGER or ZipG enable
maintaining timestamps of graph mutations, which facilitates deriving the graph
state at a selected point in time. However, these systems do not offer a dedicated API
for such snapshot derivation, delegating such logic to the system user.}

{Systems keeping past snapshots often employ some additional form
of reusing the graph structure across snapshots, to reduce memory overheads. For example, LLAMA employs
a scheme in which parts of the graph, which are identical across the snapshots, are stored
only once. Tegra uses a similar approach, with its Distributed Graph Snapshot Index.
Some systems also use persistent storage to further alleviate the issue of
maintaining multiple snapshots. An example such system is Tegra.}

{CelliQ, GraphTau,
a system by Sha et al.~, and Tegra also} support the sliding window model.
\hl{This is possible as they enable keeping past snapshots as well as
obtaining the differences between these snapshots. Thus, the user can choose
the range of past updates (e.g., incoming edges) when
computing a given graph property.} They also usually maintain indexing
structures over historical data to accelerate fetching respective
  past instances. Tegra has a particularly rich set of
  features for analyzing historical data efficiently, approaching
  in its scope offline temporal frameworks such as
  Chronos~\mbox{\cite{han2014chronos}}. Another system with a rich set of
  such features is Kineograph, the only one to support the exponential
  decay model of the visibility of past updates.

\marginpar{\vspace{-8em}\colorbox{yellow}{\textbf{R-4}}\\\colorbox{yellow}{\textbf{(2)}}}

\subsection{{Analysis of Graph Representations}}

{Most frameworks use some form of \textbf{CSR}. In certain cases,
\textbf{CSR is combined with an EL} to form a dual representation; EL is often
(but not exclusively) used in such cases as a log to store the incoming
edges, for example in GraphOne.
Certain other frameworks use \textbf{AL}, prioritizing the flexibility of graph
updates over locality of accesses.}

Most frameworks based on CSR
use blocking within neighborhoods (i.e., each neighborhood consists of a linked
list of contiguous blocks (chunks)). This enables a tradeoff between the
locality of accesses and time to perform updates. The smaller the chunks are,
the easier is to update a graph, but simultaneously traversing vertex
neighborhoods requires more random memory accesses. Larger chunks improve
locality of traversals, but require more time to update the graph structure.
{Two frameworks (Concerto and Hornet) use blocking across neighborhoods.
This may help in achieving more locality whenever processing many small
neighborhoods that fit in a block.}

A few systems use \textbf{tree based} graph representations. For
example, Sha et al.~\cite{sha2017accelerating} use a variant of \emph{packed
memory array} (PMA), which is an array with all neighborhoods (i.e.,
essentially a CSR) augmented with an implicit binary tree structure for 
edge insertions and deletions in $O(\log^2 n)$ time.

Frameworks constructed on top of a more general infrastructure use a
representation provided by the underlying system. For example,
GraphTau~\cite{iyer2016time}, built on top of Apache
Spark~\cite{zaharia2016apache}, {uses the underlying abstraction called
Resilient Distributed Datasets (RDDs)~\mbox{\cite{zaharia2012resilient, zaharia2016apache}}.
RDDs can be implemented differently, for example using HDFS files~\mbox{\cite{zaharia2012resilient}}}. {Other
frameworks use data representations that are harnessed by general
processing systems or databases, for example KV stores, tables, or general
collections.}

{All considered frameworks use some form of indexing. However, the indexes 
mostly keep the locations of vertex neighborhoods.
Such an index is usually a simple array of size~$n$, with cell~$i$
storing a pointer to the neighborhood~$N_i$; this is a standard design for
frameworks based on CSR. Whenever CSR is combined with blocking, a
corresponding framework also offers the indexing of blocks used for storing
neighborhoods contiguously. For example, this is the case for faimGraph and
LiveGraph.
Frameworks based on more complex underlying infrastructure benefit from
indexing structures offered by the underlying system. For example, Concerto
uses hash indexing offered by MySQL, and CellIQ and others can use structures
offered by Spark.
Finally, relatively few frameworks apply indexing of additional rich vertex or
edge data, such as properties or labels. This is due to the fact that streaming
frameworks, unlike graph databases, place more focus on the graph structure and
much less on rich attached data. For example, STINGER indexes edges and
vertices with given labels.}

\subsection{{Analysis of Support for Incremental Changes}}

\sethlcolor{lyellow}
{Around half of the considered frameworks support incremental changes to
accelerate global graph analytics running on top of the maintained graph
datasets. Frameworks that do not support them (e.g., faimGraph) usually put
less focus on global analytics in the streaming setting.}
\sethlcolor{yellow}
{Among systems that do support incremental computation, many are offline.
These systems offer different mechanisms for detecting which vertices must be recomputed
to update the analytics results to reflect recent graph mutations.
This includes GraphIn, EvoGraph, Tegra, Kineograph, and others.
Here, Tegra maintains is additionally able to incorporate incremental computation for
\emph{different} past snapshots, due to its focus on keeping and analyzing
historical data.}

{Some systems are online, focusing on update refinement.}
For example, GraphBolt and KickStarter both carefully track dependencies
between vertex values (that are being computed) and edge modifications. The
differences between these two {are driven by targeted classes of algorithms.
GraphBolt assumes the Bulk Synchronous Parallel
(BSP)~\mbox{\cite{valiant1990bridging}} computation and thus ensures
synchronous semantics. KickStarter instead focuses on path-based monotonic
algorithms such as SSSP. It provides different optimizations. For example, it
uses the fact that in many graph algorithms, the vertex value is simply
\emph{selected} from one single incoming edge. Unlike some other systems (e.g.,
Kineograph), GraphBolt and KickStarter enable performance gains also in the
event of edge deletions, not only insertions.
Finally, a very recent system called DZiG~\mbox{\cite{mariappan2021dzig}}
improves the incremental capabilities of GraphBolt by utilizing
the fact that in iterative graph algorithms, values of many vertices
\emph{stabilize} across iterations. This enables opportunities for annihilating
unnecessary refinements.}
\tr{{In contrast to GraphBolt, GraphInc maintains the state of \emph{all}
computations performed, and uses this state whenever possible to quickly
deliver results if a graph changes. However, the amount of information tracked
in GraphInc (i.e., all the vertex states and incoming messages) is larger than
in GraphBolt or Kickstarter.}}
RisGraph applies KickStarter's approach for incremental computation to its
design based on concurrent ingestion of fine-grained updates and queries.

{Almost} all the systems that support incremental changes focus on
\textbf{monotonic} graph algorithms, i.e., algorithms, where the computed
properties (e.g., vertex distances) are consistently either increasing or
decreasing.
{Here, GraphBolt, DZiG, and Tegra also cover \textbf{non-monotonic} algorithms, such as
Belief Propagation, Co-Training Expectation Maximization, or Collaborative
Filtering}.

\subsection{{Analysis of Offered Programming APIs and Models}}

\textbf{{Graph Mutations}}
{We first analyze the supported APIs for \textbf{graph modifications}. All
considered frameworks support a simple API for manipulating the graph, which
includes operations such as adding or removing an edge. However, some
frameworks offer more capabilities. {We identify three such frameworks:
Concerto, DeltaGraph, and GraphOne}. Concerto has special functions for
programming triggered events, i.e., events taking place automatically upon
certain specific graph modifications. DeltaGraph offers {functions for
\emph{merging} different graphs}. Finally, GraphOne {enables accessing and
analyzing the updates that are still waiting (in a special log structure) to be
ingested into the main graph structure}. This can be used to apply some form of
preprocessing of the updates, {before they are applied to the main graph
data,} or to run some analytics on the updates.}

\textbf{{Graph Analytics}}
{We also discuss supported APIs for running global analytics on the maintained
graph. First, we observe that a large fraction of frameworks do not support
developing graph analytics at all. These systems, for example faimGraph, focus
completely on graph mutations and local queries. However, other systems do
offer an API for graph analytics (e.g., PageRank) processing the main (live)
graph snapshot. These systems usually harness some existing programming model,
for example Bulk Synchronous Parallel (BSP)~\mbox{\cite{valiant1990bridging}}.
Furthermore, frameworks that enable maintaining past snapshots, for example
Tegra, also offer APIs for running analytics on such {snapshots}.
{These APIs are similar to the APIs for processing the main (live) graph
versions, \hl{with a difference} that the user also must identify the targeted
specific past snapshot}.}

\marginpar{\vspace{-2em}\colorbox{yellow}{\textbf{R-3}}\\\colorbox{yellow}{\textbf{(1)}}}

Finally, systems offering incremental changes also offer the associated APIs.
{Online systems such as GraphBolt and DZiG provide user-defined algorithm
specific functions that enable refining aggregation values. Example functions
are \texttt{propagate}, \texttt{retract}, or \texttt{repropagate}. The goal of
these functions is to appropriately implement the logic of contributing to, or
withdrawing from, vertex aggregation values.
Offline systems often provide some way to indicate which vertices must be
recomputed.} For example, GraphIn and EvoGraph make the developer responsible
for implementing a dedicated function that detects \emph{inconsistent
  vertices}, i.e., vertices that became affected by graph updates. This
  function takes as arguments a batch of incoming updates and the vertex
  property related to the graph problem being solved (e.g., a parent in the BFS
  traversal problem). Whenever any update in the batch affects a specified
  property of some vertex, this vertex is marked as inconsistent, and is
  scheduled for recomputation.
{Another example is Tegra. It offers two functions, \texttt{diff} and
\texttt{expand}.  The former returns the difference (i.e., a modified subgraph)
between two graph snapshots.  The latter expands this subgraph with its 1--hop
neighborhood. The resulting part of the graph is then scheduled for
recomputation. A similar approach is used in other systems based on the
underlying Spark infrastructure, i.e., CellIQ.}

\tr{Overall, as of now, there are no {established} programming models for dynamic graph
analysis. Most frameworks, for example GraphInc, {fall back} to a model used
for static graph processing (most often the vertex-centric
  model~\cite{malewicz2010pregel, kalavri2017high}), and make the dynamic
  nature of the graph transparent to the developer. Another recent example is
  GraphBolt that offers the Bulk Synchronous Parallel
  (BSP)~\cite{valiant1990bridging} programming model and combines it with
  incremental updates to be able to solve certain graph problems on dynamic
  graphs. 
Some engines, however, extend an existing model for static graph
processing. For example, GraphIn extends the gather-apply-scatter (GAS)
paradigm~\mbox{\cite{low2014graphlab}} to enable reacting to incremental
updates.
Specifically, the key part of this Incremental Gather Apply Scatter (I-GAS) 
is an API that enables the user to specify how to
construct the \emph{inconsistency graph} i.e., a part of the processed graph
that must be recomputed in order to appropriately update the desired results
(for a specific graph problem such as BFS or PageRank).
For this, the user implements a designated method that takes
as input the batch of next graph updates, and uses this information
to construct a list of vertices and/or edges, for which a given property
(e.g., the rank) must be recomputed.
This also includes a user-defined function that acts as a heuristic to
check if a static full recomputation is cheaper in expectation than an
incremental pass. It is the users responsibility to ensure that correctness is
guaranteed in this model, for example by conservatively marking vertices
inconsistent. Graph updates can consist of both inserts and removals. They
are applied in batches and exposed to the user automatically by a list of
inconsistent vertices for which properties (e.g., vertex degree) have been
changed by the update. Therefore, queries are always computed on the most
recent graph state.}

\all{Finally, certain systems offer a \textbf{novel} model for harnessing the
dynamic nature of streaming graphs. An example is Tegra~\cite{iyer2019tegra}, a
recent design that offers a Timelapse abstraction and an ICE model that,
together, enable retrieving past graph snapshots and using them when deriving
different structural graph properties.}

\subsection{Supported Types of Graph Updates}
\vspaceSQ{-0.25em}

Different systems support different forms of graph updates. The most widespread
update is \textbf{edge insertion}, offered by all the considered systems.
Second, \textbf{edge deletions} are supported by most frameworks.  Finally, a
system can also explicitly enable \textbf{adding} or \textbf{removing} a
specified \textbf{vertex}. In the latter, a given vertex is removed with its
adjacent edges.

\subsection{{Distributed Designs}}

{Almost all the distributed frameworks rely on underlying existing backend
infrastructure such as Spark (CelliQ, GraphTau, Tegra, ZipG, iGraph, Sprouter),
CouchDB (work by Mondal et al.), or Giraph (GraphInc). Two frameworks that
offer specialized distributed implementations are Kineograph and Concerto.
Streaming frameworks rely on distribution mostly to enable scaling to larger
datasets (by distributing a larger graph instance over multiple nodes) and to
increase the throughput of graph queries (by distributing computation and
update ingestion over multiple nodes). Furthermore, streaming frameworks rely on mature
backends for effective fault tolerance.}

\subsection{{Computation vs.~Storage}}

\marginpar{\vspace{4em}\colorbox{yellow}{\textbf{R-3}}\\\colorbox{yellow}{\textbf{(2)}}}

{Some systems focus primarily on \textbf{computation} aspects of dynamic
graph processing. For example, KickStarter offers an interesting model for
incremental computation, while storage is outside its focus. \hl{Similarly, DZiG
and GraphBolt focus on incremental computation, extending KickStarter's
capabilities by -- respectively -- targeting BSP programs and by
harnessing certain properties of such programs for more performance gains.} Contrarily,
systems such as Aspen focus on \textbf{storage}, usually by providing elaborate
graph representations.
Some systems, such as Tegra, come with enhancements into both aspects.}

\subsection{{Analysis of Relations to Theoretical Models }}

{First, despite the similarity of names, the (theoretical) field of
\emph{streaming graph algorithms} is \emph{not} well connected to graph
streaming frameworks: the focus of the former are fast algorithms operating
with tight memory constraints that \emph{by assumption} prevent from keeping
the whole graph in memory, which is not often the case for the latter.
Similarly, \emph{graph sketching} focuses on approximate algorithms in a
streaming setting, which is of little interest to streaming frameworks. 
On the other hand, the (theoretical) settings of \emph{dynamic graph
algorithms} and \emph{parallel dynamic graph algorithms} are similar to that of
the streaming frameworks. Their common assumption is that the whole maintained
graph is available for queries (in-memory), which is also common for the
streaming frameworks. Moreover, the \emph{batch dynamic} model is even closer,
as it explicitly assumes that edge updates arrive in batches, which reflects a
common optimization in the streaming frameworks. We conclude that future
developments in streaming frameworks could benefit from deepened understanding
of the above mentioned theoretical areas. For example, one could use the recent
parallel batch dynamic graph connectivity
algorithm~\mbox{\cite{acar2019parallel}} in the implementation of any streaming
framework, for more efficient connected components problem solution.}

\iftr

\section{Discussion of Selected Frameworks}

We now provide general descriptions about selected frameworks,
for readers interested in some specific systems.

\subsection{STINGER~\cite{ediger2012stinger} And Its Variants}
\vspaceSQ{-0.25em}

STINGER~\cite{ediger2012stinger} is a data structure and a corresponding
software package.
It adapts and extends the \emph{{CSR format}} to support graph updates.
Contrarily to the static CSR design, where IDs of the neighbors of a given
vertex are stored contiguously, neighbor IDs in STINGER are divided into
contiguous {blocks} of a pre-selected size. These blocks form a \emph{linked
list}, i.e., STINGER uses the blocking design. The block size is identical for
all the blocks except for the last blocks in each list. One neighbor vertex
ID~$u$ in the neighborhood of a vertex~$v$ corresponds to one edge $(v,u)$.
STINGER supports both vertices and edges with different \emph{types}. One
vertex can have adjacent edges of different types. One block always contains
edges of {one type only}.  Besides the associated neighbor vertex ID and type,
each edge has its weight and two time stamps. {The time stamps can be used
in algorithms to filter edges, for example based on the insertion time.} In
addition to this, each edge block contains certain metadata, for example lowest
and highest time stamps in a given block.  Moreover, STINGER provides the edge
type array (ETA) \emph{index data structure}. ETA contains pointers to all
blocks with edges of a {given type} {to accelerate algorithms that operate
on specific edge types.}

To increase \emph{{parallelism}}, STINGER updates a graph in \emph{batches}.
For graphs that are not scale-free, a batch of around 100,000 updates is first
sorted so that updates to different vertices are grouped. In the process,
deletions may be separated from insertions (they can also be processed in
parallel with insertions). For scale-free graphs, there is no sorting phase
{since a small number of vertices will face many updates which leads to
workload imbalance. Instead, each update is processed in parallel.} \emph{Fine
locking} on single edges is used for synchronization of updates to the
neighborhood of the same vertex.  To insert an edge or to verify if an edge
exists, one traverses a selected list of blocks, taking $O(d)$ time.
Consequently, inserting an edge into $N_v$ takes $O(d_v)$ work and depth.
STINGER is optimized for the {Cray XMT supercomputing systems} that allow for
massive thread-level parallelism.  Still, it can also be executed on general
multi-core commodity servers.  {Contrarily to other works, STINGER and its
variants does not provide a framework but a library to operate on the data
structure. Therefore, the user is in full control, for example to determine
when updates are applied and what programming model is used.}


\textbf{DISTINGER~\cite{feng2015distinger}} is a distributed version of STINGER
that targets \emph{``shared-nothing'' commodity clusters}.  DISTINGER inherits
the STINGER design, with the following modifications.  First, a designated
\emph{master} process is used to interact between the DISTINGER instance and
the outside world. The master process maps external (application-level) vertex
IDs to the internal IDs used by DISTINGER. The master process maintains a list
of \emph{slave} processes and it assigns incoming  queries and updates to
slaves that maintain the associated part of the processed graph.  Each slave
maintains and is responsible for updating a portion of the vertices together
with edges attached to each of these vertices. The graph is partitioned with a
simple hash-based scheme. The inter-process communication uses
MPI~\cite{fompi-paper, advancedMPI} with established optimizations such as
message batching or overlap of computation and communication.


\textbf{cuSTINGER~\cite{green2016custinger}} extends STINGER for CUDA GPUs.
The main design change is to replace lists of edge blocks with \emph{contiguous
adjacency arrays}, {i.e. a single adjacency array for each vertex.}
Moreover, contrarily to STINGER, cuSTINGER \emph{always separately
processes updates and deletions}, to better utilize \emph{massive parallelism in GPUs}.
cuSTINGER offers several ``meta-data modes'': based on the user's needs, the
framework can support only unweighted edges, weighted edges without any
additional associated data, or edges with weights, types, and additional data
such as time stamps.  However, the paper focuses on
unweighted graphs that do not use time stamps and types, and the exact GPU
design of the last two modes is unclear~\cite{green2016custinger}.

\subsection{Work by Mondal et al.~\cite{mondal2012managing}}

A system by Mondal et al.~\cite{mondal2012managing} focuses on data
replication, graph partitioning, and load balancing.  As such, the system is
distributed: on each compute node, a replication manager decides locally (based
on analyzing graph queries) what vertex is replicated and what compute nodes
store its copies.  The main contribution is the definition of a \emph{fairness
criterion} which denotes that at least a certain configurable fraction of
neighboring vertices must be replicated on some compute node.  This approach
reduces pressure on network bandwidth and improves latency for queries that
need to fetch neighborhoods (common in social network analysis).  The framework
stores the data on Apache CouchDB~\cite{apache_couchdb}, an in-memory key-value
store. No detailed information how the data is represented is given.
%

\subsection{LLAMA~\cite{macko2015llama}}
\vspaceSQ{-0.25em}

LLAMA~\cite{macko2015llama} {(\emph{L}inked-node analytics using
\emph{LA}rge \emph{M}ultiversioned \emph{A}rrays)} -- similarly to STINGER --
digests graph updates in batches. It differs from STINGER in that each such
batch generates a new \emph{snapshot} of graph data using a copy-on-write
approach.  Specifically, the graph in LLAMA is represented using {a variant
of CSR that relies on \emph{large multiversioned arrays}.  Contrarily to CSR,
the array that maps vertices to per-vertex structures is divided into smaller
parts, so called data pages.  Each part can belong to a different snapshot and
contains pointers to the single edge array that stores graph edges.
To create a new snapshot, new data pages and a new edge array are allocated
that hold the delta that represents the update. This design points to older
snapshots and thus shares some data pages and parts of the edge array
among all snapshots, enabling lightweight updates.}
For example, if there is a batch with edge insertions into the neighborhood
of vertex~$v$, this batch may become a part of $v$'s adjacency list within a
new snapshot, but only represents the update and relies on the old graph data.
{Contiguous allocations are used for all data structures to improve
allocation and access time.}

LLAMA also focuses on out-of-memory graph processing. For this,
snapshots can be persisted on disk and mapped to memory using \emph{mmap}. The
system is implemented as a library, such that users are responsible to ingest
graph updates and can use a programming model of their choice.

LLAMA does not impose any specific programming model. Instead, if offers a
simple API to iterate over the neighbors of a given vertex~$v$ (most recent
ones, or the ones belonging to a given snapshot).

\subsection{GraphIn~\cite{sengupta2016graphin}}
\vspaceSQ{-0.35em}

GraphIn~\cite{sengupta2016graphin} uses a \emph{hybrid} dynamic data
structure. First, it uses an AM (in the \emph{CSR format}) to
store the adjacency data. This part is \emph{static} and is not modified when
updates arrive. Second, incremental graph updates are stored in dedicated \emph{edge
lists}. Every now and then, the AM with graph structure and the edge lists with
updates are merged to update the structure of AM. Such a design maximizes
performance and the amount of used parallelism when accessing the graph
structure that is mostly stored in the CSR format.

\subsection{GraphTau~\cite{iyer2016time}}
\vspaceSQ{-0.35em}


GraphTau~\cite{iyer2016time} is a framework based on Apache Spark and its data
model called resilient distributed datasets (RDD)~\cite{zaharia2016apache}.
RDDs are read-only, immutable, partitioned collections of data sets that can be
modified by different operators (e.g., map, reduce, filter, and join).
Similarly to GraphX~\cite{gonzalez2014graphx}, GraphTau exploits RDDs and
stores a graph snapshot (called a GraphStream) using \emph{two RDDs: an RDD for
storing vertices and edges}. Due to the snapshots, the framework offers fault
tolerance by replaying the processing of respective data streams. Different
operations allow to receive data form multiple sources (including graph
databases such as Neo4j and Titan) and to include unstructured and tabular data
(e.g., from RDBMS).
To maximize parallelism
when ingesting updates, it applies the snapshot scheme: graph workloads run
concurrently with graph updates using different snapshots.

{
	GraphTau only enables using the window sliding model.
	It provides options to write custom
	iterative and window algorithms by defining a directed acyclic graph (DAG) of
	operations. The underlying Apache Spark framework analyzes the DAG and
	processes the data in parallel on a compute cluster.
	For example, it is possible to write a function
	that explicitly handles sub-graphs that are not part of the graph any more due to the shift of the sliding window.
	The work focuses on iterative algorithms and stops the next iteration when an update arrives even when the algorithm has not converged yet.
	This is not an issue since the implemented algorithms (PageRank and CC) can reuse the previous result and converge on the updated snapshot.
	In GraphTau, graph updates can consist of both inserts and removals.
	They are applied in batches and exposed to the program automatically by the new graph snapshot.
	Therefore, queries are always computed on the most recent graph for the selected window.
}

\subsection{faimGraph~\cite{winter2018faimgraph}}
\vspaceSQ{-0.35em}

faimGraph~\cite{winter2018faimgraph} (\emph{f}ully-dynamic,
\emph{a}utonomous, \emph{i}ndependent \emph{m}anagement of \emph{graphs}) is a
library for graph processing on a single GPU with focus on fully-dynamic edge
and vertex updates (add, remove) { - contrarily, some GPU frameworks \mbox{\cite{winter2017autonomous, sha2017accelerating}}
focus only on edge updates.
It allocates a single
block of memory on the GPU to prevent memory fragmentation.
A memory manager autonomously handles data
management without round-trips to the CPU, enabling fast initialization
and efficient updates since no intervention from the host is required.
Generally, the GPU memory is partitioned into vertex data, edge data and management data structures such as
index queues which keep track of free memory.
Also, the algorithms that run on the graph operate on this allocated memory.
The vertex data and the edge data grow from opposite sides of the memory region to not restrict the amount of vertices and edges.
Vertices are stored in dedicated vertex \emph{data blocks} that can also
contain user-defined properties and meta information.
For example, vertices store their according host identifier since the host can dynamically create vertices with arbitrary identifiers and vertices are therefore
identified on the GPU using their memory offset.
To store edges, the library implements a combination of the linked list and adjacency array resulting in \emph{pages that form a linked list}.
This enables the growth and shrink of edge lists and also optimizes memory locality.
Further, properties can be stored together with edges.
The design does not
return free memory to the device, but keeps it allocated as it might be used
during graph processing - so the parallel use of the GPU for other processing
is limited.  In such cases, faimGraph can be reinitialized to claim memory (or
expand memory if needed).}
Updates can be received from the device or from the host.
Further,
%
%
faimGraph relies on a bulk update scheme, where queries cannot be interleaved
with updates. However, the library supports exploiting parallelism of the GPU by running
updates in parallel.
faimGraph mainly presents a new data structure and therefore does not enforce a certain programming model.

\subsection{Hornet~\cite{busato2018hornet}}
\vspaceSQ{-0.35em}

%

Hornet~\cite{busato2018hornet} is a data structure and associated system that
focuses on efficient batch updates (inserting, deleting, and updating vertices
and edges), and more effective memory utilization by requiring no re-allocation
and no re-initialization of used data structures during computation.  To
achieve this, Hornet implements its own memory manager. The graph is maintained
using an AL: vertices are stored in an array, with pointers pointing to the
associated adjacency list. The lists are (transparently to the user) stored in
\emph{blocks} that can hold edges in counts that are powers of two. The allocation of
specific edge lists to specific blocks is resolved by the system.  Finally,
$B^+$ trees are used to maintain the blocks efficiently and to keep track of
empty space.

Hornet implements the bulk update scheme in which bulk updates and graph
queries alternate.
%
%
The bulk update exploits parallelism for efficient usage of the GPU resources.
{No specific programming model is enforced.}

\subsection{GraphOne~\cite{kumar2019graphone}}
\vspaceSQ{-0.35em}

GraphOne~\cite{kumar2019graphone} focuses on the parallel
\emph{efficient} execution of both global graph algorithms (such as PageRank) and
{stream analytics while supporting high velocity
streaming graph updates.
To achieve this goal, the graph updates are first appended to an
edge list. If this edge list exceeds a
certain archiving threshold, the updates are moved as a batch in parallel from the edge list to
the adjacency list.
Only a small amount of overlapping data must be kept both in the edge list and the adjacency list
to ensure no interruption of already running graph algorithms.
Similarly to faimGraph \mbox{\cite{winter2018faimgraph}}, the adjacency list consists of chained, cache-aligned blocks to increase locality.
Further, high degree vertices store their edges in page-aligned memory to reduce chaining and their memory footprint.
This design provides different advantages:
First, it exploits the fast edge list for immediate updates
and stream processing, and provides snapshots of the adjacency list for long
running graph analytics.
Second, two ways to access
the graph are offered (stream or batch analysis), allowing to select the most suitable
way for a given algorithm.
Third, multiple snapshots of the adjacency list can be
created in a lightweight way, such that queries are processed immediately when they
arrive.
Since deletes are applied by marking the according edges or vertices to not affect snapshots,
a compaction phase removes stale data.
The graph data store allows to implement vertex-centric, edge-centric and Sliding Window algorithms
- contrarily to other solutions which mostly support only the vertex-centric model.
Also, graph updates are written periodically to disk for persistence. Since the data is not persisted
immediately, some recent data might get lost in case of an unexpected shutdown, such that a stream
broker might be required.}


\subsection{Aspen~\cite{dhulipala2019low}}
\vspaceSQ{-0.35em}

%
The Aspen framework~\cite{dhulipala2019low} uses a novel
data structure called the \emph{C-tree} to store graph structures.  A C-tree is
based on a \emph{purely-functional compressed search tree}.  A functional
search tree is a search tree data structure that can be expressed only by mathematical
functions, which makes the data structure immutable (since a mathematical
function must always return the same result for the same input, independently
of any state associated with the data structure).  Furthermore, functional
search trees offer lightweight snapshots, provably efficient running times, and
they facilitate concurrent processing of queries and updates.
Now, the C-tree extends purely-functional search trees: it overcomes the poor
space usage and low locality.  Elements represented by the tree are stored in
\emph{chunks} and each chunk is stored contiguously in an array, leading to improved
locality.  To improve the space usage, chunks can be compressed by applying
difference encoding, since each block stores a sorted set of integers.
%

%
A graph is represented as a tree-of-trees: A purely-functional tree
stores the set of vertices (vertex-tree) and each vertex stores the edges in
its own C-tree (edge-tree). Additional information is stored in the vertex-tree
such that basic graph structural properties, such as the total number of edges and vertices,
can be queried in constant time. Similarly, the trees can be augmented to store
properties (such as weights), but it is omitted in the described work.  For
algorithms that operate on the whole graph (such as BFS), it is possible to
precompute a \emph{flat snapshot}: instead of accessing all vertices by
querying the vertex-tree, an array is used to directly store the pointers to
the vertices. This approach requires an initial overhead, but
reduces access time to edges and
ultimately decreases runtimes of various algorithms.
Similarly to Aspen, Tegra~\cite{iyer2019tegra} and the work by Sha et
al.~\cite{sha2017accelerating} also use trees to represent the graph.

{
	No specific programming model is enforced.
	The API allows any number of parallel readers and a single writer.
	No reader or writer is ever blocked and the framework guarantees strict serializability.
	The update routines allow to both add and remove edges or vertices.
	They are applied in batches and not exposed to running algorithms.
	Instead, algorithms run on an immutable snapshot.
}

%

\subsection{Tegra~\cite{iyer2019tegra}}
\vspaceSQ{-0.35em}

Tegra~\cite{iyer2019tegra} enables graph analysis based on graph updates that
are a part of any window of time.  {This implies that Tegra must store the
full history of the graph, in contrast to most systems that often store only
one state (and the snapshots, on which graph algorithms are running).
Therefore, this system faces different challenges: it must be able to share
graph data among different windows and share state between parallel running
queries.  To achieve these goals, Tegra relies on a novel computation model,
the Incremental Computation by entity Expansion (ICE) model: Many graph
algorithms run iteratively and converge to a solution, allowing to reuse
certain parts of the previous solution when the graph is updated.  Others
\mbox{\cite{sengupta2017evograph,cai2012facilitating,suzumura2014towards,sengupta2016graphin,shi2016tornado}}
have already focused on such algorithms, but are often restricted to graph
expansion (i.e. no removals are allowed) to guarantee correctness.  ICE extends
this approach and recomputes graph algorithms on the subgraphs that are
affected by the recomputation.  Therefore, also removals of vertices and edges
can be taken into account.  Since the tracking of state and the following
recomputation might lead to high overhead, a cost model is used and the
framework switches to full recomputation if needed.}


To support the ICE model, the core data structure of Tegra is an adaptive radix
tree - a tree data structure that enables efficient updates and range scans.
It allows to map a graph efficiently by storing it in \emph{two} trees (a
vertex tree and an edge tree) and create lightweight snapshots by generating a
new root node that holds the differences.  For scaling, the graph is
partitioned (by the hash of the vertex ID) among compute nodes.  Users can
interface with Tegra by the given API and can manually create new snapshots of
the graph.  The system can also automatically create snapshots when a certain
limit of changes is reached.  Therefore, queries and updates (that can be
ingested from main memory or graph databases) run concurrently.  The framework
also stores the changes that happened in-between snapshots, allowing to restore
any state and apply computations on any window.  Since the snapshots take a lot
of memory, they are written to disk using the last recently used policy.  The
framework is implemented on top of Apache Spark~\cite{zaharia2016apache} that
handles scheduling and work distribution. 


\subsection{Apache Flink~\cite{carbone2015apache}}
\vspaceSQ{-0.35em}

%
Apache Flink~\cite{carbone2015apache} is a general purpose streaming system
for streaming \emph{and} batch computations. 
These two concepts are usually considered different, but Flink
treats them similarly.
%
%
Two user APIs are available for implementation: the \texttt{DataSet} API for
batch processing and the \texttt{DataStream} API for unbounded stream
processing.  A variety of custom operators can be implemented, allowing to
maintain computation state, define iterative dataflows, compute over a stream
window, and implement algorithms from the Bulk Synchronous Parallel
model~\cite{valiant1990bridging}.  Both APIs generate programs that are
represented as a directed acyclic graph of operators connected by data streams.
%
%
%
%
%
Since operators can keep state and the system makes no assumption over the
input streams, it is suited for graph streaming for rich data (edge and vertex
properties), and it enables the user to update the graph and execute a broad
range of graph algorithms.

\subsection{Others}
\vspaceSQ{-0.35em}

Other streaming frameworks come with similar design tradeoffs and
features~\cite{fairbanks2013statistical, king2016dynamic, vora2017kickstarter,
aridhi2017bladyg, zhang2017sub, wang2018rstream, mariappan2019graphbolt,
joaquimhourglass, sallinen2019incremental, ji2019low, hartmann2017analyzing}.
We now briefly describe examples, providing a starting point for further
reading.
\textbf{GraphInc}~\cite{cai2012facilitating} is a framework built on top of
Giraph~\cite{martella2015practical} that enables the developer to develop
programs using the vertex-centric abstraction, which is then executed by the
runtime over dynamic graphs.
\textbf{UNICORN}~\cite{suzumura2014towards} is a system that relies on
InfoSphere, a large-scale, distributed data stream processing middleware
developed at IBM Research. 
\textbf{DeltaGraph}~\cite{dexter2016lazy} is a Haskell library for graph
processing, which performs graph updates \emph{lazily}.
%
%
\textbf{iGraph}~\cite{ju2016igraph} is a system implemented on top of Apache
Spark~\cite{zaharia2016apache} and GraphX~\cite{gonzalez2014graphx} that
focuses on hash-based vertex-cut partitioning strategies for dynamic graphs,
and proposes to use the vertex-centric programming model for such graphs.
However, it is unclear on the details of developing different graph algorithms
with the proposed approach.
\textbf{EvoGraph}~\cite{sengupta2017evograph} is a simple extension of GraphIn.
Whenever a batch of updates arrives, EvoGraph decides whether to use an
incremental form of updating its structure, similar to that in GraphIn, or
whether to recompute the maintained graph stored as an AM. 
\textbf{Sprouter}~\cite{abughofa2018sprouter} is another system built on top of
Spark.
\textbf{PAST}~\cite{ding2019storing} is a framework for processing
\emph{spatio-temporal graphs} with \emph{up to 100 trillion edges} that track
people, locations, and their connections. It relies on the underlying Cassandra
storage~\cite{lakshman2010cassandra}.

\fi

\section{Graph Databases}
\label{sec:gdb}
\vspaceSQ{-0.25em}

{Graph databases such as Neo4j~\mbox{\cite{neo4j_book}} were introduced
to alleviate performance overheads of querying graphs
maintained as tables in relational databases; these overheads have been
caused by the need to conduct many expensive joins when, for example,
traversing a graph.}

Streaming graph frameworks, similarly to graph databases, maintain a
dynamically changing graph dataset under a series of updates and queries to the
graph data. However, there are certain crucial differences that we now discuss.
We refer the reader to a recent survey on the latter class of
systems~\cite{besta2019demystifying}, which provides details of native graph
databases such as Neo4j~\cite{neo4j_book}, RDF stores~\cite{rdf_links}, and
other types of NoSQL stores used for managing graphs. 
\tr{In the following, we exclude RDF streaming designs as we identify them to
be strongly related to the domain of database systems, and point the reader to
respective publications for more details~\cite{goasdoue2013efficient,
broekstra2002sesame, komazec2012sparkwave, calbimonte2010enabling}.}

\subsection{Graph Databases vs.~Graph Streaming Systems}
\vspaceSQ{-0.35em}

\tr{We compare graph databases and graph streaming frameworks mostly according
to our taxonomy, but we also touch on other aspects such as key targeted
workloads and their characteristics.}

{\textbf{Targeted Workloads }}
{Graph databases have traditionally focused on simple fine graph queries or
updates, related to both the graph structure (e.g., verify if two vertices are
connected) and the rich attached data (e.g., fetch the value of a given
property)~\mbox{\cite{ldbc_snb_specification}}. Another important class are
``business intelligence'' complex queries (e.g., fetch all vertices modeling
cars, sorted by production year)~\mbox{\cite{early_ldbc_paper}}. Only recently,
there has been interest in augmenting graph databases with capabilities to run
global analytics such as PageRank~\mbox{\cite{capotua2015graphalytics}}.}
{In contrast, streaming frameworks focus on fine updates and queries, and on
global analytics, but \emph{not} on complex business intelligence queries.}
These frameworks put more focus on \emph{high velocity updates} that
can be rapidly ingested into the maintained.
{Next, of key interest are queries into the \emph{structure} of the
adjacency of vertices. This is often in contrast to graph databases, where many
queries focus on the rich data attached to edges and vertices.}
{These differences are reflected in all the following design aspects.}

{\textbf{Ingesting Updates }}
{Graph databases can use \emph{many} different underlying designs (RDBMS
style engines, native graph databases, KV stores, document stores, and
others~\mbox{\cite{besta2019demystifying}}), which means they may use different
schemes for ingesting updates.}
\sethlcolor{lyellow}{However, a certain general difference between graph streaming frameworks
and graph databases is that graph databases often include transactional
support with ACID properties~\mbox{\cite{besta2019demystifying, han2016providing}},
while very few streaming frameworks supports transactions and the ACID
  semantics of transactions.} 
\sethlcolor{yellow} {While most graph databases offer ACID, an example that
does not is Cray Graph Engine~\mbox{\cite{besta2019demystifying}}.}
The streaming graph updates, even if sometimes
  they also referred to as transactions~\cite{zhu2019livegraph}, are usually
  ``lightweight'': single edge insertions or deletions, rather than arbitrary
  pattern matching queries common in graph database workloads. 
{Overall, streaming frameworks focus on lightweight methods for fast and
scalable ingestion of incoming updates, which includes optimizations such as
batching of updates.}

{\textbf{Graph Models and Representations }}
Graph databases usually deal with \textbf{complex and rich graph models} (such
as the Labeled Property Graph~\cite{gdb_query_language_Angles} or Resource
Description Framework~\cite{rdf_links}) where both vertices and edges may be of
different types and may be associated with {arbitrary} rich properties such as
pictures, strings, arrays of integers, or even data blobs.
In contrast, graph data models in streaming frameworks are usually simple,
without support for arbitrary attached properties. {This reflects the fact
that the main focus in streaming frameworks is to investigate the structure of
the maintained graph and its changes, and usually not rich attached data.} 
{This is also reflected by the associated indexing structures. While graph
database systems maintain complex distributed index structures to
accelerate different forms of queries over the rich attached data, streaming
frameworks use simple index structures, most often only pointers to each vertex
neighborhood, and very rarely additional structures pointing to edges/vertices
with, e.g., common labels (an example streaming framework with such indexes
is STINGER).}

{\textbf{Data Distribution }}
{Another interesting observation is support for \emph{data replication} and
\emph{data sharding}. These two concepts refer to, respectively, the ability to
replicate the maintained graph to more than one server (to accelerate certain
read queries), and to partition the same single graph into several servers (to
enable storing large graphs fully in-memory and to accelerate different types of
accesses).
Interestingly, streaming frameworks that enable distributed computation
also support the more powerful but also more complex data sharding.  Contrarily,
while many distributed data stores used as graph databases (e.g., document
  stores) enable sharding as well, the class of ``native'' graph databases do
  not always support sharding. For example, the well-known
  Neo4j~\mbox{\cite{neo4j_book}} graph databases only recently added support
  for sharding for \emph{some} of its queries.}

{\textbf{Keeping Historical Data }}
{We observe that streaming frameworks often offer dedicated support for
maintaining historical data, starting from simple forms such as 
dedicated edge insertion timestamps (e.g., in STINGER), to rich forms
such as full historical data in a form of snapshots and different
optimizations to minimize storage overheads (e.g., in Tegra). In contrast,
graph databases most often do not offer such dedicated schemes. However,
the generality of the used graph models facilitates maintaining such
information at the user level (e.g., the user can use a timestamp
label and/or property attached to each vertex or edge).}

{\textbf{Incremental Changes }}
{We do not know of any graph databases that offer explicit dedicated
support for incremental changes. However, as most of such systems do
not offer open source implementations, confirming this is hard.
However, many streaming frameworks offer strong support for incremental
changes, both in the form of its architecture and computational model
tuned for this purpose, and its offered programming API. This is because
incremental changes specifically target accelerating global graph
analytics such as PageRank. These analytics have always been of
key focus for streaming frameworks, and only recently became
a relevant use case for graph databases~\mbox{\cite{capotua2015graphalytics}}.}

{\textbf{Programming APIs and Models }}
Despite a lack of agreement on a single language for querying graph
databases, all the languages (e.g., SPARQL~\cite{perez2009semantics},
Gremlin~\cite{rodriguez2015gremlin}, Cypher~\cite{francis2018cypher,
holzschuher2013performance}, and SQL~\cite{date1987guide}) provide rich support
for pattern matching queries~\cite{ldbc_snb_specification} or
business intelligence queries~\cite{early_ldbc_paper}.
{On the other hand, streaming frameworks do not offer such support.
However, they do come with rich APIs for global graph analytics.}

{\textbf{Summary }}
{In summary, graph databases and streaming frameworks, despite different
shared characteristics, are mostly complementary designs. Graph databases focus
on rich data models and complex business intelligence workloads, while
streaming frameworks' central interest are lightweight models and very fast
update ingestion rates and global analytics. This can be seen in, for example,
the design of the GraphTau framework, which explicitly offers an interface to
load data for analytics \emph{from a graph database}. Thus, using both systems
together may often help to combine their advantages. Simultaneously, the gap
between these two system classes is slowly shrinking, especially from the side
of graph databases, where focus on global analytics and more performance can be
seen in recent designs~\mbox{\cite{capotua2015graphalytics}}.}

\subsection{Systems Combining Both Areas}
\vspaceSQ{-0.35em}

We describe example systems that provide features related to both graph
streaming frameworks and graph databases.

\textbf{Concerto}~\cite{lee2013views} is a distributed in-memory graph store.
The system presents features that can be found both in graph streaming
frameworks (real-time graph queries and focus on fast, concurrent ingestion of
updates) and in graph databases (triggers, ACID properties). It relies on
Sinfonia~\cite{aguilera2007sinfonia}, an infrastructure that provides a flat
memory region over a set of distributed servers. Further, it offers ACID
guarantees by distributed transactions (similar to the two-phase commit
protocol) and writing logs to disk.  The transactions are only short living for
small operations such as reading and writing memory blocks; no transactions are
available that consist of multiple updates.  The graph data is stored by
Sinfonia directly within in-memory objects that make up a data structure
similar to an adjacency list.  This data structure can also hold arbitrary
properties.

\textbf{ZipG}~\cite{khandelwal2017zipg} is a framework with
focus on memory-efficient storage. It builds on
Succint~\cite{agarwal2015succinct}, a data store that supports random access to
\emph{compressed} unstructured data. ZipG exploits this feature and stores the
graph in two files. The vertex file consists of the vertices that form the
graph. Each row in the file contains the data related to one vertex, including
the vertex properties. The edge file contains the edges stored in the graph. A
single record in the edge file holds all edges of a particular type (e.g., a
relationship or a comment in a social network) that are incident to a vertex.
Further, this record contains all the properties of these edges. To enable fast
access to the properties, metadata (e.g., lengths of different records, and
offsets to the positions of different records) are also maintained by ZipG
files. Succint compresses these files and creates immutable logs that are kept
in main memory for fast access. Updates to the graph are stored in a single log
store and compressed after a threshold is exceeded, allowing to run updates and
queries concurrently. Pointers to the information on updates are managed such
that logs do not have to be scanned during a query. Contrary to traditional
graph databases, the system does not offer strict consistency or transactions.

Finally, \textbf{LiveGraph}~\cite{zhu2019livegraph} targets both transactional
graph data management and graph analytics. Similarly to graph databases, it
implements the property graph model and supports transactions, and similarly to
analytics frameworks, it handles long running tasks that access the whole graph.
For high performance, the system focuses on sequential data accesses.  Vertices
are stored in an array of vertex blocks on which updates are secured by a lock
and applied using copy-on-write.  For edges, a novel graph data structure is
presented, called transactional edge log.  Similar to an adjacency list there
is a list of edges per vertex, but the data structure keeps all insertions,
deletions and updates as edge log entries appended to the list.  The data is
stored in blocks, consisting of a header, edge log entries of fixed size and
property entries (stored separately from the edge log entries).  Each edge log
entry stores the incident vertex, a create time and an update time.  During a
transaction, the reader receives a time stamp and reads only the data for which
the create time is smaller than the given time stamp. Also the update time must
be considered to omit stale data.  Data is read starting from a tail pointer so
a reader sees the updates first (no need to scan the old data).  Further
optimizations are applied, e.g., a Bloom filter allows to check quickly for
existing edges.  For an update, a writer must acquire a lock of the vertex.
New data is appended on the tail of the edge log entries.  Since the
transaction edge log grows over time, a compression scheme is applied which is
non-blocking for readers.  The system guarantees persistence by writing data
into a log and keeps changes locally until the commit phase,
guaranteeing snapshot isolated transactions.

\section{Performance Analysis}
\label{sec:eval}

{We now summarize key insights about performance of the described frameworks.
We focus on (1) identifying the \emph{fastest} frameworks, and on (2)
understanding the performance effects of various design choices.
Due to space constraints, we refer the reader to respective publications for
the details of the evaluation setup. For concreteness, we report specific
performance numbers, but the general performance patterns of the analyzed
effects are similar for other input datasets and hardware architectures used in
respective works.}
{Our key source of data is a recent excellent broad analysis
accompanying the evaluation of the DZiG processing
system~\mbox{\cite{mariappan2021dzig}}.}

{\textbf{Summary of performance-oriented goals }}
{Two main performance goals of the studied frameworks are (1) maximizing the
throughput of ingested updates, usually expressed in millions of inserted (or
deleted) edges per second, and (2) accelerating graph analytics running on top
of the maintained graph.
Some systems (e.g.,
faimGraph~\mbox{\cite{winter2018faimgraph}}) only focus on maximizing the raw update
rate.
However, most systems attempt to maximizing the performance in both (1) \emph{and}
(2). Here, certain systems offer incremental changes (e.g.,
GraphBolt~\mbox{\cite{mariappan2019graphbolt}} or DZiG~\mbox{\cite{mariappan2021dzig}}) while
others do not offer this capability, instead focusing on enhancing the
schemes for incorporating graph mutations efficiently in the 
graph structures (e.g., Aspen~\mbox{\cite{dhulipala2019low}} or
GraphOne~\mbox{\cite{kumar2019graphone}}).
}

\marginpar{\vspace{13em}\colorbox{yellow}{\textbf{R-4}}\\\colorbox{yellow}{\textbf{(3)}}}

\marginpar{\vspace{3em}\colorbox{yellow}{\textbf{R-4}}\\\colorbox{yellow}{\textbf{(3)}}}

{Different results indicate that the former significantly outperform the
latter when considering both (1) and (2) at the same time (i.e., in the
\textbf{end-to-end runtime} comparisons of graph analytics such as PageRank or SSSP
\emph{and} simultaneous graph mutations)~\mbox{\cite{mariappan2021dzig}}.}
Here, we summarize the analysis in~\mbox{\cite{mariappan2021dzig}}, which
considers the following dimensions: a targeted graph problem (PageRank with
batched mutations, SSSP with batched mutations, and plain mutations), the size
of graph mutation batches (1, 10, 100, 1k, 10k), and a framework (DZiG, GraphBolt,
Aspen, GraphOne, LLAMA, STINGER).
Now, \hl{in the DZiG analysis~\mbox{\cite{mariappan2021dzig}}, for \emph{plain
mutations},} Aspen is the fastest \hl{on the above batch sizes (e.g., on a
32-core machine, Aspen achieves runtimes of 1e-4, 3e-4, 2e-3,
6e-3, 7e-3 for batches of 1, 10, 100, 1k, and 10k, respectively)}. Yet, for
a combination of both mutations and graph analytics, the frameworks featuring
dependency-driven incremental computation (GraphBolt, DZiG) outperform all
other comparison targets, regardless of batch sizes and
targeted problems. \hl{For example, for batch size 100, GraphBolt/DZiG use
11.7s/11.2s for PR and both take 0.06s for SSSP. Aspen takes 29.8s for PR and
3.31s for SSSP.}

{Systems such as Aspen come with more potential for the highest
performance of \textbf{raw graph updates}~\mbox{\cite{mariappan2021dzig}}.
These frameworks still try to minimize performance penalties when running
graph analytics, compared to the running times of \emph{static} graph
processing frameworks.
%
The highest performance of raw updates reported in the literature, without
considering analytics, belongs to the GPU based
faimGraph~\mbox{\cite{winter2018faimgraph}}. It achieves processing rates of
nearly 200M edge updates / second \hl{(for batch size 1M, on several graphs
such as coAuthorsD, on an NVIDIA Geforce GTX Titan Xp)}.}

\marginpar{\vspace{-2em}\colorbox{yellow}{\textbf{R-4}}\\\colorbox{yellow}{\textbf{(3)}}}

\marginpar{\vspace{7em}\colorbox{yellow}{\textbf{R-4}}\\\colorbox{yellow}{\textbf{(3)}}}

{We also summarize performance patterns of techniques for \textbf{incremental
computation}, using existing detailed
analyses~\mbox{\cite{mariappan2019graphbolt, mariappan2021dzig}}.  First,
dependency tracking (the online approach)
systematically outperforms restarting computation upon graph mutations (the
offline approach)~\mbox{\cite{mariappan2019graphbolt}}. Within the class of
dependency tracking, differences between respective schemes depend on design
details and targeted algorithms. For example, as expected, KickStarter
outperforms GraphBolt on a non-BSP SSSP problem \hl{(consistent speedups of
$\approx$7$\times$ or more on the Twitter graph, for batch sizes of 1, 10, 100,
1k, 10k, on a single-socket 32-core machine)}, because GraphBolt, being tuned
for BSP programs, ensures synchronous semantics, which is unnecessary for SSSP. 
Moreover, a very recent design indicates further opportunities for speedups
within the class of dependency driven designs. Specifically, one can also
utilize the fact that, in iterative graph algorithms, vertex values often
\emph{stabilize} after several iterations. This enables pruning unnecessary
updates, and deliver speedups over other tuned dependency-driven systems that
do not consider this effect~\mbox{\cite{mariappan2021dzig}}. We expect that
this direction will be further explored in future works, for example
by considering complex non-iterative and non-path-based graph mining
workloads.}

\iftr
\vspaceSQ{-0.35em}
\subsection{Specific Streaming Solutions}
\vspaceSQ{-0.5em}

There are works on streaming and dynamic graphs that focus on solving a
specific graph problem in a dynamic setting. Details of such solutions are
outside the core focus of this survey. We outline them as a reference point for
the reader.
First, different designs target effective \textbf{partitioning of streaming
graph datasets}~\cite{petroni2015hdrf, nicoara2015hermes, yang2012towards,
firth2017loom, firth2016workload, filippidou2015online, huang2016leopard,
shi2017partitioning, hoang2019cusp, mccrabb2019dredge, pacaci2019experimental}.
Second, different works focus on solving a \textbf{specific graph problem in a
streaming setting}. Targeted problems include graph
clustering~\cite{hassani2016hastream}, mining periodic
cliques~\cite{qin2019mining}, search for persistent
communities~\cite{li2018persistent, riedy2013multithreaded}, tracking
conductance~\cite{galhotra2015tracking}, event pattern~\cite{namaki2017event}
and subgraph~\cite{mondal2016casqd} discovery, solving ego-centric
queries~\cite{mondal2014eagr}, pattern
detection~\cite{choudhury2015selectivity, gao2016toward, shao2018efficient,
kim2018turboflux, li2019time, sun2017join, choudhury2018percolator,
gao2014continuous}, densest subgraph identification~\cite{jin2018quickpoint},
frequent subgraph mining~\cite{aslay2018mining}, dense subgraph
detection~\cite{ma2017fast}, construction and querying of knowledge
graphs~\cite{choudhury2017nous}, stream summarization~\cite{gou2019fast}, graph
sparsification~\cite{ahn2009graph, besta2019slim}, $k$-core
maintenance~\cite{aksu2014distributed}, shortest
paths~\cite{srinivasan2018shared}, Betweenness
Centrality~\cite{hayashi2015fully, tripathy2018scaling, solomonik2017scaling},
Triangle Counting~\cite{makkar2017exact}, Katz
Centrality~\cite{van2018scalable}, mincuts~\cite{kogan2015sketching,
gianinazzi2018communication} Connected Components~\cite{mccoll2013new}, or
PageRank~\cite{guo2017parallel, coimbra2018graphbolt}.

\fi

\vspaceSQ{-0.25em}
\section{Challenges}
\vspaceSQ{-0.5em}

Many research challenges related to streaming graph frameworks are
similar to those in graph databases~\cite{besta2019demystifying}.
First, 
\tr{one should identify the most beneficial design choices for different use
cases in the domain of streaming and dynamic graph processing.  As shown in
this paper,} 
existing systems support numerous forms of data organization and
types of graph representations, and it is unclear how to match these different
schemes for different workload scenarios. A strongly related challenge,
similarly to that in graph databases, is a high-performance system design for
supporting both OLAP and OLTP style workloads.
One can also try to accelerate different graph analytics problems
in the streaming setting, for example graph coloring~\cite{besta2020high}.

Second, while there is no consensus on a standard language for querying graph
databases, even less is established for streaming frameworks. Different systems
provide different APIs or programming abstractions~\cite{tate2014programming}.
Difficulties are intensified by a similar lack of consensus on most beneficial
techniques for update ingestion and on computation models. This area is rapidly
evolving and one should expect numerous new ideas, before a certain consensus
is reached.

Moreover, contrarily to static graph processing, little research exists into
accelerating streaming graph processing using hardware acceleration such as
FPGAs~\cite{besta2019substream, besta2019graph, de2018transformations},
high-performance networking hardware and associated
abstractions~\cite{di2019network, besta2015active, besta2018slim,
schmid2016high, besta2014fault, fompi-paper}, low-cost
atomics~\cite{nai2017graphpim, schweizer2015evaluating}, hardware
transactions~\cite{besta2015accelerating}, and others~\cite{besta2018slim,
ahn2016scalable}. One could also investigate topology-aware or
routing-aware data distribution for graph streaming, especially together 
with recent high-performance network topologies~\cite{besta2014slim,
kim2008technology} and routing~\cite{besta2019fatpaths,
lu2018multi, ghorbani2017drill, besta2020highrouting}. Finally, ensuring speedups due to 
different data modeling abstractions, such as the algebraic
abstraction~\cite{kepner2016mathematical, besta2019communication,
besta2017slimsell, kwasniewski2019red}, may be a promising direction.

{We also observe that, despite the fact that several streaming frameworks
offer distributed execution and data sharding, the highest rate of ingestion is
achieved by shared-memory single-node designs
(cf.~Section~\mbox{\ref{sec:eval}}). An interesting challenge would be to make
these designs distributed and to ensure that their ingestion rates increase
even further, proportionally to the number of used compute nodes.}

Finally, an interesting question is whether graph databases are inherently
different from streaming frameworks. While merging these two classes of systems
is an interesting {ongoing effort, reflected by systems such as
Graphflow~\mbox{\cite{kankanamge2017graphflow}}} with many potential benefits, the
difference in the associated workloads and industry requirements may be
fundamentally different for a single unified solution.

\vspaceSQ{-0.25em}
\section{Conclusion}
\vspaceSQ{-0.5em}

Streaming and dynamic graph processing is an important research field. It is
used to maintain numerous dynamic graph datasets, simultaneously ensuring
high-performance graph updates, queries, and analytics workloads. Many graph
streaming frameworks have been developed. They use different data
representations, they are based on miscellaneous design choices for fast
parallel ingestion of updates and resolution of queries, and they enable a
plethora of queries and workloads. We present the first analysis and taxonomy
of the rich landscape of streaming and dynamic graph processing.  We
crystallize a broad number of related concepts (both theoretical and
practical), we list and categorize existing systems and discuss key design
choices, we explain associated models, and we discuss related fields such as
graph databases. Our work can be used by architects, developers, and project
managers who want to select the most advantageous processing system or design,
or simply understand this broad and fast-growing field.

\iftr

{\vspace{1em}\footnotesize\textbf{Acknowledgements}
We thank Khuzaima Daudjee for useful suggestions regarding
related work. We thank PRODYNA AG (Darko Križić, Jens Nixdorf, and Christoph
Körner) for generous support, and anonymous reviewers for comments that helped
to significantly enhance the paper quality. This work was funded by Google
European Doctoral Fellowship and ETH Zurich.}

\fi

\iftr
{
\bibliographystyle{abbrv}
\bibliography{refs}

\begin{thebibliography}{100}

\bibitem{apache_giraph}
{Apache Giraph Project}.
\newblock \url{https://giraph.apache.org/}.

\bibitem{abughofa2018sprouter}
T.~Abughofa and F.~Zulkernine.
\newblock Sprouter: Dynamic graph processing over data streams at scale.
\newblock In {\em DEXA}, pages 321--328. Springer, 2018.

\bibitem{acar2019parallel}
U.~A. Acar, D.~Anderson, G.~E. Blelloch, and L.~Dhulipala.
\newblock Parallel batch-dynamic graph connectivity.
\newblock In {\em ACM SPAA}, pages 381--392, 2019.

\bibitem{Acar}
U.~A. Acar, A.~Cotter, B.~Hudson, and D.~T\"{u}rkoglu.
\newblock Parallelism in dynamic well-spaced point sets.
\newblock In {\em ACM SPAA}, pages 33--42, 2011.

\bibitem{agarwal2015succinct}
R.~Agarwal et~al.
\newblock Succinct: Enabling queries on compressed data.
\newblock In {\em NSDI}, pages 337--350, 2015.

\bibitem{aggarwal2014evolutionary}
C.~Aggarwal and K.~Subbian.
\newblock Evolutionary network analysis: A survey.
\newblock {\em ACM Computing Surveys (CSUR)}, 47(1):10, 2014.

\bibitem{aggarwal2004streaming}
G.~Aggarwal, M.~Datar, S.~Rajagopalan, and M.~Ruhl.
\newblock On the streaming model augmented with a sorting primitive.
\newblock In {\em IEEE FOCS}, pages 540--549, 2004.

\bibitem{aguilera2007sinfonia}
M.~K. Aguilera, A.~Merchant, M.~Shah, A.~Veitch, and C.~Karamanolis.
\newblock Sinfonia: a new paradigm for building scalable distributed systems.
\newblock In {\em ACM SIGOPS Op. Sys. Rev.}, 2007.

\bibitem{ahn2016scalable}
J.~Ahn, S.~Hong, S.~Yoo, O.~Mutlu, and K.~Choi.
\newblock A scalable processing-in-memory accelerator for parallel graph
  processing.
\newblock {\em ACM SIGARCH Comp. Arch. News}, 2016.

\bibitem{ahn2009graph}
K.~J. Ahn and S.~Guha.
\newblock Graph sparsification in the semi-streaming model.
\newblock In {\em ICALP}, pages 328--338. Springer, 2009.

\bibitem{ahn2012graph}
K.~J. Ahn, S.~Guha, and A.~McGregor.
\newblock Graph sketches: sparsification, spanners, and subgraphs.
\newblock In {\em ACM PODS}, pages 5--14, 2012.

\bibitem{akidau2015dataflow}
T.~Akidau, R.~Bradshaw, C.~Chambers, S.~Chernyak, R.~J.
  Fern{\'a}ndez-Moctezuma, R.~Lax, S.~McVeety, D.~Mills, F.~Perry, E.~Schmidt,
  et~al.
\newblock The dataflow model: a practical approach to balancing correctness,
  latency, and cost in massive-scale, unbounded, out-of-order data processing.
\newblock 2015.

\bibitem{aksu2014distributed}
H.~Aksu, M.~Canim, Y.-C. Chang, I.~Korpeoglu, and {\"O}.~Ulusoy.
\newblock Distributed $ k $-core view materialization and maintenance for large
  dynamic graphs.
\newblock {\em IEEE TKDE}, 26(10):2439--2452, 2014.

\bibitem{ammar2016techniques}
K.~Ammar.
\newblock Techniques and systems for large dynamic graphs.
\newblock In {\em SIGMOD'16 PhD Symposium}, pages 7--11. ACM, 2016.

\bibitem{anderson2010couchdb}
J.~C. Anderson, J.~Lehnardt, and N.~Slater.
\newblock {\em CouchDB: the definitive guide: time to relax}.
\newblock " O'Reilly Media, Inc.", 2010.

\bibitem{AndoniCKQWZ16}
A.~Andoni, J.~Chen, R.~Krauthgamer, B.~Qin, D.~P. Woodruff, and Q.~Zhang.
\newblock On sketching quadratic forms.
\newblock In {\em ACM ITCS}, pages 311--319, 2016.

\bibitem{gdb_query_language_Angles}
R.~Angles, M.~Arenas, P.~Barcel\'{o}, A.~Hogan, J.~Reutter, and D.~Vrgo\v{c}.
\newblock {Foundations of Modern Query Languages for Graph Databases}.
\newblock {\em in ACM Comput. Surv.}, 50(5):68:1--68:40, 2017.

\bibitem{apache_couchdb}
{Apache Software Foundation}.
\newblock {Apache CouchDB}.
\newblock \url{https://couchdb.apache.org/}.

\bibitem{aridhi2017bladyg}
S.~Aridhi et~al.
\newblock Bladyg: A graph processing framework for large dynamic graphs.
\newblock {\em Big data research}, 9:9--17, 2017.

\bibitem{armstrong2013linkbench}
T.~G. Armstrong, V.~Ponnekanti, D.~Borthakur, and M.~Callaghan.
\newblock Linkbench: a database benchmark based on the facebook social graph.
\newblock In {\em ACM SIGMOD}, pages 1185--1196, 2013.

\bibitem{aslay2018mining}
C.~Aslay, M.~A.~U. Nasir, G.~De~Francisci~Morales, and A.~Gionis.
\newblock Mining frequent patterns in evolving graphs.
\newblock In {\em ACM CIKM}, pages 923--932, 2018.

\bibitem{AssadiKL17}
S.~Assadi, S.~Khanna, and Y.~Li.
\newblock On estimating maximum matching size in graph streams.
\newblock {\em SODA}, 2017.

\bibitem{AssadiKLY16}
S.~Assadi, S.~Khanna, Y.~Li, and G.~Yaroslavtsev.
\newblock Maximum matchings in dynamic graph streams and the simultaneous
  communication model.
\newblock 2016.

\bibitem{batarfi2015large}
O.~Batarfi, R.~El~Shawi, A.~G. Fayoumi, R.~Nouri, A.~Barnawi, S.~Sakr, et~al.
\newblock Large scale graph processing systems: survey and an experimental
  evaluation.
\newblock {\em Cluster Computing}, 18(3):1189--1213, 2015.

\bibitem{behnezhad19}
S.~Behnezhad, M.~Derakhshan, M.~Hajiaghayi, C.~Stein, and M.~Sudan.
\newblock Fully dynamic maximal independent set with polylogarithmic update
  time.
\newblock {\em FOCS}, 2019.

\bibitem{ben2019modular}
T.~Ben-Nun, M.~Besta, S.~Huber, A.~N. Ziogas, D.~Peter, and T.~Hoefler.
\newblock A modular benchmarking infrastructure for high-performance and
  reproducible deep learning.
\newblock {\em arXiv preprint arXiv:1901.10183}, 2019.

\bibitem{besta2020high}
M.~Besta, A.~Carigiet, K.~Janda, Z.~Vonarburg-Shmaria, L.~Gianinazzi, and
  T.~Hoefler.
\newblock High-performance parallel graph coloring with strong guarantees on
  work, depth, and quality.
\newblock In {\em ACM/IEEE Supercomputing}, pages 1--17, 2020.

\bibitem{besta2020highrouting}
M.~Besta, J.~Domke, M.~Schneider, M.~Konieczny, S.~Di~Girolamo, T.~Schneider,
  A.~Singla, and T.~Hoefler.
\newblock High-performance routing with multipathing and path diversity in
  ethernet and hpc networks.
\newblock {\em IEEE TPDS}, 2020.

\bibitem{besta2019slim}
M.~Besta et~al.
\newblock Slim graph: Practical lossy graph compression for approximate graph
  processing, storage, and analytics.
\newblock 2019.

\bibitem{besta2019substream}
M.~Besta, M.~Fischer, T.~Ben-Nun, J.~De~Fine~Licht, and T.~Hoefler.
\newblock Substream-centric maximum matchings on fpga.
\newblock In {\em ACM/SIGDA FPGA}, pages 152--161, 2019.

\bibitem{besta2018slim}
M.~Besta, S.~M. Hassan, S.~Yalamanchili, R.~Ausavarungnirun, O.~Mutlu, and
  T.~Hoefler.
\newblock Slim noc: A low-diameter on-chip network topology for high energy
  efficiency and scalability.
\newblock In {\em ACM SIGPLAN Notices}, 2018.

\bibitem{besta2014fault}
M.~Besta and T.~Hoefler.
\newblock Fault tolerance for remote memory access programming models.
\newblock In {\em ACM HPDC}, pages 37--48, 2014.

\bibitem{besta2014slim}
M.~Besta and T.~Hoefler.
\newblock Slim fly: A cost effective low-diameter network topology.
\newblock In {\em ACM/IEEE Supercomputing}, pages 348--359, 2014.

\bibitem{besta2015accelerating}
M.~Besta and T.~Hoefler.
\newblock Accelerating irregular computations with hardware transactional
  memory and active messages.
\newblock In {\em ACM HPDC}, 2015.

\bibitem{besta2015active}
M.~Besta and T.~Hoefler.
\newblock Active access: A mechanism for high-performance distributed
  data-centric computations.
\newblock In {\em ACM ICS}, 2015.

\bibitem{besta2019communication}
M.~Besta, R.~Kanakagiri, H.~Mustafa, M.~Karasikov, G.~R{\"a}tsch, T.~Hoefler,
  and E.~Solomonik.
\newblock Communication-efficient jaccard similarity for high-performance
  distributed genome comparisons.
\newblock {\em arXiv preprint arXiv:1911.04200}, 2019.

\bibitem{besta2017slimsell}
M.~Besta, F.~Marending, E.~Solomonik, and T.~Hoefler.
\newblock Slimsell: A vectorizable graph representation for breadth-first
  search.
\newblock In {\em IEEE IPDPS}, pages 32--41, 2017.

\bibitem{besta2019demystifying}
M.~Besta, E.~Peter, R.~Gerstenberger, M.~Fischer, M.~Podstawski, C.~Barthels,
  G.~Alonso, and T.~Hoefler.
\newblock Demystifying graph databases: Analysis and taxonomy of data
  organization, system designs, and graph queries.
\newblock {\em arXiv preprint arXiv:1910.09017}, 2019.

\bibitem{besta2017push}
M.~Besta, M.~Podstawski, L.~Groner, E.~Solomonik, and T.~Hoefler.
\newblock To push or to pull: On reducing communication and synchronization in
  graph computations.
\newblock In {\em ACM HPDC}, 2017.

\bibitem{besta2019fatpaths}
M.~Besta, M.~Schneider, K.~Cynk, M.~Konieczny, E.~Henriksson, S.~Di~Girolamo,
  A.~Singla, and T.~Hoefler.
\newblock Fatpaths: Routing in supercomputers and data centers when shortest
  paths fall short.
\newblock In {\em ACM/IEEE Supercomputing}, 2019.

\bibitem{besta2019graph}
M.~Besta, D.~Stanojevic, J.~D.~F. Licht, T.~Ben-Nun, and T.~Hoefler.
\newblock Graph processing on fpgas: Taxonomy, survey, challenges.
\newblock {\em arXiv preprint arXiv:1903.06697}, 2019.

\bibitem{besta2018log}
M.~Besta, D.~Stanojevic, T.~Zivic, J.~Singh, M.~Hoerold, and T.~Hoefler.
\newblock Log (graph): a near-optimal high-performance graph representation.
\newblock In {\em PACT}, pages 7--1, 2018.

\bibitem{bhattacharya2019new}
S.~Bhattacharya, M.~Henzinger, and D.~Nanongkai.
\newblock A new deterministic algorithm for dynamic set cover.
\newblock {\em FOCS}, 2019.

\bibitem{bhattacharya2015space}
S.~Bhattacharya, M.~Henzinger, D.~Nanongkai, and C.~Tsourakakis.
\newblock Space-and time-efficient algorithm for maintaining dense subgraphs on
  one-pass dynamic streams.
\newblock In {\em ACM STOC}, 2015.

\bibitem{biem2010ibm}
A.~Biem, E.~Bouillet, H.~Feng, A.~Ranganathan, A.~Riabov, O.~Verscheure,
  H.~Koutsopoulos, and C.~Moran.
\newblock Ibm infosphere streams for scalable, real-time, intelligent
  transportation services.
\newblock In {\em ACM SIGMOD}, 2010.

\bibitem{boldi2004webgraph}
P.~Boldi and S.~Vigna.
\newblock The webgraph framework i: compression techniques.
\newblock In {\em ACM WWW}, pages 595--602, 2004.

\bibitem{broekstra2002sesame}
J.~Broekstra et~al.
\newblock Sesame: A generic architecture for storing and querying rdf and rdf
  schema.
\newblock In {\em ISWC}, pages 54--68. Springer, 2002.

\bibitem{BuryGMMSVZ19}
M.~Bury, E.~Grigorescu, A.~McGregor, M.~Monemizadeh, C.~Schwiegelshohn,
  S.~Vorotnikova, and S.~Zhou.
\newblock Structural results on matching estimation with applications to
  streaming.
\newblock {\em Algorithmica}, 81(1):367--392, 2019.

\bibitem{busato2018hornet}
F.~Busato, O.~Green, N.~Bombieri, and D.~A. Bader.
\newblock Hornet: An efficient data structure for dynamic sparse graphs and
  matrices on gpus.
\newblock In {\em IEEE HPEC}, pages 1--7, 2018.

\bibitem{byun2019chronograph}
J.~Byun, S.~Woo, and D.~Kim.
\newblock Chronograph: Enabling temporal graph traversals for efficient
  information diffusion analysis over time.
\newblock {\em IEEE Transactions on Knowledge and Data Engineering},
  32(3):424--437, 2019.

\bibitem{cai2012facilitating}
Z.~Cai, D.~Logothetis, and G.~Siganos.
\newblock Facilitating real-time graph mining.
\newblock In {\em ACM CloudDB}, pages 1--8, 2012.

\bibitem{calbimonte2010enabling}
J.-P. Calbimonte, O.~Corcho, and A.~J. Gray.
\newblock Enabling ontology-based access to streaming data sources.
\newblock In {\em Springer ISWC}, 2010.

\bibitem{capotua2015graphalytics}
M.~Capot{\u{a}}, T.~Hegeman, A.~Iosup, A.~Prat-P{\'e}rez, O.~Erling, and
  P.~Boncz.
\newblock Graphalytics: A big data benchmark for graph-processing platforms.
\newblock In {\em Proceedings of the GRADES'15}, pages 1--6. 2015.

\bibitem{carbone2015apache}
P.~Carbone, A.~Katsifodimos, S.~Ewen, V.~Markl, S.~Haridi, and K.~Tzoumas.
\newblock Apache flink: Stream and batch processing in a single engine.
\newblock {\em IEEE-CS Bull. Tech. Com. on Data Eng.}, 2015.

\bibitem{ChechikZhang19}
S.~Chechik and T.~Zhang.
\newblock Fully dynamic maximal independent set in expected poly-log update
  time.
\newblock {\em FOCS}, 2019.

\bibitem{cheng2012kineograph}
R.~Cheng, J.~Hong, A.~Kyrola, Y.~Miao, X.~Weng, M.~Wu, F.~Yang, L.~Zhou,
  F.~Zhao, and E.~Chen.
\newblock Kineograph: taking the pulse of a fast-changing and connected world.
\newblock In {\em ACM EuroSys}, pages 85--98, 2012.

\bibitem{ching2015one}
A.~Ching, S.~Edunov, M.~Kabiljo, D.~Logothetis, and S.~Muthukrishnan.
\newblock One trillion edges: Graph processing at facebook-scale.
\newblock {\em Proceedings of the VLDB Endowment}, 8(12):1804--1815, 2015.

\bibitem{choudhury2017nous}
S.~Choudhury, K.~Agarwal, S.~Purohit, B.~Zhang, M.~Pirrung, W.~Smith, and
  M.~Thomas.
\newblock Nous: Construction and querying of dynamic knowledge graphs.
\newblock In {\em IEEE ICDE}, pages 1563--1565, 2017.

\bibitem{choudhury2015selectivity}
S.~Choudhury, L.~Holder, G.~Chin, K.~Agarwal, and J.~Feo.
\newblock A selectivity based approach to continuous pattern detection in
  streaming graphs.
\newblock {\em arXiv preprint arXiv:1503.00849}, 2015.

\bibitem{choudhury2018percolator}
S.~Choudhury, S.~Purohit, P.~Lin, Y.~Wu, L.~Holder, and K.~Agarwal.
\newblock Percolator: Scalable pattern discovery in dynamic graphs.
\newblock In {\em ACM WSDM}, pages 759--762, 2018.

\bibitem{coimbra2018graphbolt}
M.~E. Coimbra, R.~Rosa, S.~Esteves, A.~P. Francisco, and L.~Veiga.
\newblock Graphbolt: Streaming graph approximations on big data.
\newblock {\em arXiv preprint arXiv:1810.02781}, 2018.

\bibitem{rdf_links}
R.~Cyganiak, D.~Wood, and M.~Lanthaler.
\newblock {RDF 1.1 Concepts and Abstract Syntax}.
\newblock Available at \url{https://www.w3.org/TR/rdf11-concepts/}.

\bibitem{datar2002maintaining}
M.~Datar, A.~Gionis, P.~Indyk, and R.~Motwani.
\newblock Maintaining stream statistics over sliding windows.
\newblock {\em SIAM journal on computing}, 31(6):1794--1813, 2002.

\bibitem{date1987guide}
C.~J. Date and H.~Darwen.
\newblock {\em A Guide to the SQL Standard}, volume~3.
\newblock Addison-Wesley New York, 1987.

\bibitem{davepersistent}
A.~Dave, J.~E. Gonzalez, M.~J. Franklin, and I.~Stoica.
\newblock Persistent adaptive radix trees: Efficient fine-grained updates to
  immutable data.

\bibitem{de2018transformations}
J.~de~Fine~Licht et~al.
\newblock Transformations of high-level synthesis codes for high-performance
  computing.
\newblock {\em arXiv:1805.08288}, 2018.

\bibitem{dean2008mapreduce}
J.~Dean and S.~Ghemawat.
\newblock Mapreduce: simplified data processing on large clusters.
\newblock {\em Communications of the ACM}, 51(1):107--113, 2008.

\bibitem{demetrescu2009trading}
C.~Demetrescu, I.~Finocchi, and A.~Ribichini.
\newblock Trading off space for passes in graph streaming problems.
\newblock {\em ACM TALG}, 6(1):6, 2009.

\bibitem{dexter2016lazy}
P.~Dexter, Y.~D. Liu, and K.~Chiu.
\newblock Lazy graph processing in haskell.
\newblock In {\em ACM SIGPLAN Notices}, volume~51, pages 182--192. ACM, 2016.

\bibitem{dexter2019formal}
P.~Dexter, Y.~D. Liu, and K.~Chiu.
\newblock Formal foundations of continuous graph processing.
\newblock {\em arXiv preprint arXiv:1911.10982}, 2019.

\bibitem{dhulipala2019low}
L.~Dhulipala et~al.
\newblock Low-latency graph streaming using compressed purely-functional trees.
\newblock {\em arXiv:1904.08380}, 2019.

\bibitem{di2019network}
S.~Di~Girolamo, K.~Taranov, A.~Kurth, M.~Schaffner, T.~Schneider,
  J.~Ber{\'a}nek, M.~Besta, L.~Benini, D.~Roweth, and T.~Hoefler.
\newblock Network-accelerated non-contiguous memory transfers.
\newblock {\em arXiv preprint arXiv:1908.08590}, 2019.

\bibitem{di2012protein}
L.~Di~Paola, M.~De~Ruvo, P.~Paci, D.~Santoni, and A.~Giuliani.
\newblock Protein contact networks: an emerging paradigm in chemistry.
\newblock {\em Chemical reviews}, 113(3):1598--1613, 2012.

\bibitem{ding2019storing}
M.~Ding et~al.
\newblock Storing and querying large-scale spatio-temporal graphs with
  high-throughput edge insertions.
\newblock {\em arXiv:1904.09610}, 2019.

\bibitem{doekemeijer2014survey}
N.~Doekemeijer and A.~L. Varbanescu.
\newblock A survey of parallel graph processing frameworks.
\newblock {\em Delft University of Technology}, page~21, 2014.

\bibitem{DuanHZ19}
R.~Duan, H.~He, and T.~Zhang.
\newblock Dynamic edge coloring with improved approximation.
\newblock In {\em ACM-SIAM SODA}, pages 1937--1945, 2019.

\bibitem{durfee2019parallel}
D.~Durfee, L.~Dhulipala, J.~Kulkarni, R.~Peng, S.~Sawlani, and X.~Sun.
\newblock Parallel batch-dynamic graphs: Algorithms and lower bounds.
\newblock {\em SODA}, 2020.

\bibitem{DurfeeGGP19}
D.~Durfee, Y.~Gao, G.~Goranci, and R.~Peng.
\newblock Fully dynamic spectral vertex sparsifiers and applications.
\newblock In {\em ACM STOC}, pages 914--925, 2019.

\bibitem{ediger2012stinger}
D.~Ediger, R.~McColl, J.~Riedy, and D.~A. Bader.
\newblock Stinger: High performance data structure for streaming graphs.
\newblock In {\em IEEE HPEC}, pages 1--5, 2012.

\bibitem{ldbc_snb_specification}
O.~Erling, A.~Averbuch, J.~Larriba-Pey, H.~Chafi, A.~Gubichev, A.~Prat, M.-D.
  Pham, and P.~Boncz.
\newblock {The LDBC Social Network Benchmark: Interactive Workload}.
\newblock in SIGMOD, pages 619--630, 2015.

\bibitem{EsfandiariHLMO18}
H.~Esfandiari, M.~Hajiaghayi, V.~Liaghat, M.~Monemizadeh, and K.~Onak.
\newblock Streaming algorithms for estimating the matching size in planar
  graphs and beyond.
\newblock {\em {ACM} Trans. Algorithms}, 2018.

\bibitem{fairbanks2013statistical}
J.~Fairbanks, D.~Ediger, R.~McColl, D.~A. Bader, and E.~Gilbert.
\newblock A statistical framework for streaming graph analysis.
\newblock In {\em IEEE/ACM ASONAM}, pages 341--347, 2013.

\bibitem{fard2012towards}
A.~Fard, A.~Abdolrashidi, L.~Ramaswamy, and J.~A. Miller.
\newblock Towards efficient query processing on massive time-evolving graphs.
\newblock In {\em 8th International Conference on Collaborative Computing:
  Networking, Applications and Worksharing (CollaborateCom)}, pages 567--574.
  IEEE, 2012.

\bibitem{feigenbaum2005graph}
J.~Feigenbaum, S.~Kannan, A.~McGregor, S.~Suri, and J.~Zhang.
\newblock On graph problems in a semi-streaming model.
\newblock {\em Theoretical Computer Science}, 348(2-3):207--216, 2005.

\bibitem{feng2015distinger}
G.~Feng et~al.
\newblock Distinger: A distributed graph data structure for massive dynamic
  graph processing.
\newblock In {\em IEEE Big Data}, pages 1814--1822, 2015.

\bibitem{feng2020risgraph}
G.~Feng, Z.~Ma, D.~Li, S.~Chen, X.~Zhu, W.~Han, and W.~Chen.
\newblock Risgraph: A real-time streaming system for evolving graphs to support
  sub-millisecond per-update analysis at millions ops/s.
\newblock In {\em Proceedings of the 2021 International Conference on
  Management of Data}, pages 513--527, 2021.

\bibitem{filippidou2015online}
I.~Filippidou and Y.~Kotidis.
\newblock Online and on-demand partitioning of streaming graphs.
\newblock In {\em IEEE Big Data}, pages 4--13.

\bibitem{firth2016workload}
H.~Firth and P.~Missier.
\newblock Workload-aware streaming graph partitioning.
\newblock In {\em EDBT/ICDT Workshops}. Citeseer, 2016.

\bibitem{firth2017loom}
H.~Firth, P.~Missier, and J.~Aiston.
\newblock Loom: Query-aware partitioning of online graphs.
\newblock {\em arXiv preprint arXiv:1711.06608}, 2017.

\bibitem{flajolet2007hyperloglog}
P.~Flajolet, {\'E}.~Fusy, O.~Gandouet, and F.~Meunier.
\newblock Hyperloglog: the analysis of a near-optimal cardinality estimation
  algorithm.
\newblock In {\em Discrete Mathematics and Theoretical Computer Science}, pages
  137--156, 2007.

\bibitem{ForsterG19}
S.~Forster and G.~Goranci.
\newblock Dynamic low-stretch trees via dynamic low-diameter decompositions.
\newblock In {\em ACM STOC}, pages 377--388, 2019.

\bibitem{fouquet2018enabling}
F.~Fouquet, T.~Hartmann, S.~Mosser, and M.~Cordy.
\newblock Enabling lock-free concurrent workers over temporal graphs composed
  of multiple time-series.
\newblock In {\em Proceedings of the 33rd Annual ACM Symposium on Applied
  Computing}, pages 1054--1061. ACM, 2018.

\bibitem{francis2018cypher}
N.~Francis, A.~Green, P.~Guagliardo, L.~Libkin, T.~Lindaaker, V.~Marsault,
  S.~Plantikow, M.~Rydberg, P.~Selmer, and A.~Taylor.
\newblock Cypher: An evolving query language for property graphs.
\newblock In {\em ACM SIGMOD}, pages 1433--1445, 2018.

\bibitem{galhotra2015tracking}
S.~Galhotra, A.~Bagchi, S.~Bedathur, M.~Ramanath, and V.~Jain.
\newblock Tracking the conductance of rapidly evolving topic-subgraphs.
\newblock {\em Proc. VLDB}, 8(13):2170--2181, 2015.

\bibitem{gao2016toward}
J.~Gao, C.~Zhou, and J.~X. Yu.
\newblock Toward continuous pattern detection over evolving large graph with
  snapshot isolation.
\newblock {\em The VLDB Journal—The International Journal on Very Large Data
  Bases}, 25(2):269--290, 2016.

\bibitem{gao2014continuous}
J.~Gao, C.~Zhou, J.~Zhou, and J.~X. Yu.
\newblock Continuous pattern detection over billion-edge graph using
  distributed framework.
\newblock In {\em IEEE ICDE}, pages 556--567, 2014.

\bibitem{fompi-paper}
R.~Gerstenberger et~al.
\newblock {Enabling Highly-scalable Remote Memory Access Programming with MPI-3
  One Sided}.
\newblock In {\em ACM/IEEE Supercomputing}, 2013.

\bibitem{ghorbani2017drill}
S.~Ghorbani, Z.~Yang, P.~Godfrey, Y.~Ganjali, and A.~Firoozshahian.
\newblock Drill: Micro load balancing for low-latency data center networks.
\newblock In {\em ACM SIGCOMM}, pages 225--238, 2017.

\bibitem{gianinazzi2018communication}
L.~Gianinazzi, P.~Kalvoda, A.~De~Palma, M.~Besta, and T.~Hoefler.
\newblock Communication-avoiding parallel minimum cuts and connected
  components.
\newblock In {\em ACM SIGPLAN Notices}, volume~53, pages 219--232. ACM, 2018.

\bibitem{goasdoue2013efficient}
F.~Goasdou{\'e}, I.~Manolescu, and A.~Roati{\c{s}}.
\newblock Efficient query answering against dynamic rdf databases.
\newblock In {\em ACM EDBT}, pages 299--310, 2013.

\bibitem{gonzalez2014graphx}
J.~E. Gonzalez, R.~S. Xin, A.~Dave, D.~Crankshaw, M.~J. Franklin, and
  I.~Stoica.
\newblock Graphx: Graph processing in a distributed dataflow framework.
\newblock In {\em OSDI}, 2014.

\bibitem{gou2019fast}
X.~Gou, L.~Zou, C.~Zhao, and T.~Yang.
\newblock Fast and accurate graph stream summarization.
\newblock In {\em 2019 IEEE 35th International Conference on Data Engineering
  (ICDE)}, pages 1118--1129. IEEE, 2019.

\bibitem{green2016custinger}
O.~Green and D.~A. Bader.
\newblock custinger: Supporting dynamic graph algorithms for gpus.
\newblock In {\em IEEE HPEC}, 2016.

\bibitem{advancedMPI}
W.~Gropp, T.~Hoefler, T.~Rajeev, and E.~Lusk.
\newblock {\em Using Advanced MPI: Modern Features of the Message-Passing
  Interface}.
\newblock The MIT Press, 2014.

\bibitem{GuhaM12}
S.~Guha and A.~McGregor.
\newblock Graph synopses, sketches, and streams: {A} survey.
\newblock {\em {PVLDB}}, 5(12):2030--2031, 2012.

\bibitem{GuhaMT15}
S.~Guha, A.~McGregor, and D.~Tench.
\newblock Vertex and hyperedge connectivity in dynamic graph streams.
\newblock In {\em ACM PODS}, pages 241--247, 2015.

\bibitem{guo2017parallel}
W.~Guo, Y.~Li, M.~Sha, and K.-L. Tan.
\newblock Parallel personalized pagerank on dynamic graphs.
\newblock {\em Proceedings of the VLDB Endowment}, 11(1):93--106, 2017.

\bibitem{han2015giraph}
M.~Han and K.~Daudjee.
\newblock Giraph unchained: barrierless asynchronous parallel execution in
  pregel-like graph processing systems.
\newblock {\em VLDB}, 2015.

\bibitem{han2016providing}
M.~Han and K.~Daudjee.
\newblock Providing serializability for pregel-like graph processing systems.
\newblock In {\em EDBT}, pages 77--88, 2016.

\bibitem{han2014experimental}
M.~Han, K.~Daudjee, K.~Ammar, M.~T. {\"O}zsu, X.~Wang, and T.~Jin.
\newblock An experimental comparison of pregel-like graph processing systems.
\newblock {\em Proc. VLDB}, 7(12):1047--1058, 2014.

\bibitem{han2018auxo}
W.~Han, K.~Li, S.~Chen, and W.~Chen.
\newblock Auxo: a temporal graph management system.
\newblock {\em Big Data Mining and Analytics}, 2(1):58--71, 2018.

\bibitem{han2014chronos}
W.~Han, Y.~Miao, K.~Li, M.~Wu, F.~Yang, L.~Zhou, V.~Prabhakaran, W.~Chen, and
  E.~Chen.
\newblock Chronos: a graph engine for temporal graph analysis.
\newblock In {\em 9th European Conference on Computer Systems}, page~1. ACM,
  2014.

\bibitem{hartmann2017analyzing}
T.~Hartmann, F.~Fouquet, M.~Jimenez, R.~Rouvoy, and Y.~Le~Traon.
\newblock Analyzing complex data in motion at scale with temporal graphs.
\newblock 2017.

\bibitem{hassani2016hastream}
M.~Hassani, P.~Spaus, A.~Cuzzocrea, and T.~Seidl.
\newblock I-hastream: density-based hierarchical clustering of big data streams
  and its application to big graph analytics tools.
\newblock In {\em CCGrid}, pages 656--665. IEEE, 2016.

\bibitem{hayashi2015fully}
T.~Hayashi, T.~Akiba, and Y.~Yoshida.
\newblock Fully dynamic betweenness centrality maintenance on massive networks.
\newblock {\em Proceedings of the VLDB Endowment}, 9(2):48--59, 2015.

\bibitem{hoang2019cusp}
L.~Hoang, R.~Dathathri, G.~Gill, and K.~Pingali.
\newblock Cusp: A customizable streaming edge partitioner for distributed graph
  analytics.
\newblock In {\em 2019 IEEE International Parallel and Distributed Processing
  Symposium (IPDPS)}, pages 439--450. IEEE, 2019.

\bibitem{holzschuher2013performance}
F.~Holzschuher and R.~Peinl.
\newblock Performance of graph query languages: comparison of cypher, gremlin
  and native access in neo4j.
\newblock In {\em ACM EDBT/ICDT}, 2013.

\bibitem{huang2016leopard}
J.~Huang and D.~J. Abadi.
\newblock Leopard: Lightweight edge-oriented partitioning and replication for
  dynamic graphs.
\newblock {\em Proceedings of the VLDB Endowment}, 9(7):540--551, 2016.

\bibitem{italiano2019dynamic}
G.~F. Italiano, S.~Lattanzi, V.~S. Mirrokni, and N.~Parotsidis.
\newblock Dynamic algorithms for the massively parallel computation model.
\newblock In {\em ACM SPAA}, 2019.

\bibitem{iyer2015celliq}
A.~Iyer, L.~E. Li, and I.~Stoica.
\newblock Celliq: Real-time cellular network analytics at scale.
\newblock In {\em NSDI}, 2015.

\bibitem{iyer2016time}
A.~P. Iyer, L.~E. Li, T.~Das, and I.~Stoica.
\newblock Time-evolving graph processing at scale.
\newblock In {\em ACM GRADES}, 2016.

\bibitem{iyer2019tegra}
A.~P. Iyer, Q.~Pu, K.~Patel, J.~E. Gonzalez, and I.~Stoica.
\newblock Tegra: Efficient ad-hoc analytics on evolving graphs.
\newblock In {\em NSDI}, pages 337--355, 2021.

\bibitem{ji2019low}
S.~Ji et~al.
\newblock A low-latency computing framework for time-evolving graphs.
\newblock {\em The Journal of Supercomputing}, 75(7):3673--3692, 2019.

\bibitem{jin2018quickpoint}
H.~Jin, C.~Lin, H.~Chen, and J.~Liu.
\newblock Quickpoint: Efficiently identifying densest sub-graphs in online
  social networks for event stream dissemination.
\newblock {\em IEEE Transactions on Knowledge and Data Engineering}, 2018.

\bibitem{joaquimhourglass}
P.~Joaquim.
\newblock Hourglass-incremental graph processing on heterogeneous
  infrastructures.

\bibitem{ju2016igraph}
W.~Ju, J.~Li, W.~Yu, and R.~Zhang.
\newblock igraph: an incremental data processing system for dynamic graph.
\newblock {\em Frontiers of Computer Science}, 10(3):462--476, 2016.

\bibitem{kalavri2017high}
V.~Kalavri, V.~Vlassov, and S.~Haridi.
\newblock High-level programming abstractions for distributed graph processing.
\newblock {\em IEEE TKDE}, 2017.

\bibitem{KallaugherKP18}
J.~Kallaugher, M.~Kapralov, and E.~Price.
\newblock The sketching complexity of graph and hypergraph counting.
\newblock {\em FOCS}, 2018.

\bibitem{kamburugamuve2016survey}
S.~Kamburugamuve and G.~Fox.
\newblock Survey of distributed stream processing.
\newblock {\em Bloomington: Indiana University}, 2016.

\bibitem{kane2012counting}
D.~M. Kane, K.~Mehlhorn, T.~Sauerwald, and H.~Sun.
\newblock Counting arbitrary subgraphs in data streams.
\newblock In {\em International Colloquium on Automata, Languages, and
  Programming}, pages 598--609. Springer, 2012.

\bibitem{kankanamge2017graphflow}
C.~Kankanamge, S.~Sahu, A.~Mhedbhi, J.~Chen, and S.~Salihoglu.
\newblock Graphflow: An active graph database.
\newblock In {\em ACM SIGMOD}, pages 1695--1698, 2017.

\bibitem{KapralovKS14}
M.~Kapralov, S.~Khanna, and M.~Sudan.
\newblock Approximating matching size from random streams.
\newblock {\em SODA}, 2014.

\bibitem{KMNT19}
M.~Kapralov, S.~Mitrovic, A.~Norouzi{-}Fard, and J.~Tardos.
\newblock Space efficient approximation to maximum matching size from uniform
  edge samples.
\newblock {\em SODA}, 2020.

\bibitem{KMMMNST19}
M.~Kapralov, A.~Mousavifar, C.~Musco, C.~Musco, N.~Nouri, A.~Sidford, and
  J.~Tardos.
\newblock Fast and space efficient spectral sparsification in dynamic streams.
\newblock {\em SODA}, abs/1903.12150, 2020.

\bibitem{KNST19}
M.~Kapralov, N.~Nouri, A.~Sidford, and J.~Tardos.
\newblock Dynamic streaming spectral sparsification in nearly linear time and
  space.
\newblock {\em CoRR}, abs/1903.12150, 2019.

\bibitem{KapralovW14}
M.~Kapralov and D.~P. Woodruff.
\newblock Spanners and sparsifiers in dynamic streams.
\newblock {\em PODC}, 2014.

\bibitem{karloff2010model}
H.~Karloff, S.~Suri, and S.~Vassilvitskii.
\newblock A model of computation for mapreduce.
\newblock In {\em ACM-SIAM SODA}, 2010.

\bibitem{kepner2016mathematical}
J.~Kepner, P.~Aaltonen, D.~Bader, A.~Bulu{\c{c}}, F.~Franchetti, J.~Gilbert,
  D.~Hutchison, M.~Kumar, A.~Lumsdaine, H.~Meyerhenke, et~al.
\newblock Mathematical foundations of the graphblas.
\newblock In {\em 2016 IEEE High Performance Extreme Computing Conference
  (HPEC)}, pages 1--9. IEEE, 2016.

\bibitem{khandelwal2017zipg}
A.~Khandelwal, Z.~Yang, E.~Ye, R.~Agarwal, and I.~Stoica.
\newblock Zipg: A memory-efficient graph store for interactive queries.
\newblock In {\em ACM SIGMOD}, 2017.

\bibitem{khurana2013efficient}
U.~Khurana and A.~Deshpande.
\newblock Efficient snapshot retrieval over historical graph data.
\newblock In {\em 2013 IEEE 29th International Conference on Data Engineering
  (ICDE)}, pages 997--1008. IEEE, 2013.

\bibitem{khurana2015storing}
U.~Khurana and A.~Deshpande.
\newblock Storing and analyzing historical graph data at scale.
\newblock {\em arXiv preprint arXiv:1509.08960}, 2015.

\bibitem{kim2008technology}
J.~Kim, W.~J. Dally, S.~Scott, and D.~Abts.
\newblock Technology-driven, highly-scalable dragonfly topology.
\newblock In {\em 2008 International Symposium on Computer Architecture}, pages
  77--88. IEEE, 2008.

\bibitem{kim2018turboflux}
K.~Kim, I.~Seo, W.-S. Han, J.-H. Lee, S.~Hong, H.~Chafi, H.~Shin, and G.~Jeong.
\newblock Turboflux: A fast continuous subgraph matching system for streaming
  graph data.
\newblock In {\em Proceedings of the 2018 International Conference on
  Management of Data}, pages 411--426. ACM, 2018.

\bibitem{king2016dynamic}
J.~King, T.~Gilray, R.~M. Kirby, and M.~Might.
\newblock Dynamic sparse-matrix allocation on gpus.
\newblock In {\em International Conference on High Performance Computing},
  pages 61--80. Springer, 2016.

\bibitem{kogan2015sketching}
D.~Kogan and R.~Krauthgamer.
\newblock Sketching cuts in graphs and hypergraphs.
\newblock In {\em Proceedings of the 2015 Conference on Innovations in
  Theoretical Computer Science}, pages 367--376. ACM, 2015.

\bibitem{komazec2012sparkwave}
S.~Komazec, D.~Cerri, and D.~Fensel.
\newblock Sparkwave: continuous schema-enhanced pattern matching over rdf data
  streams.
\newblock In {\em 6th ACM International Conference on Distributed Event-Based
  Systems}, pages 58--68. ACM, 2012.

\bibitem{kumar2016g}
P.~Kumar and H.~H. Huang.
\newblock G-store: high-performance graph store for trillion-edge processing.
\newblock In {\em ACM/IEEE Supercomputing}, 2016.

\bibitem{kumar2019graphone}
P.~Kumar and H.~H. Huang.
\newblock Graphone: A data store for real-time analytics on evolving graphs.
\newblock In {\em USENIX FAST}, 2019.

\bibitem{kwasniewski2019red}
G.~Kwasniewski, M.~Kabi{\'c}, M.~Besta, J.~VandeVondele, R.~Solc{\`a}, and
  T.~Hoefler.
\newblock Red-blue pebbling revisited: near optimal parallel matrix-matrix
  multiplication.
\newblock In {\em ACM/IEEE Supercomputing}, page~24. ACM, 2019.

\bibitem{lakshman2010cassandra}
A.~Lakshman and P.~Malik.
\newblock Cassandra: a decentralized structured storage system.
\newblock {\em ACM SIGOPS Operating Systems Review}, 44(2):35--40, 2010.

\bibitem{LarsenNNT19}
K.~G. Larsen, J.~Nelson, H.~L. Nguyen, and M.~Thorup.
\newblock Heavy hitters via cluster-preserving clustering.
\newblock {\em Commun. {ACM}}, 62(8):95--100, 2019.

\bibitem{lee2013views}
M.~M. Lee, I.~Roy, A.~AuYoung, V.~Talwar, K.~Jayaram, and Y.~Zhou.
\newblock Views and transactional storage for large graphs.
\newblock In {\em ACM/IFIP/USENIX Middleware}, 2013.

\bibitem{li2018persistent}
R.-H. Li, J.~Su, L.~Qin, J.~X. Yu, and Q.~Dai.
\newblock Persistent community search in temporal networks.
\newblock In {\em 2018 IEEE 34th International Conference on Data Engineering
  (ICDE)}, pages 797--808. IEEE, 2018.

\bibitem{li2019time}
Y.~Li, L.~Zou, M.~T. {\"O}zsu, and D.~Zhao.
\newblock Time constrained continuous subgraph search over streaming graphs.
\newblock In {\em 2019 IEEE 35th International Conference on Data Engineering
  (ICDE)}, pages 1082--1093. IEEE, 2019.

\bibitem{lightenberg2018tink}
W.~Lightenberg, Y.~Pei, G.~Fletcher, and M.~Pechenizkiy.
\newblock Tink: A temporal graph analytics library for apache flink.
\newblock In {\em Companion Proceedings of the The Web Conference 2018}, pages
  71--72, 2018.

\bibitem{lin2018shentu}
H.~Lin, X.~Zhu, B.~Yu, X.~Tang, W.~Xue, W.~Chen, L.~Zhang, T.~Hoefler, X.~Ma,
  X.~Liu, et~al.
\newblock Shentu: processing multi-trillion edge graphs on millions of cores in
  seconds.
\newblock In {\em ACM/IEEE Supercomputing}, page~56. IEEE Press, 2018.

\bibitem{liu2017graphene}
H.~Liu and H.~H. Huang.
\newblock Graphene: Fine-grained $\{$IO$\}$ management for graph computing.
\newblock In {\em USENIX FAST}, 2017.

\bibitem{low2014graphlab}
Y.~Low, J.~E. Gonzalez, A.~Kyrola, D.~Bickson, C.~E. Guestrin, and
  J.~Hellerstein.
\newblock Graphlab: A new framework for parallel machine learning.
\newblock {\em arXiv preprint arXiv:1408.2041}, 2014.

\bibitem{lu2018multi}
Y.~Lu, G.~Chen, B.~Li, K.~Tan, Y.~Xiong, P.~Cheng, J.~Zhang, E.~Chen, and
  T.~Moscibroda.
\newblock Multi-path transport for $\{$RDMA$\}$ in datacenters.
\newblock In {\em NSDI}, 2018.

\bibitem{ma2017fast}
S.~Ma, R.~Hu, L.~Wang, X.~Lin, and J.~Huai.
\newblock Fast computation of dense temporal subgraphs.
\newblock In {\em ICDE}, pages 361--372. IEEE, 2017.

\bibitem{maass2017mosaic}
S.~Maass, C.~Min, S.~Kashyap, W.~Kang, M.~Kumar, and T.~Kim.
\newblock Mosaic: Processing a trillion-edge graph on a single machine.
\newblock In {\em Proceedings of the Twelfth European Conference on Computer
  Systems}, pages 527--543, 2017.

\bibitem{macko2015llama}
P.~Macko, V.~J. Marathe, D.~W. Margo, and M.~I. Seltzer.
\newblock Llama: Efficient graph analytics using large multiversioned arrays.
\newblock In {\em 2015 IEEE 31st International Conference on Data Engineering},
  pages 363--374. IEEE, 2015.

\bibitem{makkar2017exact}
D.~Makkar, D.~A. Bader, and O.~Green.
\newblock Exact and parallel triangle counting in dynamic graphs.
\newblock In {\em 2017 IEEE 24th International Conference on High Performance
  Computing (HiPC)}, pages 2--12. IEEE, 2017.

\bibitem{malewicz2010pregel}
G.~Malewicz, M.~H. Austern, A.~J. Bik, J.~C. Dehnert, I.~Horn, N.~Leiser, and
  G.~Czajkowski.
\newblock Pregel: a system for large-scale graph processing.
\newblock In {\em ACM SIGMOD}, 2010.

\bibitem{mariappan2021dzig}
M.~Mariappan, J.~Che, and K.~Vora.
\newblock Dzig: sparsity-aware incremental processing of streaming graphs.
\newblock In {\em EuroSys}, 2021.

\bibitem{mariappan2019graphbolt}
M.~Mariappan and K.~Vora.
\newblock Graphbolt: Dependency-driven synchronous processing of streaming
  graphs.
\newblock In {\em ACM EuroSys}, 2019.

\bibitem{martella2015practical}
C.~Martella, R.~Shaposhnik, D.~Logothetis, and S.~Harenberg.
\newblock {\em Practical graph analytics with apache giraph}, volume~1.
\newblock Springer, 2015.

\bibitem{mccoll2013new}
R.~McColl, O.~Green, and D.~A. Bader.
\newblock A new parallel algorithm for connected components in dynamic graphs.
\newblock In {\em 20th Annual International Conference on High Performance
  Computing}, pages 246--255. IEEE, 2013.

\bibitem{mccrabb2019dredge}
A.~McCrabb, E.~Winsor, and V.~Bertacco.
\newblock Dredge: Dynamic repartitioning during dynamic graph execution.
\newblock In {\em Proceedings of the 56th Annual Design Automation Conference
  2019}, page~28. ACM, 2019.

\bibitem{mccune2015thinking}
R.~R. McCune et~al.
\newblock Thinking like a vertex: a survey of vertex-centric frameworks for
  large-scale distributed graph processing.
\newblock {\em ACM CSUR}, 2015.

\bibitem{mcgregor2014graph}
A.~McGregor.
\newblock Graph stream algorithms: a survey.
\newblock {\em ACM SIGMOD Record}, 2014.

\bibitem{mcsherry2013differential}
F.~McSherry, D.~G. Murray, R.~Isaacs, and M.~Isard.
\newblock Differential dataflow.
\newblock In {\em CIDR}, 2013.

\bibitem{miao2015immortalgraph}
Y.~Miao, W.~Han, K.~Li, M.~Wu, F.~Yang, L.~Zhou, V.~Prabhakaran, E.~Chen, and
  W.~Chen.
\newblock Immortalgraph: A system for storage and analysis of temporal graphs.
\newblock {\em ACM Transactions on Storage (TOS)}, 11(3):14, 2015.

\bibitem{michail2016introduction}
O.~Michail.
\newblock An introduction to temporal graphs: An algorithmic perspective.
\newblock {\em Internet Mathematics}, 12(4):239--280, 2016.

\bibitem{minton2014improved}
G.~T. Minton and E.~Price.
\newblock Improved concentration bounds for count-sketch.
\newblock In {\em ACM-SIAM SODA}, 2014.

\bibitem{moffitt2017temporal}
V.~Z. Moffitt and J.~Stoyanovich.
\newblock Temporal graph algebra.
\newblock In {\em Proceedings of The 16th International Symposium on Database
  Programming Languages}, pages 1--12, 2017.

\bibitem{moffitt2017towards}
V.~Z. Moffitt and J.~Stoyanovich.
\newblock Towards sequenced semantics for evolving graphs.
\newblock In {\em EDBT}, pages 446--449, 2017.

\bibitem{mondal2012managing}
J.~Mondal and A.~Deshpande.
\newblock Managing large dynamic graphs efficiently.
\newblock In {\em ACM SIGMOD}, 2012.

\bibitem{mondal2014eagr}
J.~Mondal and A.~Deshpande.
\newblock Eagr: Supporting continuous ego-centric aggregate queries over large
  dynamic graphs.
\newblock In {\em Proceedings of the 2014 ACM SIGMOD International Conference
  on Management of data}, pages 1335--1346. ACM, 2014.

\bibitem{mondal2016casqd}
J.~Mondal and A.~Deshpande.
\newblock Casqd: continuous detection of activity-based subgraph pattern
  queries on dynamic graphs.
\newblock In {\em Proceedings of the 10th ACM International Conference on
  Distributed and Event-based Systems}, pages 226--237. ACM, 2016.

\bibitem{murray2016incremental}
D.~G. Murray, F.~McSherry, M.~Isard, R.~Isaacs, P.~Barham, and M.~Abadi.
\newblock Incremental, iterative data processing with timely dataflow.
\newblock {\em Communications of the ACM}, 59(10):75--83, 2016.

\bibitem{muthukrishnan2005data}
S.~Muthukrishnan et~al.
\newblock Data streams: Algorithms and applications.
\newblock {\em Foundations and Trends{\textregistered} in Theoretical Computer
  Science}, 2005.

\bibitem{nai2017graphpim}
L.~Nai, R.~Hadidi, J.~Sim, H.~Kim, P.~Kumar, and H.~Kim.
\newblock Graphpim: Enabling instruction-level pim offloading in graph
  computing frameworks.
\newblock In {\em IEEE HPCA}, 2017.

\bibitem{namaki2017event}
M.~H. Namaki, P.~Lin, and Y.~Wu.
\newblock Event pattern discovery by keywords in graph streams.
\newblock In {\em 2017 IEEE International Conference on Big Data (Big Data)},
  pages 982--987. IEEE, 2017.

\bibitem{neuendorffer2008streaming}
S.~Neuendorffer and K.~Vissers.
\newblock {Streaming systems in FPGAs}.
\newblock In {\em Intl. Workshop on Embedded Computer Systems}, pages 147--156.
  Springer, 2008.

\bibitem{nicoara2015hermes}
D.~Nicoara, S.~Kamali, K.~Daudjee, and L.~Chen.
\newblock Hermes: Dynamic partitioning for distributed social network graph
  databases.
\newblock In {\em EDBT}, pages 25--36, 2015.

\bibitem{o2009survey}
T.~C. O’connell.
\newblock A survey of graph algorithms under extended streaming models of
  computation.
\newblock In {\em Fundamental Problems in Computing}, pages 455--476. Springer,
  2009.

\bibitem{pacaci2019experimental}
A.~Pacaci and M.~T. {\"O}zsu.
\newblock Experimental analysis of streaming algorithms for graph partitioning.
\newblock In {\em Proceedings of the 2019 International Conference on
  Management of Data}, pages 1375--1392. ACM, 2019.

\bibitem{PengS18}
P.~Peng and C.~Sohler.
\newblock Estimating graph parameters from random order streams.
\newblock 2018.

\bibitem{perez2009semantics}
J.~P{\'e}rez, M.~Arenas, and C.~Gutierrez.
\newblock Semantics and complexity of sparql.
\newblock {\em ACM TODS}, 34(3):16, 2009.

\bibitem{petroni2015hdrf}
F.~Petroni, L.~Querzoni, K.~Daudjee, S.~Kamali, and G.~Iacoboni.
\newblock {HDRF: Stream-based partitioning for power-law graphs}.
\newblock In {\em Proceedings of the 24th ACM International on Conference on
  Information and Knowledge Management}, pages 243--252. ACM, 2015.

\bibitem{pitoura2017historical}
E.~Pitoura.
\newblock Historical graphs: models, storage, processing.
\newblock In {\em European Business Intelligence and Big Data Summer School},
  pages 84--111. Springer, 2017.

\bibitem{prabhakaran2012managing}
V.~Prabhakaran, M.~Wu, X.~Weng, F.~McSherry, L.~Zhou, and M.~Haradasan.
\newblock Managing large graphs on multi-cores with graph awareness.
\newblock In {\em USENIX ATC}, 2012.

\bibitem{qin2019mining}
H.~Qin, R.-H. Li, G.~Wang, L.~Qin, Y.~Cheng, and Y.~Yuan.
\newblock Mining periodic cliques in temporal networks.
\newblock In {\em ICDE}, pages 1130--1141. IEEE, 2019.

\bibitem{ren2011querying}
C.~Ren, E.~Lo, B.~Kao, X.~Zhu, and R.~Cheng.
\newblock On querying historical evolving graph sequences.
\newblock {\em Proceedings of the VLDB Endowment}, 4(11):726--737, 2011.

\bibitem{riedy2013multithreaded}
J.~Riedy and D.~A. Bader.
\newblock Multithreaded community monitoring for massive streaming graph data.
\newblock In {\em 2013 IEEE International Symposium on Parallel \& Distributed
  Processing, Workshops and Phd Forum}, pages 1646--1655. IEEE, 2013.

\bibitem{neo4j_book}
I.~Robinson, J.~Webber, and E.~Eifrem.
\newblock Graph database internals.
\newblock In {\em Graph Databases, Second Edition}, chapter~7, pages 149--170.
  O'Relly, 2015.

\bibitem{rodriguez2015gremlin}
M.~A. Rodriguez.
\newblock The gremlin graph traversal machine and language (invited talk).
\newblock In {\em ACM DBPL}, 2015.

\bibitem{roy2015chaos}
A.~Roy, L.~Bindschaedler, J.~Malicevic, and W.~Zwaenepoel.
\newblock Chaos: Scale-out graph processing from secondary storage.
\newblock In {\em Proceedings of the 25th Symposium on Operating Systems
  Principles}, pages 410--424, 2015.

\bibitem{roy2013x}
A.~Roy, I.~Mihailovic, and W.~Zwaenepoel.
\newblock X-stream: Edge-centric graph processing using streaming partitions.
\newblock In {\em ACM SOSP}, 2013.

\bibitem{sallinen2019incremental}
S.~Sallinen, R.~Pearce, and M.~Ripeanu.
\newblock Incremental graph processing for on-line analytics.
\newblock In {\em 2019 IEEE International Parallel and Distributed Processing
  Symposium (IPDPS)}, pages 1007--1018. IEEE, 2019.

\bibitem{schmid2016high}
P.~Schmid, M.~Besta, and T.~Hoefler.
\newblock High-performance distributed rma locks.
\newblock In {\em ACM HPDC}, pages 19--30, 2016.

\bibitem{schweizer2015evaluating}
H.~Schweizer, M.~Besta, and T.~Hoefler.
\newblock Evaluating the cost of atomic operations on modern architectures.
\newblock In {\em IEEE PACT}, pages 445--456, 2015.

\bibitem{semertzidis2016time}
K.~Semertzidis and E.~Pitoura.
\newblock Time traveling in graphs using a graph database.
\newblock In {\em EDBT/ICDT Workshops}, 2016.

\bibitem{sengupta2017evograph}
D.~Sengupta and S.~L. Song.
\newblock Evograph: On-the-fly efficient mining of evolving graphs on gpu.
\newblock In {\em International Supercomputing Conference}, pages 97--119.
  Springer, 2017.

\bibitem{sengupta2016graphin}
D.~Sengupta, N.~Sundaram, X.~Zhu, T.~L. Willke, J.~Young, M.~Wolf, and
  K.~Schwan.
\newblock Graphin: An online high performance incremental graph processing
  framework.
\newblock In {\em European Conference on Parallel Processing}, pages 319--333.
  Springer, 2016.

\bibitem{sha2017accelerating}
M.~Sha, Y.~Li, B.~He, and K.-L. Tan.
\newblock Accelerating dynamic graph analytics on gpus.
\newblock {\em Proceedings of the VLDB Endowment}, 11(1):107--120, 2017.

\bibitem{shao2013trinity}
B.~Shao, H.~Wang, and Y.~Li.
\newblock Trinity: A distributed graph engine on a memory cloud.
\newblock In {\em ACM SIGMOD}, 2013.

\bibitem{shao2018efficient}
M.~Shao, J.~Li, F.~Chen, and X.~Chen.
\newblock An efficient framework for detecting evolving anomalous subgraphs in
  dynamic networks.
\newblock In {\em IEEE INFOCOM 2018-IEEE Conference on Computer
  Communications}, pages 2258--2266. IEEE, 2018.

\bibitem{sheng2018grapu}
F.~Sheng, Q.~Cao, H.~Cai, J.~Yao, and C.~Xie.
\newblock Grapu: Accelerate streaming graph analysis through preprocessing
  buffered updates.
\newblock In {\em ACM SoCC}, 2018.

\bibitem{sheng2020exploiting}
F.~Sheng, Q.~Cao, and J.~Yao.
\newblock Exploiting buffered updates for fast streaming graph analysis.
\newblock {\em IEEE Transactions on Computers}, 2020.

\bibitem{shi2016tornado}
X.~Shi, B.~Cui, Y.~Shao, and Y.~Tong.
\newblock Tornado: A system for real-time iterative analysis over evolving
  data.
\newblock In {\em ACM SIGMOD}, 2016.

\bibitem{shi2018graph}
X.~Shi, Z.~Zheng, Y.~Zhou, H.~Jin, L.~He, B.~Liu, and Q.-S. Hua.
\newblock Graph processing on gpus: A survey.
\newblock {\em ACM Computing Surveys (CSUR)}, 50(6):81, 2018.

\bibitem{shi2017partitioning}
Z.~Shi, J.~Li, P.~Guo, S.~Li, D.~Feng, and Y.~Su.
\newblock Partitioning dynamic graph asynchronously with distributed fennel.
\newblock {\em Future Generation Computer Systems}, 71:32--42, 2017.

\bibitem{shun2013ligra}
J.~Shun and G.~E. Blelloch.
\newblock Ligra: a lightweight graph processing framework for shared memory.
\newblock In {\em ACM Sigplan Notices}, volume~48, pages 135--146. ACM, 2013.

\bibitem{SimsiriTTW16}
N.~Simsiri, K.~Tangwongsan, S.~Tirthapura, and K.~Wu.
\newblock Work-efficient parallel union-find with applications to incremental
  graph connectivity.
\newblock In {\em Euro-Par}, pages 561--573, 2016.

\bibitem{solomonik2017scaling}
E.~Solomonik, M.~Besta, F.~Vella, and T.~Hoefler.
\newblock Scaling betweenness centrality using communication-efficient sparse
  matrix multiplication.
\newblock In {\em ACM/IEEE Supercomputing}, page~47, 2017.

\bibitem{srinivasan2018shared}
S.~Srinivasan, S.~Riazi, B.~Norris, S.~K. Das, and S.~Bhowmick.
\newblock A shared-memory parallel algorithm for updating single-source
  shortest paths in large dynamic networks.
\newblock In {\em HiPC}, pages 245--254. IEEE, 2018.

\bibitem{steinbauer2016dynamograph}
M.~Steinbauer and G.~Anderst-Kotsis.
\newblock Dynamograph: a distributed system for large-scale, temporal graph
  processing, its implementation and first observations.
\newblock In {\em Proceedings of the 25th International Conference Companion on
  World Wide Web}, pages 861--866, 2016.

\bibitem{sun2007graphscope}
J.~Sun, C.~Faloutsos, S.~Papadimitriou, and P.~S. Yu.
\newblock Graphscope: parameter-free mining of large time-evolving graphs.
\newblock In {\em Proceedings of the 13th ACM SIGKDD international conference
  on Knowledge discovery and data mining}, pages 687--696, 2007.

\bibitem{sun2017join}
X.~Sun, Y.~Tan, Q.~Wu, and J.~Wang.
\newblock A join-cache tree based approach for continuous temporal pattern
  detection in streaming graph.
\newblock In {\em ICSPCC}, pages 1--6. IEEE, 2017.

\bibitem{suzumura2014towards}
T.~Suzumura, S.~Nishii, and M.~Ganse.
\newblock Towards large-scale graph stream processing platform.
\newblock In {\em ACM WWW}, pages 1321--1326, 2014.

\bibitem{early_ldbc_paper}
G.~Sz\'{a}rnyas, A.~Prat-P{\'e}rez, A.~Averbuch, J.~Marton, M.~Paradies,
  M.~Kaufmann, O.~Erling, P.~Boncz, V.~Haprian, and J.~B. Antal.
\newblock {An Early Look at the LDBC Social Network Benchmark's Business
  Intelligence Workload}.
\newblock GRADES-NDA, pages 9:1--9:11, 2018.

\bibitem{tate2014programming}
A.~Tate et~al.
\newblock Programming abstractions for data locality.
\newblock PADAL Workshop 2014.

\bibitem{then2017automatic}
M.~Then, T.~Kersten, S.~G{\"u}nnemann, A.~Kemper, and T.~Neumann.
\newblock Automatic algorithm transformation for efficient multi-snapshot
  analytics on temporal graphs.
\newblock {\em Proceedings of the VLDB Endowment}, 10(8):877--888, 2017.

\bibitem{tripathy2018scaling}
A.~Tripathy and O.~Green.
\newblock Scaling betweenness centrality in dynamic graphs.
\newblock In {\em 2018 IEEE High Performance extreme Computing Conference
  (HPEC)}, pages 1--7. IEEE, 2018.

\bibitem{tseng2019batch}
T.~Tseng et~al.
\newblock Batch-parallel euler tour trees.
\newblock In {\em SIAM ALENEX}, 2019.

\bibitem{valiant1990bridging}
L.~G. Valiant.
\newblock A bridging model for parallel computation.
\newblock {\em Communications of the ACM}, 33(8):103--111, 1990.

\bibitem{danupon19}
J.~van~den Brand and D.~Nanongkai.
\newblock Dynamic approximate shortest paths and beyond: Subquadratic and
  worst-case update time.
\newblock {\em FOCS}, 2019.

\bibitem{van2018scalable}
A.~van~der Grinten, E.~Bergamini, O.~Green, D.~A. Bader, and H.~Meyerhenke.
\newblock Scalable katz ranking computation in large static and dynamic graphs.
\newblock {\em arXiv preprint arXiv:1807.03847}, 2018.

\bibitem{vaquero2014systems}
L.~M. Vaquero, F.~Cuadrado, and M.~Ripeanu.
\newblock Systems for near real-time analysis of large-scale dynamic graphs.
\newblock {\em arXiv:1410.1903}, 2014.

\bibitem{vora2014aspire}
K.~Vora et~al.
\newblock Aspire: exploiting asynchronous parallelism in iterative algorithms
  using a relaxed consistency based dsm.
\newblock {\em ACM SIGPLAN Notices}, pages 861--878, 2014.

\bibitem{vora2017kickstarter}
K.~Vora et~al.
\newblock Kickstarter: Fast and accurate computations on streaming graphs via
  trimmed approximations.
\newblock {\em ACM SIGOPS Operating Systems Review}, 2017.

\bibitem{vora2016synergistic}
K.~Vora, R.~Gupta, and G.~Xu.
\newblock Synergistic analysis of evolving graphs.
\newblock {\em ACM TACO}, 13(4):32, 2016.

\bibitem{wang2018rstream}
K.~Wang, Z.~Zuo, J.~Thorpe, T.~Q. Nguyen, and G.~H. Xu.
\newblock Rstream: marrying relational algebra with streaming for efficient
  graph mining on a single machine.
\newblock In {\em USENIX OSDI}, pages 763--782, 2018.

\bibitem{winter2017autonomous}
M.~Winter et~al.
\newblock Autonomous, independent management of dynamic graphs on gpus.
\newblock In {\em IEEE HPEC}, 2017.

\bibitem{winter2018faimgraph}
M.~Winter, D.~Mlakar, R.~Zayer, H.-P. Seidel, and M.~Steinberger.
\newblock faimgraph: high performance management of fully-dynamic graphs under
  tight memory constraints on the gpu.
\newblock In {\em ACM/IEEE Supercomputing}, 2018.

\bibitem{wu2014path}
H.~Wu, J.~Cheng, S.~Huang, Y.~Ke, Y.~Lu, and Y.~Xu.
\newblock Path problems in temporal graphs.
\newblock {\em VLDB}, 2014.

\bibitem{wu2016reachability}
H.~Wu, Y.~Huang, J.~Cheng, J.~Li, and Y.~Ke.
\newblock Reachability and time-based path queries in temporal graphs.
\newblock In {\em IEEE ICDE}, 2016.

\bibitem{wu2015gram}
M.~Wu, F.~Yang, J.~Xue, W.~Xiao, Y.~Miao, L.~Wei, H.~Lin, Y.~Dai, and L.~Zhou.
\newblock Gram: Scaling graph computation to the trillions.
\newblock In {\em Proceedings of the Sixth ACM Symposium on Cloud Computing},
  pages 408--421, 2015.

\bibitem{xiangyu2020efficient}
L.~Xiangyu, L.~Yingxiao, G.~Xiaolin, and Y.~Zhenhua.
\newblock An efficient snapshot strategy for dynamic graph storage systems to
  support historical queries.
\newblock {\em IEEE Access}, 8:90838--90846, 2020.

\bibitem{xie2015dynamic}
W.~Xie, Y.~Tian, Y.~Sismanis, A.~Balmin, and P.~J. Haas.
\newblock Dynamic interaction graphs with probabilistic edge decay.
\newblock In {\em IEEE ICDE}, pages 1143--1154, 2015.

\bibitem{yang2012towards}
S.~Yang, X.~Yan, B.~Zong, and A.~Khan.
\newblock Towards effective partition management for large graphs.
\newblock In {\em Proceedings of the 2012 ACM SIGMOD International Conference
  on Management of Data}, pages 517--528. ACM, 2012.

\bibitem{zaharia2012resilient}
M.~Zaharia, M.~Chowdhury, T.~Das, A.~Dave, J.~Ma, M.~McCauly, M.~J. Franklin,
  S.~Shenker, and I.~Stoica.
\newblock Resilient distributed datasets: A fault-tolerant abstraction for
  in-memory cluster computing.
\newblock In {\em USENIX NSDI}, 2012.

\bibitem{zaharia2016apache}
M.~Zaharia, R.~S. Xin, P.~Wendell, T.~Das, M.~Armbrust, A.~Dave, X.~Meng,
  J.~Rosen, S.~Venkataraman, M.~J. Franklin, et~al.
\newblock Apache spark: a unified engine for big data processing.
\newblock {\em Communications of the ACM}, 59(11):56--65, 2016.

\bibitem{zaki2016comprehensive}
A.~Zaki, M.~Attia, D.~Hegazy, and S.~Amin.
\newblock Comprehensive survey on dynamic graph models.
\newblock {\em International Journal of Advanced Computer Science and
  Applications}, 7(2):573--582, 2016.

\bibitem{zhang2010survey}
J.~Zhang.
\newblock A survey on streaming algorithms for massive graphs.
\newblock {\em Managing and Mining Graph Data}, pages 393--420, 2010.

\bibitem{zhang2017sub}
Y.~Zhang, R.~Chen, and H.~Chen.
\newblock Sub-millisecond stateful stream querying over fast-evolving linked
  data.
\newblock In {\em Proceedings of the 26th Symposium on Operating Systems
  Principles}, pages 614--630. ACM, 2017.

\bibitem{zhou2018fpga}
S.~Zhou, R.~Kannan, H.~Zeng, and V.~K. Prasanna.
\newblock An fpga framework for edge-centric graph processing.
\newblock In {\em Proceedings of the 15th ACM International Conference on
  Computing Frontiers}, pages 69--77. ACM, 2018.

\bibitem{zhu2019livegraph}
X.~Zhu, G.~Feng, M.~Serafini, X.~Ma, J.~Yu, L.~Xie, A.~Aboulnaga, and W.~Chen.
\newblock Livegraph: A transactional graph storage system with purely
  sequential adjacency list scans.
\newblock {\em arXiv preprint arXiv:1910.05773}, 2019.

\end{thebibliography}
}
\fi

\end{document}